\documentclass[twocolumn,tighten]{aastex631}
\usepackage{graphicx,natbib}

\shorttitle{$r$-process in M92}
\shortauthors{Kirby et al.}

\newcommand{\nstars}{35}
\newcommand{\nagb}{5}
\newcommand{\nmsto}{6}
\newcommand{\nrgb}{24}
\newcommand{\ngiant}{29}

\newcommand{\citeposs}[1]{\citeauthor{#1}'s (\citeyear{#1})}

\newcommand{\teff}{$T_{\rm eff}$}
\newcommand{\logg}{$\log g$}

\newcommand{\linemake}{\texttt{Linemake}}
\newcommand{\moog}{\texttt{MOOG}}
\newcommand{\atlas}{\texttt{ATLAS9}}
\newcommand{\leopy}{\texttt{LEO-Py}}
\newcommand{\mpfit}{\texttt{MPFIT}}

\begin{document}

\title{$r$-process Abundance Patterns in the Globular Cluster M92\footnote{The
    data presented herein were obtained at the W.~M.~Keck Observatory,
    which is operated as a scientific partnership among the California
    Institute of Technology, the University of California and the
    National Aeronautics and Space Administration. The Observatory was
    made possible by the generous financial support of the W.~M.~Keck
    Foundation.}}

\correspondingauthor{Evan N.\ Kirby}
\email{ekirby@nd.edu}

\author[0000-0001-6196-5162]{Evan N.\ Kirby}
\affiliation{Department of Physics and Astronomy, University of Notre Dame, 225 Nieuwland Science Hall, Notre Dame, IN 46556, USA}

\author[0000-0002-4863-8842]{Alexander P.\ Ji}
\affiliation{Department of Astronomy \& Astrophysics, University of Chicago, 5640 S Ellis Avenue, Chicago, IL 60637, USA}
\affiliation{Kavli Institute for Cosmological Physics, University of Chicago, Chicago, IL 60637, USA}

\author[0000-0002-9975-7833]{Mikhail Kovalev}
\affiliation{Yunnan Observatories, China Academy of Sciences, Kunming 650216, China} \affiliation{Key Laboratory for the Structure and Evolution of Celestial Objects, Chinese Academy of Sciences, Kunming 650011, China}


\begin{abstract}
Whereas light element abundance variations are a hallmark of globular clusters, there is little evidence for variation in neutron-capture elements.  A significant exception is M15, which shows a star-to-star dispersion in neutron-capture abundances of at least one order of magnitude.  The literature contains evidence both for and against a neutron-capture dispersion in M92.  We conducted an analysis of archival Keck/HIRES spectra of \nstars\ stars in M92, \ngiant\ of which are giants, which we use exclusively for our conclusions.  M92 conforms to the light element abundance variations typical of massive clusters.  Like other globular clusters, its neutron-capture abundances were generated by the $r$-process.  We confirm a star-to-star dispersion in the $r$-process.  Unlike M15, the dispersion is limited to ``first-generation'' (low Na, high Mg) stars, and the dispersion is smaller for Sr, Y, and Zr than for Ba and the lanthanides.  This is the first detection of a relation between light element and neutron-capture abundances in a globular cluster.  We propose that a source of the main $r$-process polluted the cluster shortly before or concurrently with the first generation of star formation.  The heavier $r$-process abundances were inhomogeneously distributed while the first-generation stars were forming.  The second-generation stars formed after several crossing times ($\sim 0.8$~Myr); hence, the second generation shows no $r$-process dispersion.  This scenario imposes a minimum temporal separation of 0.8~Myr between the first and second generations.
\end{abstract}


\section{Introduction}
\label{sec:intro}

Globular clusters (GCs) were once thought to be the exemplars of single stellar populations, in which all the stars had the same age and elemental composition.  Cracks in this perception appeared in the 1970s.  \citet{kra79} discussed the growing photometric and spectroscopic evidence for large star-to-star variations in the abundances of carbon and nitrogen within individual clusters.  At the time, there was a debate on the origin of the light element variations.  Were they a sign of internal mixing or primordial pollution?  \citeposs{pet80} discovery of sodium variations in M13 marked a paradigm shift in the theory of the chemical evolution of GCs.  Sodium variations were thought to be evidence of primordial pollution because sodium would not be manufactured by internal processes in the low-mass stars that are observable today.  \citeauthor{pet80}\ further saw evidence for anti-correlation between sodium and oxygen abundances in M13.  Subsequent spectroscopic observations of other GCs confirmed the anti-correlation between sodium and cyanogen \citep{cot81} and between sodium and oxygen \citep{kra92}.

Due to concerted observations of dozens of clusters \citep[e.g.,][]{car19,mas19}, we know that chemical abundance inhomogeneities are ubiquitous in globular clusters.  (See \citealt{mil22} for a review.)  In fact, \citet{car10} suggested that a GC be defined as an object that exhibits the Na--O anti-correlation.  Many clusters exhibit other element abundance variations, such as the Mg--Al anti-correlation \citep{kra97}.  Some clusters even show variations in the abundance of K \citep{coh11a,coh12,muc12}.  With just a few exceptions, K is the heaviest element that typically shows any variation.  Iron and iron-peak elements have immeasurably low dispersion in most GCs \citep[e.g.,][]{wil12}.  The major exceptions are $\omega$~Centauri, M54, and Terzan 5.  The first two are thought to be the nuclei of accreted or accreting dwarf galaxies \citep{bek03}, but Terzan~5 is more mysterious \citep[e.g.,][]{ori13}.  A few more GCs have small Fe abundance variations \citep[][among others]{mar09,mar15,sim13,joh15}.  These GCs can often be identified by a splitting of the red giant branch (RGB) in the color--magnitude diagram \citep{mil17}.

The light element abundance variations do not fit well into theories of GC formation.  High-temperature hydrogen burning neatly explains the pattern of C, N, O, Na, Mg, and Al abundances \citep{den90,lan95,cav96}.  However, no theory manages to fit this nucleosynthetic process into a framework of star formation.  It seems likely that the stars enhanced in Na and Al and depleted in O and Mg formed after the stars with ``normal'' abundance patterns.  However, each candidate source for high-temperature hydrogen burning, such as asymptotic giant branch (AGB) stars or fast-rotating massive stars, fails to explain some observable property of GCs \citep{bas18}.

An even rarer phenomenon is variation in neutron-capture elements.  The prototypical GC for this phenomenon is M15.  \citet{sne97} discovered a strong correlation between Ba and Eu abundances in M15.  The presence of a correlation implies that that both elements vary from star to star.  Furthermore, they were probably generated in the same nucleosynthetic event or events.  The [Eu/Ba] ratio indicated that the elements were created in the $r$-process rather than the $s$-process.  This discovery added new complexity to the study of GCs because $s$-process variation might be explained by AGB stars, a potential site of high-temperature hydrogen burning that could explain the light element abundance variations.  Instead, there must be an additional nucleosynthetic site to explain the $r$-process abundance variations.  \citet{sne00a}, \citet{ots06}, \citet{sob11}, and \citet{wor13} confirmed the $r$-process variation in M15.

The neutron-capture abundances in M15 do not have any apparent connection to the light element abundances.  In other words, stars with enhanced Na or depleted Mg do not preferentially show high or low abundances of the $r$-process elements.  This observation further rules out a direct connection between the source of the light elements and the source of the $r$-process.  Furthermore, it means that the abundance variations in the $r$-process and/or the light elements are not the result of temporal evolution.  For example, consider that Na became enhanced and Mg became depleted over time in M15.  If the abundances of Ba and Eu also grew over time, they should show a correlation with Na and an anti-correlation with Mg.  However, no such patterns are observed.

The difficulty in explaining the abundance trends in M15 is reflected in the dearth of models to describe it.  \citet{tsu14} proposed that a neutron star merger polluted M15 long after the cluster finished forming stars.  The stars closest to the neutron star merger became polluted with the $r$-process, whereas those farther away received a lower ``dose.''  This theory explains why the $r$-process variations do not correlate with the light element abundances.  The theory also presented an observational test.  Stars that evolve to the present-day RGB would undergo the first dredge-up, which would deplete the surface abundances of externally polluted material.  \citet{kir20} showed that stars on the main sequence have the same average Ba abundance as stars on the RGB, which ruled out an external pollution scenario.  \citet{tar21} proposed instead that a lanthanide-rich event polluted the cluster while it was still forming.  In this scenario, the event happens external to the cluster, and the stars must form quickly so that the cluster gas does not mix and homogenize before the star formation finishes.

M15 remained unique in showing $r$-process abundance variations until \citet{roe11b} reported that M92 also showed correlated variations in the abundances of Y, Zr, La, and Eu, among other neutron-capture abundances.  Their results were based on medium-resolution spectra from the WIYN/Hydra multi-object spectrograph \citep{bar94}.  Like M15, they also showed abundance ratios indicative of the $r$-process, and they did not correlate with the light element abundance variations.

It is interesting that M92 was the next cluster to show $r$-process variations because it is similar to M15 in several respects.  Both clusters have nearly the same metallicity (${\rm [Fe/H]} = -2.4$).  M15's luminosity is just one magnitude brighter than M92 \citep{van91,dur93,har96,van16}.  M15 is core-collapsed, whereas M92 has a high central concentration but falls shy of being core-collapsed \citep{tra95,mcl05}.

\begin{deluxetable*}{llcccccc}
\tablecolumns{8}
\tablewidth{0pt}
\tablecaption{Coordinates and Photometry\label{tab:phot}}
\tablehead{\colhead{Star} & \colhead{Gaia Source ID} & \colhead{RA (J2000)} & \colhead{Dec (J2000)} & \colhead{$G_0$} & \colhead{$({\rm BP}-{\rm RP})_0$} & \colhead{$({\rm BP}-K)_0$} & \colhead{$v_r$ (km~s$^{-1}$)}}
\startdata
III-13          & 1360410343884596096  &  $17^{\rm h}17^{\rm m}21\fs 71$ & $+43^{\circ}12\arcmin 53\farcs 4$ & 11.56 & 1.56                  & 3.39\tablenotemark{a} & $-110.652$ \\
VII-18          & 1360384647101112832  &  $17^{\rm h}16^{\rm m}37\fs 48$ & $+43^{\circ}06\arcmin 15\farcs 6$ & 11.68 & 1.50                  & 3.25\tablenotemark{a} & $-118.552$ \\
X-49            & 1360404399652677760  &  $17^{\rm h}17^{\rm m}12\fs 80$ & $+43^{\circ}05\arcmin 41\farcs 8$ & 11.77 & 1.47                  & 3.22\tablenotemark{a} & $-132.417$ \\
III-65          & 1360405877121517568  &  $17^{\rm h}17^{\rm m}14\fs 12$ & $+43^{\circ}10\arcmin 46\farcs 2$ & 12.01 & 1.41                  & 3.09\tablenotemark{a} & $-118.964$ \\
VII-122         & 1360405262943998208  &  $17^{\rm h}16^{\rm m}57\fs 37$ & $+43^{\circ}07\arcmin 23\farcs 7$ & 12.02 & 1.41                  & 3.09\tablenotemark{a} & $-125.579$ \\
II-53           & 1360405808402036608  &  $17^{\rm h}17^{\rm m}13\fs 06$ & $+43^{\circ}09\arcmin 48\farcs 3$ & 12.04 & 1.40                  & 3.08\tablenotemark{a} & $-124.641$ \\
XII-8*          & 1360218208524710272  &  $17^{\rm h}17^{\rm m}31\fs 71$ & $+43^{\circ}05\arcmin 41\farcs 4$ & 12.36 & 1.31                  & 2.90\tablenotemark{a} & $-119.156$ \\
V-45            & 1360408522818465536  &  $17^{\rm h}16^{\rm m}49\fs 85$ & $+43^{\circ}10\arcmin 41\farcs 2$ & 12.42 & 1.31                  & 2.91\tablenotemark{a} & $-121.735$ \\
XI-19           & 1360216765415718528  &  $17^{\rm h}17^{\rm m}18\fs 74$ & $+43^{\circ}04\arcmin 50\farcs 9$ & 12.48 & 1.30                  & 2.88\tablenotemark{a} & $-117.273$ \\
XI-80           & 1360404605811200256  &  $17^{\rm h}17^{\rm m}14\fs 67$ & $+43^{\circ}06\arcmin 24\farcs 7$ & 12.58 & 1.28                  & 2.86\tablenotemark{a} & $-125.783$ \\
II-70*          & 1360405812699772416  &  $17^{\rm h}17^{\rm m}16\fs 53$ & $+43^{\circ}10\arcmin 44\farcs 9$ & 12.70 & 1.25                  & 2.77\tablenotemark{a} & $-111.765$ \\
IV-94*          & 1360408694620020480  &  $17^{\rm h}17^{\rm m}05\fs 87$ & $+43^{\circ}10\arcmin 17\farcs 2$ & 12.71 & 1.24                  & 2.75\tablenotemark{a} & $-119.801$ \\
I-67            & 1360404811969647616  &  $17^{\rm h}17^{\rm m}21\fs 23$ & $+43^{\circ}08\arcmin 27\farcs 0$ & 12.96 & 1.21                  & 2.72\tablenotemark{a} & $-121.490$ \\
III-82          & 1360408728979771776  &  $17^{\rm h}17^{\rm m}08\fs 06$ & $+43^{\circ}10\arcmin 44\farcs 9$ & 12.96 & 1.22                  & 2.75\tablenotemark{a} & $-121.772$ \\
IV-10           & 1360409695350418176  &  $17^{\rm h}16^{\rm m}57\fs 72$ & $+43^{\circ}14\arcmin 11\farcs 4$ & 13.06 & 1.21                  & 2.73\tablenotemark{a} & $-118.071$ \\
XII-34*         & 1360404747547336448  &  $17^{\rm h}17^{\rm m}21\fs 57$ & $+43^{\circ}07\arcmin 40\farcs 8$ & 13.08 & 1.15                  & 2.58\tablenotemark{a} & $-115.400$ \\
IV-79           & 1360408385382348672  &  $17^{\rm h}17^{\rm m}00\fs 80$ & $+43^{\circ}10\arcmin 25\farcs 1$ & 13.11 & 1.20                  & 2.70\tablenotemark{a} & $-121.686$ \\
IX-13           & 1360404163432211456  &  $17^{\rm h}16^{\rm m}56\fs 12$ & $+43^{\circ}04\arcmin 07\farcs 3$ & 13.66 & 1.11                  & 2.51\tablenotemark{a} & $-126.466$ \\
VIII-24*        & 1360404953706355200  &  $17^{\rm h}16^{\rm m}50\fs 34$ & $+43^{\circ}05\arcmin 53\farcs 1$ & 13.79 & 1.03                  & 2.31\tablenotemark{a} & $-119.054$ \\
X-20            & 1360216662336507648  &  $17^{\rm h}17^{\rm m}13\fs 33$ & $+43^{\circ}04\arcmin 13\farcs 4$ & 15.30 & 0.95                  & 2.17\tablenotemark{a} & $-118.018$ \\
S2710           & 1360404330933201536  &  $17^{\rm h}17^{\rm m}15\fs 71$ & $+43^{\circ}05\arcmin 32\farcs 4$ & 15.32 & 0.95                  & 2.16\tablenotemark{a} & $-120.568$ \\
VI-90           & 1360408355320576128  &  $17^{\rm h}16^{\rm m}54\fs 21$ & $+43^{\circ}09\arcmin 21\farcs 1$ & 15.42 & 0.95                  & 2.15\tablenotemark{a} & $-126.224$ \\
S2265           & 1360216593617025152  &  $17^{\rm h}17^{\rm m}13\fs 61$ & $+43^{\circ}03\arcmin 35\farcs 7$ & 15.79 & 0.93                  & 2.09\tablenotemark{a} & $-120.836$ \\
VIII-45         & 1360405159864786304  &  $17^{\rm h}16^{\rm m}53\fs 32$ & $+43^{\circ}06\arcmin 34\farcs 5$ & 15.86 & 0.92                  & 2.12\tablenotemark{a} & $-123.652$ \\
VII-28          & 1360384750180330496  &  $17^{\rm h}16^{\rm m}40\fs 79$ & $+43^{\circ}07\arcmin 07\farcs 6$ & 15.92 & 0.91\tablenotemark{a} & 2.17                  & $-121.582$ \\
G17181\_0638    & 1360404507029158656  &  $17^{\rm h}17^{\rm m}18\fs 13$ & $+43^{\circ}06\arcmin 38\farcs 1$ & 16.32 & 0.89\tablenotemark{a} & 2.08                  & $-125.134$ \\
C17333\_0832    & 1360406053217363456  &  $17^{\rm h}17^{\rm m}33\fs 32$ & $+43^{\circ}08\arcmin 32\farcs 8$ & 16.81 & 0.87\tablenotemark{a} & 2.05                  & $-120.012$ \\
S3108           & 1360405808402045184  &  $17^{\rm h}17^{\rm m}15\fs 24$ & $+43^{\circ}09\arcmin 53\farcs 9$ & 16.83 & 0.85\tablenotemark{a} & 2.00                  & $-118.782$ \\
S652            & 1360404232151697152  &  $17^{\rm h}16^{\rm m}57\fs 89$ & $+43^{\circ}04\arcmin 40\farcs 4$ & 16.85 & 0.89\tablenotemark{a} & 2.00                  & $-127.370$ \\
S19             & 1360384750180328704  &  $17^{\rm h}16^{\rm m}41\fs 26$ & $+43^{\circ}07\arcmin 08\farcs 5$ & 17.66 & 0.75\tablenotemark{a} & \nodata               & $-130.271$ \\
D21             & 1360216215659937280  &  $17^{\rm h}16^{\rm m}58\fs 78$ & $+43^{\circ}02\arcmin 04\farcs 1$ & 17.71 & 0.71\tablenotemark{a} & 2.60                  & $-127.547$ \\
S3880           & 1360406220716655616  &  $17^{\rm h}17^{\rm m}25\fs 68$ & $+43^{\circ}08\arcmin 16\farcs 1$ & 17.79 & 0.71\tablenotemark{a} & \nodata               & $-115.078$ \\
S4038           & 1360406255076061440  &  $17^{\rm h}17^{\rm m}32\fs 99$ & $+43^{\circ}09\arcmin 21\farcs 1$ & 17.80 & 0.68\tablenotemark{a} & \nodata               & $-123.615$ \\
S61             & 1360408144861994368  &  $17^{\rm h}16^{\rm m}45\fs 28$ & $+43^{\circ}07\arcmin 00\farcs 8$ & 18.11 & 0.62\tablenotemark{a} & \nodata               & $-124.981$ \\
S162            & 1360408179227211776  &  $17^{\rm h}16^{\rm m}48\fs 11$ & $+43^{\circ}07\arcmin 35\farcs 6$ & 18.13 & 0.57\tablenotemark{a} & \nodata               & $-123.328$ \\
\enddata
\tablenotetext{*}{AGB star}
\tablenotetext{a}{This is the color used for calculating \teff.}
\end{deluxetable*}

\citet{roe11a} showed evidence that M5 and NGC~3201 also exhibit $r$-process dispersions.  He also found marginal evidence for $r$-process dispersion in M3 and M13.  That study was based on the quantification of correlations between different $r$-process abundances because any correlation between two or more elements implies that there is a dispersion among them.  \citeauthor{roe11a}'s study was based on a compilation of mostly high-resolution spectroscopic abundances from the literature.  In some GCs, including M3, M5, and M13, the spectra came from multiple, heterogeneous studies.

In a study of 12 stars in M92, \citet{coh11b} could not confirm \citeposs{roe11b} observation of $r$-process variations in M92.  \citeauthor{coh11b} used Keck/HIRES \citep{vog94} spectra.  She showed that neutron-capture absorption line strengths of stars with similar effective temperatures and surface gravities were similar to each other.  As a result, she concluded that M92 does not exhibit any star-to-star $r$-process dispersion.  She attributed her discrepant result to the higher spectral resolution and signal-to-noise ratio (S/N) of the Keck/HIRES spectra compared to \citeauthor{roe11b}'s WIYN/Hydra spectra.

\citet{roe15} also found that spurious correlations between neutron-capture abundance ratios could be introduced by errors in atmospheric parameters, especially effective temperature.  The unphysical correlations persist even when the neutron-capture abundances are normalized to iron.  For example, errors in effective temperature would cause an apparent correlation between [Eu/Fe] and [La/Fe] among a group of stars, even if the abundances of Fe, La, and Eu were constant.  \citet{roe15} suggested that earlier reports of neutron-capture dispersions in all ``classical'' globular clusters except for M15 were potentially the result of this spurious correlation.

We revisit the question of $r$-process abundance variation in M92.  We analyzed all \nstars\ stars available in the Keck/HIRES archive.  Our study uses higher-resolution spectra than those of \citet{roe11b}, and it has a larger sample size than that of \citet{coh11b}.  We describe the spectra in Section~\ref{sec:obs}.  We assign stellar parameters, like effective temperature, to the stars in Section~\ref{sec:parameters}.  Section~\ref{sec:msmts} describes the procedure by which we measure abundances, and Section~\ref{sec:trends} points out some trends in the abundances.  We propose a scenario for the $r$-process abundance variation in Section~\ref{sec:discussion}, and we summarize our study in Section~\ref{sec:summary}.


\section{Observations}
\label{sec:obs}

\subsection{Archival Spectra}
\label{sec:KOA}

Nearly all of the data from this project come from the Keck Observatory Archive (KOA)\@.  We queried the archive using its web interface on 2020 February 5.  We searched for all publicly available HIRES spectra within 30~arcmin of the center of M92.  We also queried the archive again on 2023 July 18 to confirm that no additional spectra were added since our original query.  The KOA provides both raw and extracted HIRES spectra.  We used the extracted (one-dimensional, wavelength-calibrated, sky-subtracted) spectra provided in FITS format on the archive.   

We paired each spectrum with photometry from \textit{Gaia} Data Release 3 \citep{gaiadr3} and 2MASS \citep{skr06}.  The coordinates given in the metadata from the KOA are not always precise enough to give an unambiguous match to a star in the \textit{Gaia} catalog ($G$, BP, and RP magnitudes) and 2MASS catalog ($K_s$ magnitude).  In some cases, we looked up the star in the SIMBAD astronomical database \citep{wen00} by the identifier assigned by the HIRES observer.  We used the coordinates of the star listed in SIMBAD to match to \textit{Gaia} and 2MASS\@.  \textit{Gaia} measured a significantly different proper motion for one star in the HIRES archive (S1521) than the other M92 stars.  We excluded S1521 from further analysis because we assumed that it is not a cluster member, and we verified that the other stars in our sample have proper motions consistent with cluster membership.  We also excluded S4375 because it is faint ($G=18.80$).

Table~\ref{tab:phot} lists the coordinates of the stars along with photometry.  The last column of Table~\ref{tab:phot} gives the $({\rm BP} - K_s)_0$ color, where available.  The magnitudes and photometry in the table were corrected for extinction and reddening assuming $E(B-V) = 0.0191$ \citep{sch11}.  The \textit{Gaia} magnitudes were corrected following the color--extinction relations provided by the \citet{gaia18}.  The $K_s$ magnitudes were corrected assuming $A_K/E(B-V) = 0.310$ \citep{fit99}.

Table~\ref{tab:obs} gives the observing log organized by star.  Most stars were observed on multiple dates.  The name of the principal investigator (PI) and exposure times for each exposure are given for each date.  The first row for each star includes the total exposure time and the S/N per pixel.  The S/N is calculated from the pixels between 4650~\AA\ and 4800~\AA\ in the continuum-divided spectrum.  We sigma clipped this spectral range by excluding pixels more than 3.0 standard deviations above the continuum or 0.5 standard deviations below the continuum.  The asymmetry of the sigma clipping is intended to exclude strong absorption lines from the measurement of S/N\@.  Finally, the S/N is calculated as the inverse of the median absolute deviation of the spectrum from its mean.

\begin{figure}[ht!]
\centering
\includegraphics[width=\linewidth]{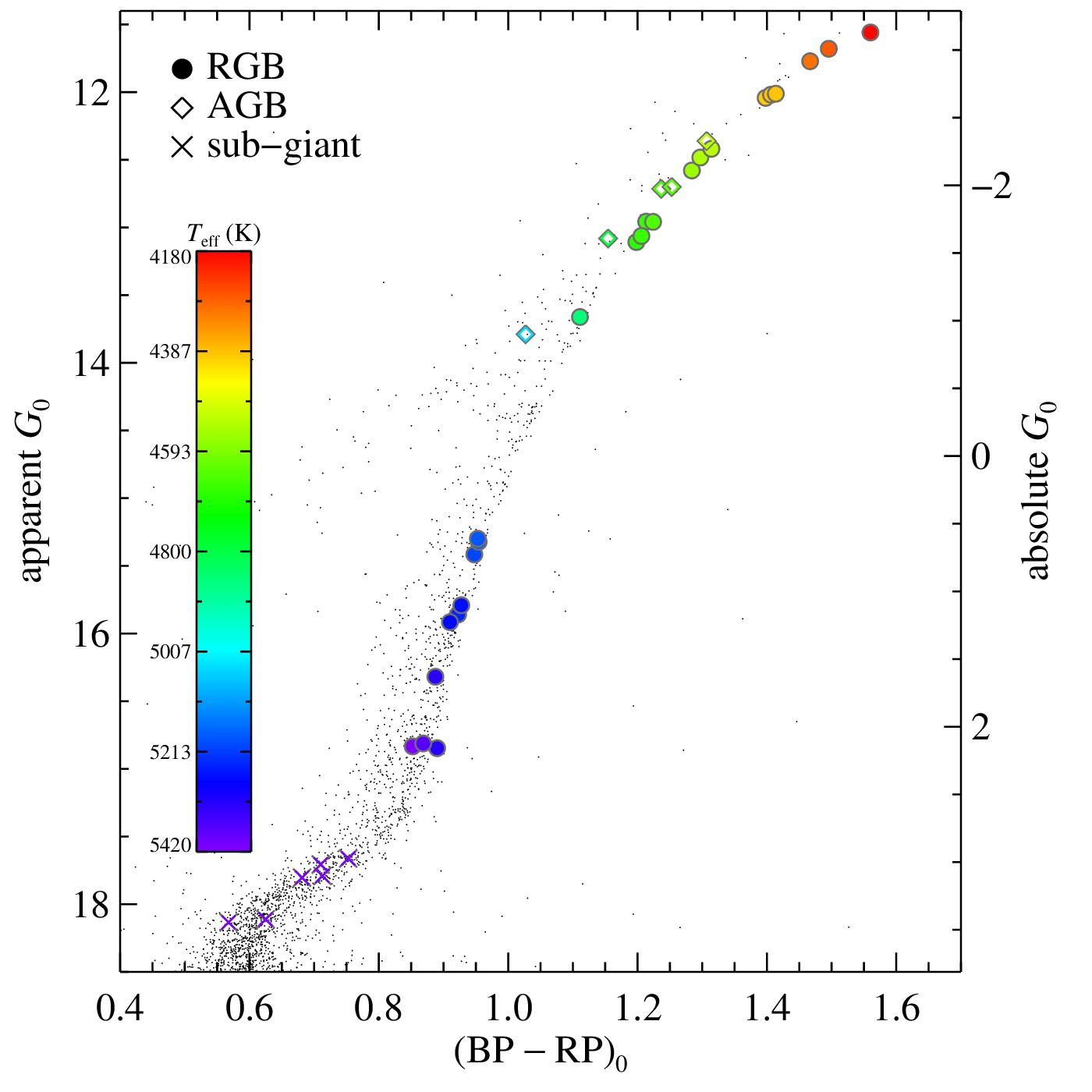}
\caption{The {\it Gaia} CMD for M92.  Colors and magnitudes are corrected for reddening and extinction.  The right axis gives the absolute $G_0$ magnitude.  The color scale gives the effective temperature (Section~\ref{sec:parameters}).  The sub-giants are all hotter than 5420~K, and they are shown with the darkest shade of purple to give the RGB more color range.  The symbol shape depicts the evolutionary state of the star, which is determined based on the location in the CMD\@.\label{fig:cmd}}
\end{figure}

Figure~\ref{fig:cmd} shows the {\it Gaia} color--magnitude diagram (CMD) for M92.  The figure shows the stars with HIRES spectra in rainbow colors.  The colors correspond to the effective temperatures (Section~\ref{sec:parameters}).  The symbol shape corresponds to evolutionary state.  We determined the evolutionary state by close inspection of the CMD\@.  A sufficiently zoomed-in CMD shows a clear delineation of the AGB from the RGB\@.  (This distinction is not readily apparent in the zoomed-out CMD shown in Figure~\ref{fig:cmd}.)  Our sample contains \nstars\ stars: \nrgb\ RGB stars, \nagb\ AGB stars, and \nmsto\ sub-giants.  While we report the abundances of the sub-giants (Section~\ref{sec:abund}), they do not play a role in our discussion (Section~\ref{sec:discussion}) because their abundances have larger uncertainties.  In the tables and figure legends in this paper, AGB stars are indicated with asterisks next to their names.  The small points in Figure~\ref{fig:cmd} show stars within 7~arcmin of the cluster center and with proper motions within 1~mas/yr of the median proper motion.
The right axis of the figure gives absolute magnitudes assuming an apparent distance modulus $(m-M)_V = 14.74$ \citep{van16}, which we corrected for extinction to $(m-M)_0 = 14.69$.

\newpage
\startlongtable
\begin{deluxetable*}{lclclll}
\tablecolumns{7}
\tablewidth{0pt}
\tablecaption{Archival Observations\label{tab:obs}}
\tablehead{\colhead{Star} & \colhead{Date} & \colhead{PI} & \colhead{Slit Width ($\arcsec$)} & \colhead{Individual Exp. (s)} & \colhead{Tot.\ Exp.\ Time (min.)} & \colhead{S/N (pix$^{-1}$)}}
\startdata
III-13          & 2005 May 29 & M.\ Bolte       & 0.861 & $3 \times 1200$                                    &  60 &  85 \\
VII-18          & 1997 May 12 & J.\ Cohen       & 0.861 & $1800$                                             &  57 &  95 \\
$\,$            & 2002 Sep 27 & J.\ Cohen       & 1.148 & $2 \times 400$                                     &     &     \\
$\,$            & 2003 Jun 24 & J.\ Cohen       & 1.148 & $400$                                              &     &     \\
$\,$            & 2006 Apr 18 & J.\ Cohen       & 1.148 & $400$                                              &     &     \\
X-49            & 2002 Sep 27 & J.\ Cohen       & 1.148 & $2 \times 400$                                     &  13 &  82 \\
III-65          & 2002 Sep 28 & J.\ Cohen       & 1.148 & $3 \times 500$                                     & 195 & 114 \\
$\,$            & 2002 Sep 29 & J.\ Cohen       & 1.148 & $500$                                              &     &     \\
$\,$            & 2005 May 30 & J.\ Cohen       & 0.861 & $3 \times 1800$                                    &     &     \\
$\,$            & 2006 Apr 17 & J.\ Cohen       & 1.148 & $400 + 3 \times 900$                               &     &     \\
$\,$            & 2006 Apr 18 & J.\ Cohen       & 1.148 & $2 \times 600$                                     &     &     \\
VII-122         & 2002 Sep 30 & J.\ Cohen       & 1.148 & $2 \times 500$                                     &  20 &  97 \\
$\,$            & 2003 Aug 21 & J.\ Cohen       & 1.148 & $200$                                              &     &     \\
II-53           & 2002 May 01 & J.\ Cohen       & 1.148 & $2 \times 600$                                     & 120 & 127 \\
$\,$            & 2002 May 03 & J.\ Cohen       & 0.861 & $3 \times 400 + 600$                               &     &     \\
$\,$            & 2002 Sep 27 & J.\ Cohen       & 1.148 & $2 \times 400$                                     &     &     \\
$\,$            & 2006 Apr 17 & J.\ Cohen       & 1.148 & $400 + 5 \times 600$                               &     &     \\
XII-8*          & 2005 May 29 & M.\ Bolte       & 0.861 & $2 \times 1800$                                    & 140 & 181 \\
$\,$            & 2005 May 30 & M.\ Bolte       & 0.861 & $1800$                                             &     &     \\
$\,$            & 2006 Apr 18 & M.\ Bolte       & 1.148 & $900$                                              &     &     \\
$\,$            & 2009 May 10 & M.\ Bolte       & 1.148 & $3 \times 300$                                     &     &     \\
$\,$            & 2011 Jun 06 & M.\ Bolte       & 1.148 & $1200$                                             &     &     \\
V-45            & 2005 May 30 & F.\ Chaffee     & 0.861 & $2 \times 1800$                                    & 165 & 180 \\
$\,$            & 2005 Jul 04 & F.\ Chaffee     & 0.861 & $3 \times 1800$                                    &     &     \\
$\,$            & 2009 May 10 & F.\ Chaffee     & 1.148 & $3 \times 300$                                     &     &     \\
XI-19           & 2002 Sep 30 & J.\ Cohen       & 1.148 & $2 \times 500$                                     & 127 & 151 \\
$\,$            & 2005 May 30 & J.\ Cohen       & 0.861 & $3 \times 1800$                                    &     &     \\
$\,$            & 2011 Jun 06 & J.\ Cohen       & 1.148 & $1200$                                             &     &     \\
XI-80           & 2002 Sep 29 & J.\ Cohen       & 1.148 & $2 \times 500$                                     &  42 & 118 \\
$\,$            & 2011 Jun 06 & J.\ Cohen       & 1.148 & $1500$                                             &     &     \\
II-70*          & 2002 May 01 & J.\ Cohen       & 1.148 & $300$                                              &  25 &  99 \\
$\,$            & 2002 Sep 27 & J.\ Cohen       & 1.148 & $2 \times 600$                                     &     &     \\
IV-94*          & 2002 Sep 27 & J.\ Cohen       & 1.148 & $2 \times 600$                                     &  37 &  98 \\
$\,$            & 2003 Jun 26 & J.\ Cohen       & 1.148 & $400$                                              &     &     \\
$\,$            & 2008 Sep 23 & J.\ Cohen       & 1.148 & $600$                                              &     &     \\
I-67            & 2002 Sep 28 & J.\ Cohen       & 1.148 & $2 \times 700$                                     &  23 & 102 \\
III-82          & 2002 Sep 28 & J.\ Cohen       & 1.148 & $2 \times 700$                                     & 100 & 152 \\
$\,$            & 2003 Jun 24 & J.\ Cohen       & 1.148 & $2 \times 500$                                     &     &     \\
$\,$            & 2006 Apr 18 & J.\ Cohen       & 1.148 & $4 \times 900$                                     &     &     \\
IV-10           & 2011 Aug 07 & J.\ Cohen       & 1.148 & $500 + 1200$                                       &  28 &  71 \\
XII-34*         & 2003 Jun 25 & J.\ Cohen       & 1.148 & $2 \times 600$                                     &  53 & 141 \\
$\,$            & 2011 Aug 06 & J.\ Cohen       & 1.148 & $800 + 1200$                                       &     &     \\
IV-79           & 2011 Aug 07 & J.\ Cohen       & 1.148 & $2 \times 1200$                                    &  40 & 112 \\
IX-13           & 2002 Sep 29 & J.\ Cohen       & 1.148 & $4 \times 500$                                     &  33 &  88 \\
VIII-24*        & 2002 Sep 30 & J.\ Cohen       & 1.148 & $5 \times 600$                                     &  50 & 106 \\
X-20            & 2008 Jun 15 & J.\ Cohen       & 1.148 & $1000$                                             &  17 &  47 \\
S2710           & 2008 Aug 20 & J.\ Cohen       & 1.148 & $3 \times 1500$                                    &  75 &  71 \\
VI-90           & 2008 Aug 20 & J.\ Cohen       & 1.148 & $2 \times 1500$                                    &  50 &  74 \\
S2265           & 2008 Aug 20 & J.\ Cohen       & 1.148 & $3 \times 1800$                                    &  90 &  76 \\
VIII-45         & 2008 Jul 07 & J.\ Cohen       & 1.148 & $3 \times 1800$                                    &  90 &  85 \\
VII-28          & 2008 Jun 10 & J.\ Cohen       & 1.148 & $2 \times 1800$                                    &  90 &  79 \\
$\,$            & 2008 Jun 11 & J.\ Cohen       & 1.148 & $1800$                                             &     &     \\
G17181\_0638    & 2003 Jun 25 & J.\ Cohen       & 1.148 & $2 \times 1200$                                    & 120 &  50 \\
$\,$            & 2003 Jun 26 & J.\ Cohen       & 1.148 & $4 \times 1200$                                    &     &     \\
C17333\_0832    & 2003 Jun 26 & J.\ Cohen       & 1.148 & $2 \times 1200$                                    & 300 &  69 \\
$\,$            & 2003 Aug 23 & J.\ Cohen       & 1.148 & $6 \times 1200$                                    &     &     \\
$\,$            & 2009 Aug 27 & J.\ Cohen       & 1.148 & $1200 + 4 \times 1800$                             &     &     \\
S3108           & 2008 Sep 21 & J.\ Cohen       & 1.148 & $2 \times 1800$                                    & 150 &  57 \\
$\,$            & 2008 Sep 22 & J.\ Cohen       & 1.148 & $2 \times 1800$                                    &     &     \\
$\,$            & 2008 Sep 23 & J.\ Cohen       & 1.148 & $1800$                                             &     &     \\
S652            & 2008 Jun 15 & J.\ Cohen       & 1.148 & $4 \times 1800$                                    & 150 &  52 \\
$\,$            & 2008 Jul 06 & J.\ Cohen       & 1.148 & $1800$                                             &     &     \\
S19             & 2008 Jun 10 & J.\ Cohen       & 1.148 & $7 \times 1800$                                    & 210 &  51 \\
D21             & 2007 Sep 06 & J.\ Cohen       & 1.148 & $2 \times 1800$                                    & 210 &  52 \\
$\,$            & 2008 Jun 10 & J.\ Cohen       & 1.148 & $5 \times 1800$                                    &     &     \\
S3880           & 2008 Jun 11 & J.\ Cohen       & 1.148 & $7 \times 1800$                                    & 210 &  48 \\
S4038           & 2008 Jun 11 & J.\ Cohen       & 1.148 & $4 \times 1800$                                    & 240 &  56 \\
$\,$            & 2008 Jun 14 & J.\ Cohen       & 1.148 & $4 \times 1800$                                    &     &     \\
S61             & 2008 Jul 06 & J.\ Cohen       & 1.148 & $9 \times 1800$                                    & 577 &  65 \\
$\,$            & 2011 Jun 04 & J.\ Cohen       & 1.148 & $6 \times 1800 + 2200$                             &     &     \\
$\,$            & 2011 Jun 06 & J.\ Cohen       & 1.148 & $3 \times 1800$                                    &     &     \\
S162            & 2008 Jul 07 & J.\ Cohen       & 1.148 & $8 \times 1800$                                    & 448 &  58 \\
$\,$            & 2011 Jun 04 & J.\ Cohen       & 1.148 & $1652 + 2 \times 1800$                             &     &     \\
$\,$            & 2011 Jun 06 & J.\ Cohen       & 1.148 & $4 \times 1800$                                    &     &     \\
\enddata
\end{deluxetable*}
\vspace{18pt}
\clearpage

\subsection{New Spectrum}
\label{sec:CPS}

One star, X-20, drew our attention for what appeared to be an unusually large potassium abundance (see Section~\ref{sec:K}).  We asked members of the California Planet Survey \citep{how10} to observe X-20 in early 2021.  A.\ Howard and H.\ Isaacson observed the star on 2021 March 26.  They reduced the spectrum following the procedure described by \citeauthor{how10}

\subsection{Preparation of the Spectra}
\label{sec:prep}

The KOA provides extracted spectra for each echelle order of each exposure.  The spectra are in the heliocentric reference frame.  Most stars had multiple exposures.  In order to make the spectra suitable for measuring equivalent widths (EWs; Section~\ref{sec:ew}), we combined all of the echelle orders and all of the exposures for each star into a single one-dimensional spectrum.  Stacking all of the component spectra required that we re-bin them onto a common wavelength array.  We used a logarithmically spaced wavelength array that encompassed the minimum and maximum wavelengths of all of the individual exposures.  The bin size was chosen to be as fine as the finest wavelength bin in all of the individual orders.  This ensured that we did not lose any information during the resampling.

\subsubsection{Continuum normalization}
\label{sec:cont}

The KOA provides spectra in units of counts divided by the flat field.  The spectra are not flux-calibrated.  The flux of one echelle order is not necessarily continuous with the flux in adjacent orders.  Therefore, we continuum-normalized each echelle order within each exposure.

Continuum normalization of stellar spectra is notoriously sensitive to S/N and the depths of absorption lines.  One approach to normalization is to exclude absorption lines from the continuum determination by sigma clipping, but that approach is especially sensitive to S/N\@.  Instead, we modeled the absorption lines with a synthetic spectrum.

We synthesized a spectrum using the LTE code \moog\ \citep{sne73,sne12}.  We used \atlas\ model atmospheres \citep{kur93,kir11d} with the effective temperatures and surface gravities determined in Section~\ref{sec:parameters}.  We assumed a metallicity of ${\rm [M/H]} = -2.4$ and an $\alpha$ enhancement of ${\rm [\alpha/M]} = +0.3$.  The microturbulent velocity was calculated from an empirical, linear relation with surface gravity \citep{kir09}.

The line list was compiled from several sources.  We used the August 2021 version of \linemake\footnote{\url{https://github.com/vmplacco/linemake}} \citep{pla21a,pla21b} for the spectral range 3100--4100~\AA\@.  \linemake\ is a compilation of atomic and molecular lines from a variety of references, which are listed on the \linemake\ website.  We used all of the atomic and molecular lines available in \linemake.  We used the line lists of \citet{kir08a,kir15a} and \citet{esc19} for the spectral range 4100-9100~\AA\@.  These lists were based on large line databases.  Some oscillator strengths were adjusted to match the spectra of the Sun and Arcturus.

For each echelle order, we divided the observed spectrum by the synthetic spectrum.  \moog\ synthesizes spectra that are already continuum-normalized.  Therefore, the quotient is the observed spectrum with the absorption lines divided out.  We fit a spline to this quotient.  We used a breakpoint spacing of 500 pixels, and we performed symmetric $1.5\sigma$ clipping.  This spline fit is the continuum.  We divided the original observed echelle order by this continuum.

\subsubsection{Spectral coaddition}
\label{sec:stacking}

We rebinned each continuum-normalized echelle order onto the common wavelength array using the \texttt{x\_specrebin} function in J.\ X.\ Prochaska's \texttt{XIDL} software library\footnote{\url{https://github.com/profxj/xidl}}.  We coadded all of the individual echelle orders with inverse variance weighting.  We repeated this process for all of the exposures of each star.

The KOA provides vacuum wavelengths.  We converted the wavelength arrays to their air equivalents.

The slit widths for the individual exposures were either 0.861'' or 1.148'' (see Table~\ref{tab:obs}).  Some stars were observed with a combination of those two slit widths.  In principle, stacking spectra of different slit widths leads to a superposition of spectra at different spectral resolutions.  In practice, the spectral resolution depends on both the slit width and the seeing.  We did not attempt to match the resolution of the individual spectra before stacking.  Therefore, the line spread functions are not simple Gaussians.  We revisit this issue in our discussion of the measurement of EWs (Section~\ref{sec:ew}).

\subsubsection{Radial velocity determination}
\label{sec:vr}

We needed to measure the radial velocity $v_r$ of each star in order to put the spectra in the rest frame.  We chose star III-65 as a reference star because of its high S/N, large wavelength range, and cool temperature, which gives strong absorption lines.  We measured $v_r = -118.964$~km~s$^{-1}$ for III-65 by examination of the observed wavelengths of several narrow absorption lines.  Then, we cross-correlated a 200~\AA\ region of the spectrum of each star with the spectrum of III-65.  The region was 4200--4400~\AA, 4400--4600~\AA, or 4650--4850~\AA, depending on the wavelength coverage of the spectrum.  Table~\ref{tab:phot} includes the resulting values of $v_r$, which are all consistent with cluster membership.  We shifted all the spectra into their rest frames accordingly.


\section{Stellar Parameters}
\label{sec:parameters}

We measured abundances of elements and molecules through a combination of EWs (Section~\ref{sec:ew}) and spectral synthesis (Section~\ref{sec:synth}).  Both approaches require an estimation of stellar parameters (Section~\ref{sec:parameters}), such as effective temperature ($T_{\rm eff}$), surface gravity ($\log g$), microturbulent velocity ($v_t$), metallicity ([M/H]), and $\alpha$ enhancement ([$\alpha$/Fe]).  We use the notation [M/H] to refer to the abundance of all elements heavier than He, assuming that their abundances are scaled to the solar composition \citep{asp09}.  The $\alpha$ elements (O, Ne, Mg, Si, S, Ar, Ca, and Ti) are additionally scaled by [$\alpha$/Fe].

\subsection{Effective temperature}
\label{sec:teff}

Stellar parameters can be derived from photometry or spectroscopy.  For example, $T_{\rm eff}$ can be measured from the photometric color of a star or by balancing the spectroscopic excitation equilibrium of one or more atomic species.  The spectroscopic approach can be inaccurate when assuming LTE, especially for red giants \citep[e.g.,][]{fre13}, which comprise the majority of our sample.

Photometric colors can be tied to (semi-)direct measurements of temperature, which are not subject to non-LTE effects.  The infrared flux method (IRFM) is a semi-direct method of determining stellar temperatures.  It uses infrared flux and a model atmosphere to infer the star's angular diameter.  The flux and diameter together give $T_{\rm eff}$.  \citet{gon09} provided empirical calibrations between 2MASS colors and temperatures derived from the IRFM for several hundred stars.  \citet{muc21} extended these calibrations to {\it Gaia} colors and combinations of {\it Gaia} and 2MASS\@.  The $({\rm BP} - K_s)_0$ color has the highest precision ($\sigma_{T_{\rm eff}} = 52$~K) because it has longest dynamic range of the different filter combinations.  Therefore, we adopted the $({\rm BP} - K_s)_0$ color temperature for stars brighter than about $G_0 = 16$.  For fainter stars, the uncertainty in the $K_s$ magnitude becomes the dominant error term in the temperature.  Therefore, for fainter stars, we used the $({\rm BP} - {\rm RP})_0$ color temperature ($\sigma_{T_{\rm eff}} = 83$~K)\@.  We used \citeauthor{muc21}'s color--temperature relation for dwarfs for stars marked ``sub-giant'' in Figure~\ref{fig:cmd} and the relation for giants otherwise.  The relations require an assumption of metallicity, for which we used ${\rm [M/H]} = -2.41$, which is approximately the average [\ion{Fe}{2}/H] of M92 that we eventually derived.  Footnotes in Table~\ref{tab:phot} indicate which color was used to derive $T_{\rm eff}$ for each star.

We estimated the uncertainty on $T_{\rm eff}$ by applying standard (uncorrelated) error propagation to the color--temperature polynomial.  The relations also have an intrinsic scatter, which we added in quadrature to the random error.

\subsection{Surface gravity}
\label{sec:logg}

Like temperature, surface gravity can also be determined photometrically or spectroscopically.  The spectroscopic method typically attempts to equalize the abundance derived from neutral and ionized absorption lines of the same element.  However, overionization affects neutral absorption lines in an LTE analysis \citep{the99}.

For stars of known distance, it is generally preferable to use photometric estimates of surface gravity.  The surface gravity can be estimated from the Stefan--Boltzmann law in combination with the definition of surface gravity:

\begin{equation}
g = \frac{4 \pi G M \sigma_{\rm SB} T_{\rm eff}^4}{L} \label{eq:g}
\end{equation}

\noindent
We assumed $0.75~M_{\sun}$, which is appropriate for ancient, metal-poor stars that have evolved past the main sequence turn-off.  We calculated the luminosity from the extinction-corrected {\it Gaia} $G_0$ magnitude, the distance modulus $(m-M)_0 = 14.69$ (see Section~\ref{sec:KOA}), and bolometric corrections from \citet{and18}.

We estimated the uncertainty on $\log g$ using standard propagation of error.  We assumed that the uncertainty on the stellar mass was $0.1~M_{\sun}$.  We further assumed that the errors on $M$, $T_{\rm eff}$, and $G_0$ are uncorrelated.

\begin{deluxetable}{lccc}
\tablecolumns{4}
\tablewidth{0pt}
\tablecaption{Model Atmosphere Parameters\label{tab:atmpars}}
\tablehead{\colhead{Star} & \colhead{\teff~(K)} & \colhead{\logg~(cm~s$^{-1}$)} & \colhead{$v_t$~(km~s$^{-1}$)}}
\startdata
III-13          & $4187 \pm  51$ & $0.48 \pm 0.10$ & $2.95 \pm 0.21$ \\
VII-18          & $4275 \pm  50$ & $0.59 \pm 0.10$ & $2.96 \pm 0.15$ \\
X-49            & $4301 \pm  50$ & $0.64 \pm 0.10$ & $2.58 \pm 0.11$ \\
III-65          & $4392 \pm  51$ & $0.79 \pm 0.10$ & $2.56 \pm 0.10$ \\
VII-122         & $4388 \pm  51$ & $0.79 \pm 0.10$ & $2.68 \pm 0.10$ \\
II-53           & $4399 \pm  51$ & $0.81 \pm 0.10$ & $2.67 \pm 0.10$ \\
XII-8*          & $4530 \pm  51$ & $1.01 \pm 0.10$ & $2.85 \pm 0.17$ \\
V-45            & $4528 \pm  52$ & $1.03 \pm 0.10$ & $2.42 \pm 0.06$ \\
XI-19           & $4546 \pm  51$ & $1.07 \pm 0.10$ & $2.54 \pm 0.07$ \\
XI-80           & $4562 \pm  51$ & $1.12 \pm 0.10$ & $2.52 \pm 0.10$ \\
II-70*          & $4636 \pm  52$ & $1.20 \pm 0.10$ & $2.47 \pm 0.09$ \\
IV-94*          & $4652 \pm  52$ & $1.22 \pm 0.10$ & $2.62 \pm 0.10$ \\
I-67            & $4675 \pm  51$ & $1.33 \pm 0.10$ & $2.37 \pm 0.09$ \\
III-82          & $4648 \pm  52$ & $1.31 \pm 0.10$ & $2.35 \pm 0.08$ \\
IV-10           & $4673 \pm  51$ & $1.37 \pm 0.10$ & $2.38 \pm 0.08$ \\
XII-34*         & $4793 \pm  52$ & $1.43 \pm 0.10$ & $2.57 \pm 0.09$ \\
IV-79           & $4690 \pm  52$ & $1.39 \pm 0.10$ & $2.30 \pm 0.08$ \\
IX-13           & $4854 \pm  53$ & $1.70 \pm 0.10$ & $2.23 \pm 0.09$ \\
VIII-24*        & $5043 \pm  55$ & $1.83 \pm 0.10$ & $2.34 \pm 0.12$ \\
X-20            & $5185 \pm  63$ & $2.50 \pm 0.10$ & $1.87 \pm 0.09$ \\
S2710           & $5191 \pm  60$ & $2.51 \pm 0.10$ & $1.88 \pm 0.08$ \\
VI-90           & $5206 \pm  65$ & $2.55 \pm 0.10$ & $1.90 \pm 0.11$ \\
S2265           & $5269 \pm  73$ & $2.73 \pm 0.10$ & $1.78 \pm 0.09$ \\
VIII-45         & $5239 \pm  72$ & $2.74 \pm 0.10$ & $1.85 \pm 0.10$ \\
VII-28          & $5276 \pm  83$ & $2.78 \pm 0.10$ & $1.78 \pm 0.09$ \\
G17181\_0638    & $5328 \pm  84$ & $2.96 \pm 0.10$ & $2.01 \pm 0.22$ \\
C17333\_0832    & $5373 \pm  83$ & $3.17 \pm 0.10$ & $1.67 \pm 0.10$ \\
S3108           & $5413 \pm  93$ & $3.20 \pm 0.10$ & $1.89 \pm 0.11$ \\
S652            & $5322 \pm  83$ & $3.17 \pm 0.10$ & $1.84 \pm 0.18$ \\
S19             & $5793 \pm  66$ & $3.66 \pm 0.10$ & $1.81 \pm 0.14$ \\
D21             & $5954 \pm  74$ & $3.73 \pm 0.10$ & $1.51 \pm 0.12$ \\
S3880           & $5944 \pm  77$ & $3.76 \pm 0.10$ & $1.63 \pm 0.19$ \\
S4038           & $6069 \pm  85$ & $3.80 \pm 0.10$ & $1.86 \pm 0.33$ \\
S61             & $6308 \pm  92$ & $4.00 \pm 0.10$ & $1.25 \pm 0.39$ \\
S162            & $6573 \pm  98$ & $4.08 \pm 0.10$ & $2.00 \pm 0.37$ \\
\enddata
\end{deluxetable}

Table~\ref{tab:atmpars} gives the model atmosphere parameters for each model: $T_{\rm eff}$, $\log g$, and $v_t$ (see Section~\ref{sec:abund}).  The table does not show [M/H] or [$\alpha$/Fe] because these values are the same for every star.


\section{Abundance Measurements}
\label{sec:msmts}

\subsection{Line List}
\label{sec:linelist}

We used the same line list as \citet{ji20}.  The list is based mostly on \linemake\ \citep{pla21a,pla21b} with updates as described by \citeauthor{ji20}  Some lines have hyperfine structure, which we also take from \linemake.

\subsection{Equivalent Widths}
\label{sec:ew}

We developed a graphical utility called \texttt{hiresspec} to analyze each spectrum.  The software fits Gaussian profiles to each absorption line in the line list using \mpfit\ \citep{mar12}.  The user examines each fit and has the option to change the fit by altering the wavelength bounds considered by \mpfit\ or the placement of the continuum.  We gave special treatment to the two Mg~b lines in the line list (5173~\AA\ and 5184~\AA).  Because they often display obvious damping wings, we fit them with Voigt rather than Gaussian profiles.  The EW for each line was calculated analytically from the amplitude and Doppler width (and Lorentzian width for the Voigt profile) determined in the fit.  \mpfit\ also returns the $1\sigma$ uncertainty on EW from the diagonal elements of the covariance matrix.

\begin{deluxetable*}{lccrrrrcrrr}
\tablecolumns{11}
\tablewidth{0pt}
\tablecaption{Line List with Equivalent Widths and Abundances\label{tab:ew}}
\tablehead{ & & & & \multicolumn{3}{c}{III-13} &  & \multicolumn{3}{c}{VII-18} \\ \cline{5-7} \cline{9-11}
\colhead{Species} & \colhead{Wavelength} & \colhead{Excitation Potential} & \colhead{$\log gf$} & \colhead{EW} & \colhead{abundance} & \colhead{weight} &  & \colhead{EW} & \colhead{abundance} & \colhead{weight} \\
 & (\AA) & (eV) & & (m\AA) & & & & (m\AA) & & }
\startdata
\ion{O}{1}  & 6363.78 & 0.020 & $-10.190$ & \nodata  & \nodata  & \nodata  & & $<  1.5$ & $< 6.15$ & \nodata   \\
\ion{Na}{1} & 5682.63 & 2.100 & $ -0.710$ & $  24.2$ & $  4.25$ & $ 75.25$ & & $  18.9$ & $  4.19$ & $ 38.05$  \\
\ion{Na}{1} & 5688.20 & 2.100 & $ -0.410$ & $  44.1$ & $  4.27$ & $ 70.81$ & & $  31.4$ & $  4.14$ & $ 38.06$  \\
\ion{Na}{1} & 5889.95 & 0.000 & $  0.110$ & \nodata  & \nodata  & \nodata  & & $ 408.0$ & $  4.48$ & $ 19.66$  \\
\ion{Na}{1} & 5895.92 & 0.000 & $ -0.190$ & \nodata  & \nodata  & \nodata  & & $ 341.9$ & $  4.33$ & $  7.61$  \\
\ion{Mg}{1} & 3986.75 & 4.350 & $ -1.060$ & $  65.2$ & $  5.38$ & $  8.84$ & & \nodata  & \nodata  & \nodata   \\
\ion{Mg}{1} & 4167.27 & 4.350 & $ -0.740$ & $ 120.5$ & $  6.05$ & $  7.07$ & & $  74.8$ & $  5.39$ & $ 20.69$  \\
\ion{Mg}{1} & 4571.10 & 0.000 & $ -5.620$ & $ 166.7$ & $  5.29$ & $  1.50$ & & $ 134.4$ & $  5.24$ & $  1.53$  \\
\ion{Mg}{1} & 4702.99 & 4.350 & $ -0.440$ & $ 150.6$ & $  5.77$ & $  6.85$ & & $  91.3$ & $  5.06$ & $ 20.19$  \\
\ion{Mg}{1} & 5172.68 & 2.710 & $ -0.390$ & $ 347.0$ & $  5.36$ & $  6.56$ & & $ 280.0$ & $  5.06$ & $ 10.40$  \\
\enddata
\tablecomments{(This table is available in its entirety in a machine-readable form in the online journal.  A portion is shown here for guidance regarding its form and content.)}
\end{deluxetable*}

We computed upper limits on EW for undetected lines.  The upper limit was the strength of the Gaussian that was stronger than the observed spectrum by $3\sigma$.

Table~\ref{tab:ew} gives the atomic data (wavelength, excitation potential, and oscillator strength) for each line.  It also gives each line's EW measured for each star.  Only ten absorption lines and two stars are shown in the manuscript.  The online version of the table includes all of the absorption lines and all of the stars.

\subsubsection{Potassium}

\begin{figure}
\centering
\includegraphics[width=\linewidth]{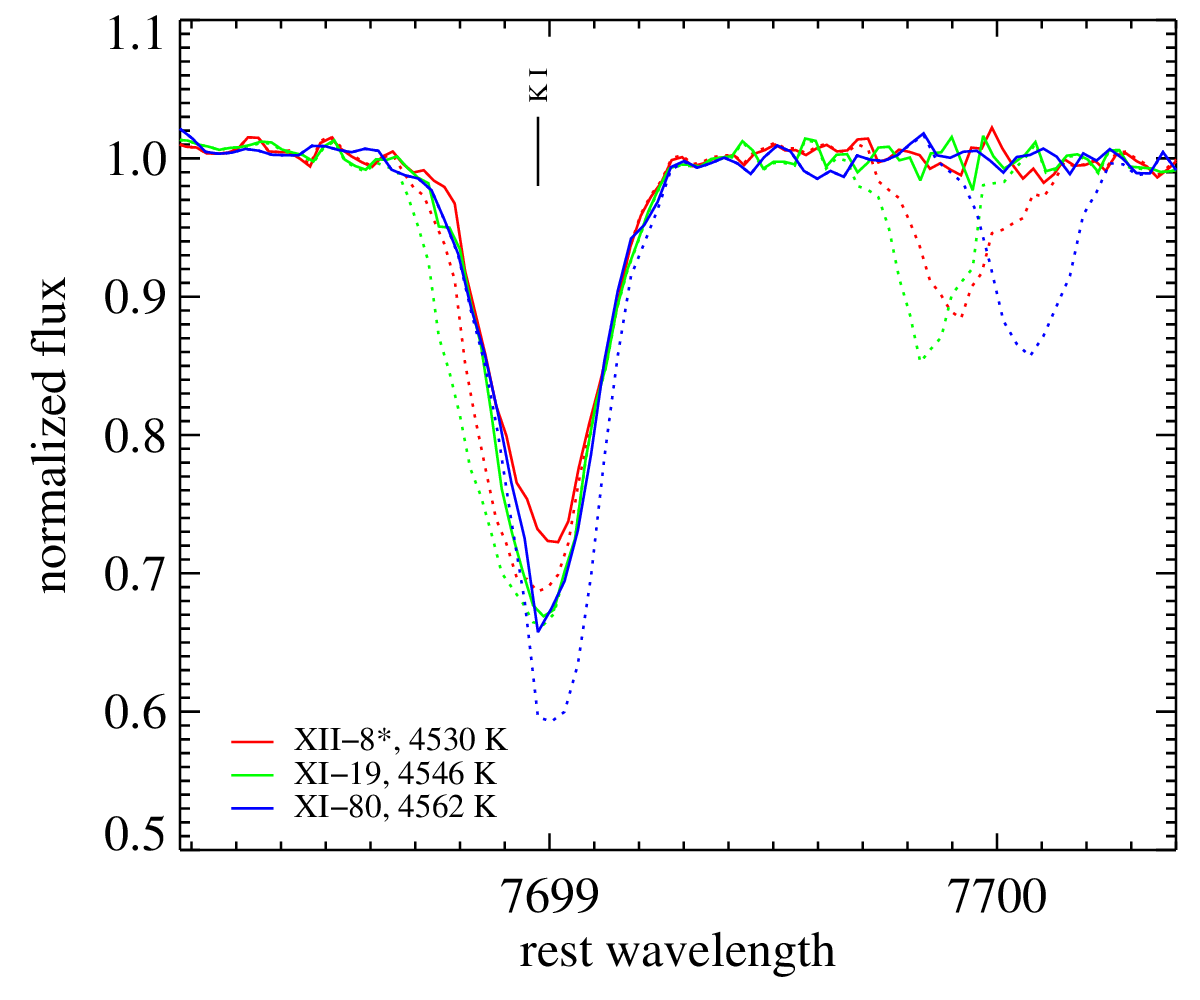}
\caption{Spectra around the resonance line \ion{K}{1}$\;\lambda 7699$.  The stars shown have effective temperatures within 32~K of each other.  The dotted (solid) spectra are uncorrected (corrected) for telluric O$_2$ absorption.\label{fig:Kspectra}}
\end{figure}

We measured potassium from the resonance line \ion{K}{1}$\;\lambda 7699$.  This region of the spectrum lies in the red tail of the telluric A band.  The most prominent telluric feature that affects the potassium line is an electronic transition of O$_2$ at a vacuum wavelength of 7697.96~\AA\@.  This line arises from the $^PQ$ branch in the $0-0$ transition of the $b\,{}^1 \Sigma_g^+ \rightarrow X\,{}^3 \Sigma_g^-$ series of O$_2$ in rotational level $K=31$ \citep{bab48}.  Fortuitously, the line is always accompanied by the $P$ branch of the same transition at 7698.98~\AA, a wavelength free of stellar features.  The strength ratio of the lines is a constant because they arise from the same electronic energy level, and the lines are weak (typically 30~m\AA) and therefore on the linear portion of the curve of growth.  The strength ratio is the ratio of their oscillator strengths ($gf$).  We computed the oscillator strengths from their Einstein coefficients and level degeneracies given by the HITRAN molecular database \citep{yu14,gor22}: $gf(7697.96)/gf(7698.99) = 0.959$.

We decontaminated the \ion{K}{1} line by subtracting the bluer of the O$_2$ lines.  First, we modeled the redder O$_2$ line, which does not overlap stellar features, as a Gaussian.  The Gaussian width is generally narrower than the width of the \ion{K}{1} line because the broadening effects in the Earth's atmosphere are less than in the star.  We fit a Gaussian centered at the observed (geocentric) wavelength, but we allowed the central wavelength, strength, and FWHM to vary.  Then, we created a model of the bluer O$_2$ line by shifting the Gaussian in wavelength and multiplying the strength by $0.959$.  We subtracted both Gaussian models from the observed spectrum.

Figure~\ref{fig:Kspectra} shows three examples of the potassium spectral region.  The dotted spectra are shown before O$_2$ subtraction, and the solid spectra are subtracted.  We discuss potassium abundances further in Section~\ref{sec:K}.

\subsection{Determination of Abundances}
\label{sec:abund}

We used \atlas\ \citep{kur93} model atmospheres in conjunction with \moog\ \citep{sne73} to compute an abundance for each EW\@.  We computed the model atmosphere by interpolating between the grid points of \citeposs{kir11d} grid of \atlas\ models.  We used $T_{\rm eff}$ and $\log g$ as determined in Sections~\ref{sec:teff} and \ref{sec:logg}.  We started with an initial assumption of $v_t$ based on a linear relation with $\log g$ \citep{kir09}.  We assumed ${\rm [M/H]} = -2.41$ and ${\rm [\alpha/Fe]} = +0.41$.  These values are consistent with past abundance estimates for M92 \citep{sne91} as well as our estimates.  Because M92 has been shown repeatedly to have no metallicity variation \citep{sne91,she96,coh97,sne00b,she01,coh11b}, we kept [M/H] fixed at $-2.41$.\footnote{\citet{kin98} found that sub-giants in M92 have higher Fe abundances than giants.  However, our measurements of heavy element abundances in sub-giant stars are consistent with those of giants in M92.}  Some of the $\alpha$ elements, especially Mg, are known to have large variations, but we also kept [$\alpha$/Fe] fixed to keep the analysis as consistent as possible between stars.

We calculated the abundance of each absorption line with the 2021 update to \moog\footnote{\url{http://www.github.com/alexji/moog17scat}}.  This version includes \citeposs{sob11} update, which separates the source function from the Planck function to allow Rayleigh scattering to be a significant source of opacity, which is important at $\lambda < 4500$~\AA\ in metal-poor giants, such as the majority of our M92 sample.   Table~\ref{tab:ew} includes the abundance for each line.  The abundances are corrected for non-LTE effects (Section~\ref{sec:nlte}) where appropriate.

We also computed abundances from upper limits on EWs.  For atomic species with no detected lines and more than one upper limit, we took the abundance from the absorption line with the most stringent abundance limit.  For upper limits, Table~\ref{tab:ew} includes only the upper limit used in computing the limit on abundance.

Some absorption lines display isotopic splitting, which we took from \linemake.  For the transition metals, we used the solar system isotope distribution \citep{asp09}.  For neutron-capture elements, we used the solar $r$-process isotope distribution \citep{sim04,sne08}.

We refined the initial guess at $v_t$ by minimizing the trend of the abundance of \ion{Fe}{1} lines with reduced width (${\rm RW} = {\rm EW}/{\lambda}$).  We considered only lines with $\log {\rm RW} < -4.5$ because strong lines tend to have damping wings, which may not be well represented by our Gaussian fits.  Furthermore, very strong lines (like very weak lines) are not very sensitive to $v_t$ and are therefore not very useful at determining $v_t$.  We fit a line with least-squares regression to abundance vs.\ $\log {\rm RW}$.  We used \mpfit\ to minimize the slope of this line by varying $v_t$.  Table~\ref{tab:atmpars} gives the final values of $v_t$ for each star.

We also propagated the uncertainty on the EW (described in Section~\ref{sec:ew}) to the uncertainty on the abundance of each line.  We ran \moog\ twice: once with the EWs perturbed upward by the EW uncertainty and another time with the EWs perturbed downward.  We took the propagated error on abundance ($e_i$) on absorption line $i$ as the average of the absolute values of the changes in abundance from the upward and downward perturbations.

\begin{deluxetable*}{lcccc}
\tablecolumns{5}
\tablewidth{0pt}
\tablecaption{Abundance Trends\label{tab:trends}}
\tablehead{\colhead{Star} & \colhead{[\ion{Fe}{1}/H]} & \colhead{$d\epsilon/d\log {\rm RW}$} & \colhead{$d\epsilon/d({\rm EP})$} & \colhead{$\epsilon$(\ion{Fe}{1}) $-$ $\epsilon$(\ion{Fe}{2})}}
\startdata
III-13          & $-2.60$ & $+0.00$ & $-0.02$ & $-0.11$ \\
VII-18          & $-2.65$ & $-0.00$ & $+0.01$ & $-0.22$ \\
X-49            & $-2.56$ & $+0.00$ & $+0.00$ & $-0.10$ \\
III-65          & $-2.55$ & $+0.00$ & $-0.03$ & $-0.10$ \\
VII-122         & $-2.60$ & $-0.00$ & $+0.00$ & $-0.12$ \\
II-53           & $-2.59$ & $+0.00$ & $-0.02$ & $-0.10$ \\
XII-8*          & $-2.71$ & $-0.00$ & $-0.05$ & $-0.14$ \\
V-45            & $-2.61$ & $+0.01$ & $-0.05$ & $-0.07$ \\
XI-19           & $-2.49$ & $-0.00$ & $-0.04$ & $-0.04$ \\
XI-80           & $-2.52$ & $-0.00$ & $-0.03$ & $-0.05$ \\
II-70*          & $-2.58$ & $-0.00$ & $-0.04$ & $-0.01$ \\
IV-94*          & $-2.62$ & $+0.00$ & $-0.03$ & $-0.06$ \\
I-67            & $-2.65$ & $+0.00$ & $-0.04$ & $-0.05$ \\
III-82          & $-2.53$ & $+0.00$ & $-0.04$ & $-0.07$ \\
IV-10           & $-2.44$ & $+0.00$ & $-0.03$ & $-0.00$ \\
XII-34*         & $-2.52$ & $-0.00$ & $-0.05$ & $-0.04$ \\
IV-79           & $-2.54$ & $+0.00$ & $-0.04$ & $-0.04$ \\
IX-13           & $-2.41$ & $+0.00$ & $-0.04$ & $+0.01$ \\
VIII-24*        & $-2.47$ & $-0.00$ & $-0.05$ & $-0.03$ \\
X-20            & $-2.36$ & $-0.00$ & $-0.05$ & $+0.06$ \\
S2710           & $-2.36$ & $+0.00$ & $-0.05$ & $+0.05$ \\
VI-90           & $-2.43$ & $+0.00$ & $-0.05$ & $+0.01$ \\
S2265           & $-2.42$ & $+0.00$ & $-0.06$ & $+0.01$ \\
VIII-45         & $-2.49$ & $-0.00$ & $-0.04$ & $-0.03$ \\
VII-28          & $-2.38$ & $+0.00$ & $-0.05$ & $-0.01$ \\
G17181\_0638    & $-2.50$ & $+0.00$ & $-0.05$ & $+0.02$ \\
C17333\_0832    & $-2.41$ & $+0.00$ & $-0.04$ & $-0.03$ \\
S3108           & $-2.47$ & $-0.00$ & $-0.05$ & $-0.00$ \\
S652            & $-2.46$ & $+0.00$ & $-0.04$ & $-0.08$ \\
S19             & $-2.43$ & $+0.00$ & $-0.06$ & $-0.01$ \\
D21             & $-2.29$ & $-0.00$ & $-0.07$ & $+0.03$ \\
S3880           & $-2.46$ & $+0.00$ & $-0.02$ & $-0.13$ \\
S4038           & $-2.45$ & $+0.00$ & $-0.00$ & $-0.03$ \\
S61             & $-2.60$ & $-0.00$ & $-0.05$ & $-0.11$ \\
S162            & $-2.33$ & $-0.00$ & $-0.07$ & $+0.07$ \\
\enddata
\end{deluxetable*}

Table~\ref{tab:trends} gives the resulting value of [Fe/H] measured from \ion{Fe}{1} lines with $\log {\rm RW} < -4.5$.  The table also gives the slope of the abundances measured from \ion{Fe}{1} lines vs.\ $\log {\rm RW}$\@.  This slope is a diagnosis of microturbulence, as discussed above.  The slopes are essentially zero because $v_t$ was tuned to minimize these slopes.  The third column of the table gives the slope of abundance vs.\ excitation potential (EP, in units of eV)\@.  This slope is a diagnosis of $T_{\rm eff}$.  Positive (negative) slopes indicate that the value of $T_{\rm eff}$ is too low (high).  Some deviation from zero is expected because of the deficiencies in the assumption of LTE\@.  The last column of the table is the difference in the mean iron abundance between neutral and singly ionized Fe lines.  This difference is a diagnosis of $\log g$.  Positive (negative) differences indicate that $\log g$ is too low (high).  As with $T_{\rm eff}$, some deviation from zero is expected in an LTE analysis, especially due to the non-LTE overionization effect in cool giants \citep{the99}.

\subsection{Synthesis}
\label{sec:synth}

Some elements are better suited to abundance measurements by spectral synthesis than by EWs.  This is especially true for measurements from molecular features (like CH), intrinsically broad atomic features (like \ion{Li}{1}$\;\lambda6707$), and features contaminated by blends.  We computed abundances of Li and C by spectral synthesis.

We used \linemake\ to create line lists for spectral synthesis.  For Li, the wavelength range was 6704.7--6710.9~\AA, and the Li was purely ${}^7$Li (no ${}^6$Li).  For C, the wavelength range was 4273.9--4333.0~\AA\@.  The ${}^{12}{\rm C}/{}^{13}{\rm C}$ ratio changes with evolution along the RGB\@.  We used $\log g$ as a proxy for evolution, and we used the following prescription \citep{kel01,kir15a} for the isotope ratio:

\begin{eqnarray}
\begin{array}{ll}
^{12}{\rm C}/^{13}{\rm C} = 50 & ~~~{\rm if}~\log g > 2.7 \\
^{12}{\rm C}/^{13}{\rm C} = 63\,\log g - 120 & ~~~{\rm if}~2.0 < \log g \le 2.7 \\
^{12}{\rm C}/^{13}{\rm C} = 6 & ~~~{\rm if}~\log g \le 2.0 \\ \label{eq:cratio}
\end{array}
\end{eqnarray}

We used \mpfit\ to find the best-fitting abundance.  We then calculated a continuum correction.  For Li, the continuum correction was the median of the residual (the observed spectrum divided by the best-fit spectrum).  For the CH feature, which spans a longer wavelength range, the continuum correction was a spline with a breakpoint spacing of 10~\AA\ and sigma clipping at $\pm 2\sigma$.  We divided the observed spectrum by the continuum correction and then re-measured the abundance.  We iterated the continuum correction until the abundance changed by less than 0.001 between iterations.

\begin{figure}
\includegraphics[width=\linewidth]{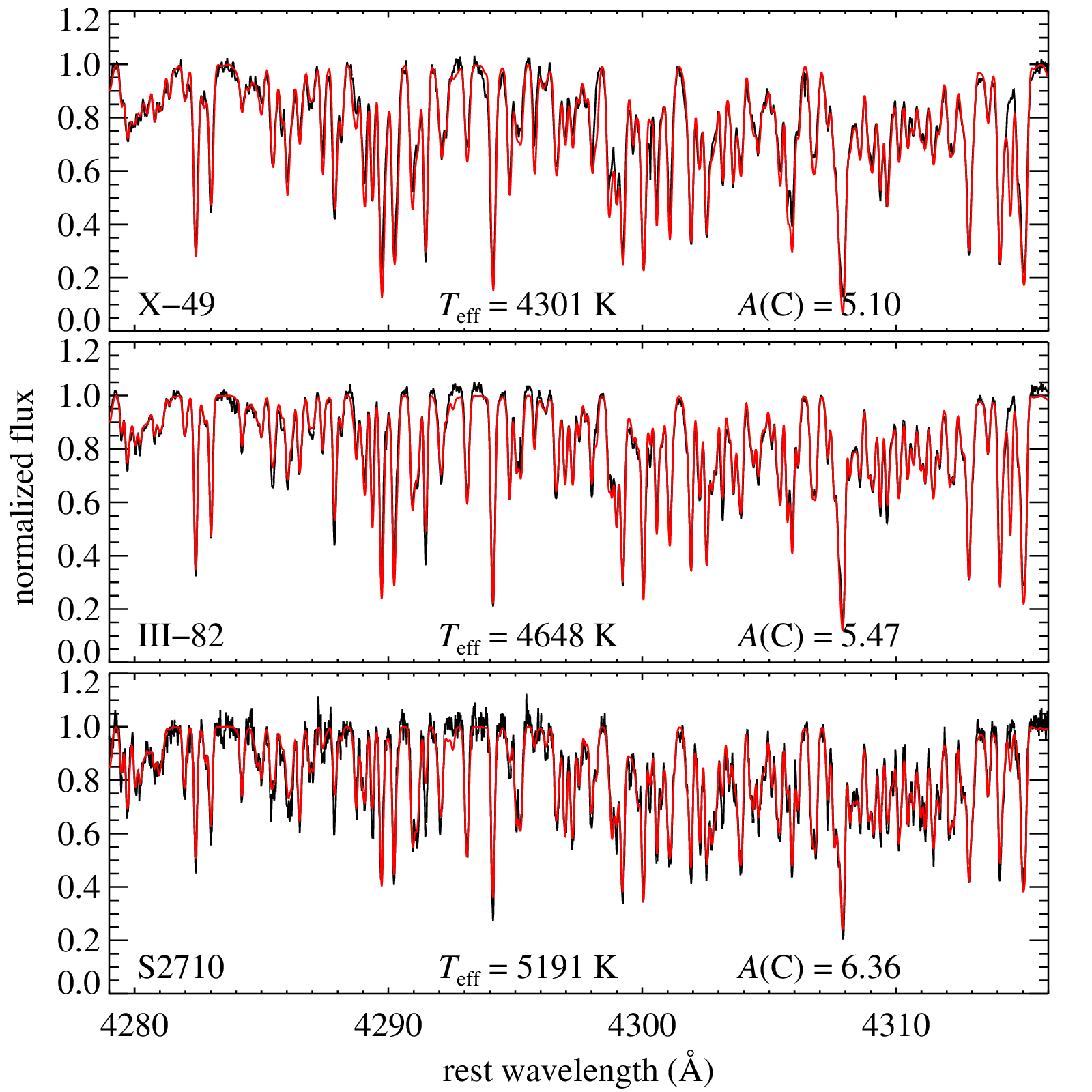}
\caption{Syntheses of a CH feature (red) compared to the observed spectra (black).  These spectra sample three different stellar luminosities in M92.\label{fig:synthc}}
\end{figure}

Figure~\ref{fig:synthc} shows the syntheses for the CH feature for stars of various effective temperatures.

\subsection{Non-LTE Corrections}
\label{sec:nlte}

We considered non-LTE corrections for most lines in the line list.  These corrections are so large for some elements, like Al and K, that the corrections are required for any meaningful interpretation of the abundances.  The corrections for some elements, like Mg, are moderate.  Finally, the corrections for some elements, like O (when measured from the forbidden lines), are negligible.

We measured \ion{Li}{1} abundances from the marginally resolved doublet at 6708~\AA\@.  The line shape and a blend with a nearby \ion{Fe}{1} line required that we synthesize the absorption line rather than fit a simple Gaussian (see Section~\ref{sec:synth}).   \citet{lin09li} computed non-LTE corrections to this feature, and we applied them to the abundance computed from the synthesis.

We considered only the [\ion{O}{1}]$\;\lambda\lambda$6300,6364 forbidden doublet.  The non-LTE corrections are negligibly small for both lines \citep{ama15,ber21o}.  The main uncertainty arises from the oscillator strengths \citep[e.g.,][]{sto00}.

We measured \ion{Na}{1} abundances from the Na~D resonance doublet and another doublet at 5683 and 5688~\AA\@.  We applied \citeposs{lin11na} corrections to each of these lines.

For \ion{Mg}{1}, we applied the corrections computed by \citet{ber17mg}.  We used the online interface provided by the MPIA database of NLTE corrections\footnote{\url{https://nlte.mpia.de/}}.  The interface interpolates NLTE corrections from a pre-computed grid based on stellar parameters.  We chose the corrections computed with the 1-D MARCS spherical model atmospheres \citep{gus08}.  The correction grids for any other options for model atmospheres did not include the full range of stellar parameters spanned by our sample.  The typical correction was $+0.2$~dex, in the sense that the non-LTE abundance is higher than the LTE abundance.

\citet{nor17} computed 1-D NLTE corrections to \ion{Al}{1} lines, including both lines used in our study.  We read the corrections for \ion{Al}{1}$\;\lambda$3962 from their Figure~13 based on each star's $T_{\rm eff}$ and $\log g$.  They did not provide a similar plot for \ion{Al}{1}$\;\lambda$3944, but the corrections for both lines are similar.  Therefore, we applied the corrections for the redder line to the bluer line.  Corrections ranged from $+0.5$~dex for the base of the RGB to $+1.1$~dex for the coolest giants.

The MPIA database showed that corrections for \ion{Si}{1} \citep{ber13si} are negligible.  Furthermore, the corrections to \ion{Ca}{1} are very small \citep{mas07}.  We applied no corrections to these elements.

We used the \ion{K}{1}$\;\lambda$7699 corrections of \citet{reg19}.  We read the corrections appropriate for each star's $T_{\rm eff}$ and $\log g$ from their Figure~12.  Although the non-LTE correction for this line is very large in warm, metal-rich giants, the corrections at low metallicity are typically $-0.2$~dex.

Non-LTE corrections for the transition metals are complicated.  For example, \citet{ber11ti} gave \ion{Ti}{1} corrections in the range $+0.4$ to $+1.0$~dex.  \ion{Ti}{2} corrections are typically less than $+0.1$~dex, as would be expected for ionized lines in red giants.  As a trial, we applied these corrections to the Ti lines in our study.  For stars at the tip of the RGB, the difference between \ion{Ti}{1} and \ion{Ti}{2} abundances decreased, and the slope of \ion{Ti}{1} abundances with excitation potential approached zero.  However, both of these diagnostics worsened after applying the non-LTE correction for the majority of the stars in our sample, with the difference between neutral and ionized abundances as large as $+0.7$~dex.  Similarly, we applied \citeposs{ber12fe} corrections to \ion{Fe}{1} and \ion{Fe}{2} lines.  The corrections are typically $+0.1$~dex, much smaller than for Ti.  The corrections did not improve the slope of \ion{Fe}{1} abundances with excitation potential, and they exacerbated the differences between \ion{Fe}{1} and \ion{Fe}{2} abundances.  Therefore, we chose not to apply non-LTE corrections to transition metals.  Importantly, our line list excludes commonly used \ion{Mn}{1} resonance lines because they are particularly affected by non-LTE corrections \citep{ber19mn}.  All of the Mn transitions used here have excitation potentials of at least 2~eV\@.

We also considered non-LTE corrections for neutron-capture elements.  For example, \citet{ber12sr} computed corrections for \ion{Sr}{1} and \ion{Sr}{2}, and \citet{kor15} computed corrections for \ion{Ba}{2}.  The typical corrections for ionized lines in our study are less than 0.1~dex.  Therefore, we applied no corrections to ionized lines of neutron-capture elements.  However, \ion{Sr}{1}$\;\lambda$4607 requires corrections of about 0.4~dex.  However, \citet{ber12sr} did not compute corrections for $\log g < 2.2$.  As a result, we omitted \ion{Sr}{1} from our study.  All of the Sr abundances reported here are based on \ion{Sr}{2}\@.

\subsection{Additional Luminosity-Based Corrections}
\label{sec:lumcorr}

\begin{figure*}
\centering
\hfill
\includegraphics[width=0.45\linewidth]{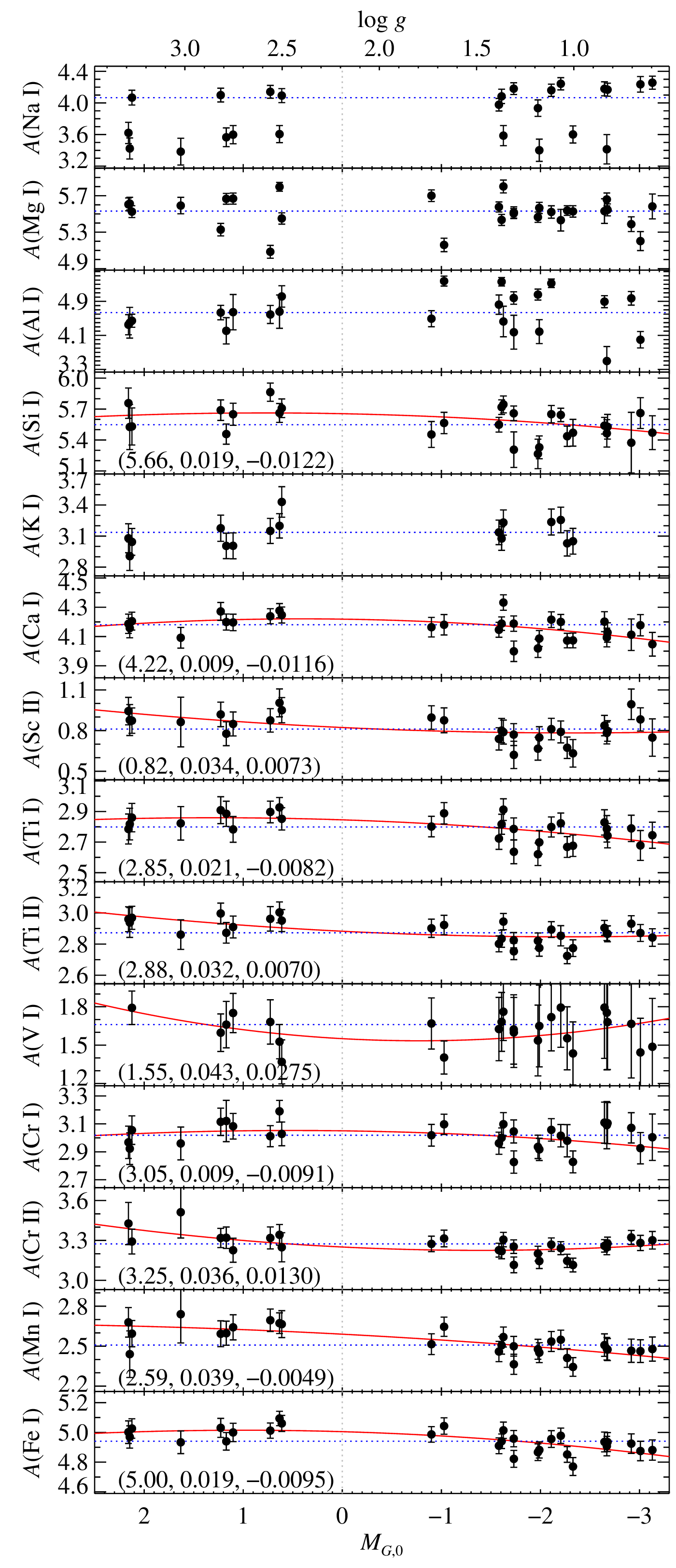}
\hfill
\includegraphics[width=0.45\linewidth]{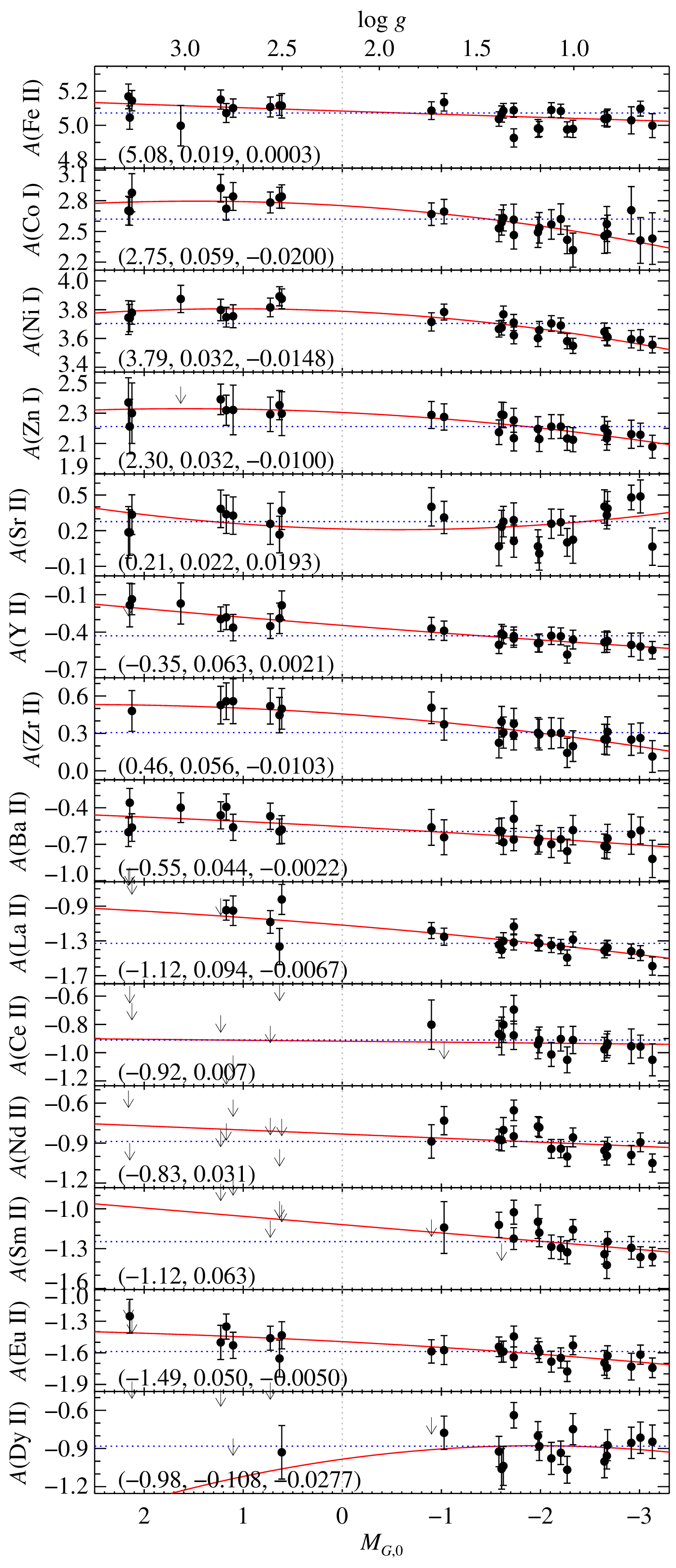}
\hfill
\caption{Trends of abundance with luminosity (bottom axis) or surface gravity (top axis).  The red lines show fits to the trend, as described in Section~\ref{sec:lumcorr}.  For the remainder of this work, the abundances are de-trended such that the correction at $M_{G,0} = 0$ (dotted gray line) is zero.  Elements without red trend lines were not corrected.  The dashed blue lines show the median abundances before the luminosity correction.  The polynomial terms of the trend are shown in parentheses.\label{fig:lumcorr}}
\end{figure*}

The abundances of some elements display a trend with the stellar luminosity (or surface gravity) even after the application of non-LTE corrections.  Figure~\ref{fig:lumcorr} shows these trends.  Some elements, especially carbon, are expected to show a trend with luminosity.  We discuss the carbon trend with luminosity in Section~\ref{sec:lumtrend}.

Trends with luminosity for heavier elements could result from atomic diffusion, including gravitational settling and radiative acceleration (or levitation).  The effect could be strong for very metal-poor globular clusters, like M92 \citep{ric02}.  However, it is expected to be strongest at the main sequence turn-off.

\begin{rotatetable*}
\begin{deluxetable*}{lrcccrcccrcccrcccrcccrcc}
\centerwidetable
\tabletypesize{\scriptsize}
\tablecolumns{24}
\tablewidth{0pt}
\tablecaption{Abundances\label{tab:abund1}}
\tablehead{ & \multicolumn{3}{c}{III-13         } & & \multicolumn{3}{c}{VII-18         } & & \multicolumn{3}{c}{X-49           } & & \multicolumn{3}{c}{III-65         } & & \multicolumn{3}{c}{VII-122        } & & \multicolumn{3}{c}{II-53          } \\ \cline{2-4}   \cline{6-8}   \cline{10-12} \cline{14-16} \cline{18-20} \cline{22-24}
\colhead{Element} & \colhead{$N$} & \colhead{[X/Fe]} & \colhead{error} & & \colhead{$N$} & \colhead{[X/Fe]} & \colhead{error} & & \colhead{$N$} & \colhead{[X/Fe]} & \colhead{error} & & \colhead{$N$} & \colhead{[X/Fe]} & \colhead{error} & & \colhead{$N$} & \colhead{[X/Fe]} & \colhead{error} & & \colhead{$N$} & \colhead{[X/Fe]} & \colhead{error}}
\startdata
\protect\ion{Fe}{1}         &        23 & $ -2.46$ & $0.07 $ &   &    28 & $ -2.48$ & $0.07 $ &   &    23 & $ -2.44$ & $0.07 $ &   &    21 & $ -2.44$ & $0.06 $ &   &    35 & $ -2.47$ & $0.07 $ &   &    35 & $ -2.45$ & $0.07 $ \\
\protect\ion{Fe}{2}         &        22 & $ -2.45$ & $0.07 $ &   &    19 & $ -2.35$ & $0.04 $ &   &    24 & $ -2.42$ & $0.08 $ &   &    22 & $ -2.41$ & $0.05 $ &   &    23 & $ -2.41$ & $0.05 $ &   &    26 & $ -2.41$ & $0.04 $ \\
\protect $A({\rm Li})$      &           & \nodata  & \nodata &   &       & \nodata  & \nodata &   &       & \nodata  & \nodata &   &       & \nodata  & \nodata &   &       & \nodata  & \nodata &   &       & $<+1.52$ & \nodata \\
\protect CH                 &       syn & $ -0.98$ & $0.11 $ &     & syn & $ -0.98$ & $0.11 $ &     & syn & $ -0.89$ & $0.11 $ &     & syn & $ -0.82$ & $0.12 $ &     & syn & $ -0.50$ & $0.11 $ &     & syn & $ -0.80$ & $0.12 $ \\
\protect\ion{O}{1}          &         0 & \nodata  & \nodata &   &     1 & $<-0.06$ & \nodata &   &     0 & \nodata  & \nodata &   &     0 & \nodata  & \nodata &   &     2 & $ +0.84$ & $0.11 $ &   &     1 & $ +0.37$ & $0.12 $ \\
\protect\ion{Na}{1}         &         2 & $ +0.48$ & $0.08 $ &   &     4 & $ +0.48$ & $0.09 $ &   &     0 & \nodata  & \nodata &   &     2 & $ +0.37$ & $0.08 $ &   &     2 & $ -0.35$ & $0.14 $ &   &     4 & $ +0.39$ & $0.08 $ \\
\protect\ion{Mg}{1}         &         8 & $ +0.45$ & $0.12 $ &   &     6 & $ +0.08$ & $0.09 $ &   &     4 & $ +0.23$ & $0.06 $ &   &     8 & $ +0.39$ & $0.05 $ &   &     8 & $ +0.53$ & $0.06 $ &   &     7 & $ +0.38$ & $0.12 $ \\
\protect\ion{Al}{1}         &         0 & \nodata  & \nodata &   &     2 & $ +0.03$ & $0.15 $ &   &     2 & $ +0.96$ & $0.13 $ &   &     0 & \nodata  & \nodata &   &     2 & $ -0.48$ & $0.32 $ &   &     2 & $ +0.89$ & $0.10 $ \\
\protect\ion{Si}{1}         &         7 & $ +0.60$ & $0.16 $ &   &     6 & $ +0.80$ & $0.15 $ &   &     2 & $ +0.46$ & $0.28 $ &   &     7 & $ +0.60$ & $0.12 $ &   &     4 & $ +0.56$ & $0.12 $ &   &     8 & $ +0.61$ & $0.08 $ \\
\protect\ion{K}{1}          &         0 & \nodata  & \nodata &   &     0 & \nodata  & \nodata &   &     0 & \nodata  & \nodata &   &     0 & \nodata  & \nodata &   &     0 & \nodata  & \nodata &   &     0 & \nodata  & \nodata \\
\protect\ion{Ca}{1}         &        11 & $ +0.32$ & $0.06 $ &   &    16 & $ +0.45$ & $0.05 $ &   &     5 & $ +0.34$ & $0.07 $ &   &    12 & $ +0.34$ & $0.04 $ &   &    14 & $ +0.33$ & $0.05 $ &   &    15 & $ +0.41$ & $0.05 $ \\
\protect\ion{Sc}{2}         &         7 & $ +0.08$ & $0.13 $ &   &     5 & $ +0.12$ & $0.07 $ &   &     6 & $ +0.30$ & $0.11 $ &   &     6 & $ +0.09$ & $0.07 $ &   &     8 & $ +0.08$ & $0.07 $ &   &     7 & $ +0.14$ & $0.06 $ \\
\protect\ion{Ti}{1}         &        25 & $ +0.41$ & $0.05 $ &   &    29 & $ +0.35$ & $0.07 $ &   &    23 & $ +0.41$ & $0.05 $ &   &    21 & $ +0.35$ & $0.05 $ &   &    24 & $ +0.42$ & $0.06 $ &   &    28 & $ +0.44$ & $0.05 $ \\
\protect\ion{Ti}{2}         &        31 & $ +0.37$ & $0.06 $ &   &    27 & $ +0.30$ & $0.04 $ &   &    23 & $ +0.43$ & $0.07 $ &   &    40 & $ +0.36$ & $0.05 $ &   &    34 & $ +0.37$ & $0.04 $ &   &    32 & $ +0.40$ & $0.04 $ \\
\protect\ion{V}{1}          &         2 & $ -0.11$ & $0.36 $ &   &     2 & $ -0.13$ & $0.24 $ &   &     2 & $ +0.06$ & $0.40 $ &   &     2 & $ +0.11$ & $0.35 $ &   &     2 & $ +0.21$ & $0.43 $ &   &     2 & $ +0.23$ & $0.38 $ \\
\protect\ion{Cr}{1}         &        16 & $ -0.05$ & $0.13 $ &   &     8 & $ -0.12$ & $0.07 $ &   &    10 & $ -0.03$ & $0.07 $ &   &    14 & $ +0.00$ & $0.12 $ &   &    12 & $ +0.01$ & $0.15 $ &   &    15 & $ +0.00$ & $0.12 $ \\
\protect\ion{Cr}{2}         &         5 & $ +0.09$ & $0.08 $ &   &     4 & $ -0.02$ & $0.06 $ &   &     6 & $ +0.09$ & $0.08 $ &   &     6 & $ +0.05$ & $0.06 $ &   &     5 & $ +0.02$ & $0.06 $ &   &     6 & $ +0.04$ & $0.05 $ \\
\protect\ion{Mn}{1}         &         5 & $ -0.32$ & $0.06 $ &   &     6 & $ -0.33$ & $0.06 $ &   &     6 & $ -0.37$ & $0.06 $ &   &     6 & $ -0.37$ & $0.05 $ &   &     6 & $ -0.34$ & $0.06 $ &   &     6 & $ -0.34$ & $0.05 $ \\
\protect\ion{Co}{1}         &         7 & $ +0.29$ & $0.22 $ &   &     6 & $ +0.26$ & $0.20 $ &   &     6 & $ +0.50$ & $0.20 $ &   &     6 & $ +0.23$ & $0.15 $ &   &     6 & $ +0.36$ & $0.13 $ &   &     6 & $ +0.21$ & $0.14 $ \\
\protect\ion{Ni}{1}         &        14 & $ +0.05$ & $0.04 $ &   &    16 & $ +0.08$ & $0.06 $ &   &    14 & $ +0.03$ & $0.04 $ &   &    12 & $ +0.03$ & $0.04 $ &   &    15 & $ +0.06$ & $0.04 $ &   &    15 & $ +0.06$ & $0.04 $ \\
\protect\ion{Zn}{1}         &         2 & $ +0.18$ & $0.09 $ &   &     2 & $ +0.26$ & $0.09 $ &   &     2 & $ +0.22$ & $0.10 $ &   &     2 & $ +0.21$ & $0.08 $ &   &     2 & $ +0.20$ & $0.09 $ &   &     2 & $ +0.24$ & $0.08 $ \\
\protect\ion{Sr}{2}         &         2 & $ -0.48$ & $0.13 $ &   &     1 & $ -0.14$ & $0.12 $ &   &     2 & $ -0.07$ & $0.10 $ &   &     1 & $ -0.16$ & $0.13 $ &   &     2 & $ -0.20$ & $0.10 $ &   &     1 & $ -0.13$ & $0.12 $ \\
\protect\ion{Y}{2}          &         7 & $ -0.13$ & $0.07 $ &   &     6 & $ -0.21$ & $0.10 $ &   &     7 & $ -0.13$ & $0.10 $ &   &     6 & $ -0.12$ & $0.06 $ &   &     8 & $ -0.12$ & $0.08 $ &   &     7 & $ -0.13$ & $0.07 $ \\
\protect\ion{Zr}{2}         &         1 & $ +0.26$ & $0.12 $ &   &     1 & $ +0.29$ & $0.11 $ &   &     1 & $ +0.34$ & $0.13 $ &   &     1 & $ +0.36$ & $0.11 $ &   &     1 & $ +0.31$ & $0.12 $ &   &     1 & $ +0.31$ & $0.11 $ \\
\protect\ion{Ba}{2}         &         2 & $ -0.40$ & $0.14 $ &   &     3 & $ -0.27$ & $0.09 $ &   &     2 & $ -0.23$ & $0.15 $ &   &     2 & $ -0.29$ & $0.10 $ &   &     5 & $ -0.36$ & $0.08 $ &   &     3 & $ -0.35$ & $0.08 $ \\
\protect\ion{La}{2}         &         4 & $ +0.11$ & $0.10 $ &   &     3 & $ +0.15$ & $0.08 $ &   &     4 & $ +0.23$ & $0.09 $ &   &     3 & $ +0.23$ & $0.08 $ &   &     4 & $ +0.23$ & $0.07 $ &   &     3 & $ +0.20$ & $0.07 $ \\
\protect\ion{Ce}{2}         &         5 & $ -0.16$ & $0.11 $ &   &     5 & $ -0.17$ & $0.07 $ &   &     4 & $ -0.10$ & $0.13 $ &   &     7 & $ -0.09$ & $0.08 $ &   &     5 & $ -0.10$ & $0.08 $ &   &     6 & $ -0.12$ & $0.07 $ \\
\protect\ion{Nd}{2}         &        25 & $ +0.07$ & $0.07 $ &   &    23 & $ +0.13$ & $0.06 $ &   &    19 & $ +0.10$ & $0.08 $ &   &    26 & $ +0.14$ & $0.06 $ &   &    20 & $ +0.08$ & $0.06 $ &   &    24 & $ +0.12$ & $0.05 $ \\
\protect\ion{Sm}{2}         &         9 & $ +0.32$ & $0.07 $ &   &     5 & $ +0.21$ & $0.07 $ &   &     4 & $ +0.35$ & $0.09 $ &   &     8 & $ +0.37$ & $0.06 $ &   &     3 & $ +0.20$ & $0.09 $ &   &     7 & $ +0.28$ & $0.06 $ \\
\protect\ion{Eu}{2}         &         2 & $ +0.39$ & $0.09 $ &   &     2 & $ +0.41$ & $0.08 $ &   &     2 & $ +0.35$ & $0.13 $ &   &     2 & $ +0.43$ & $0.09 $ &   &     2 & $ +0.32$ & $0.10 $ &   &     1 & $ +0.37$ & $0.11 $ \\
\protect\ion{Dy}{2}         &         1 & $ +0.43$ & $0.12 $ &   &     1 & $ +0.36$ & $0.11 $ &   &     2 & $ +0.38$ & $0.13 $ &   &     1 & $ +0.34$ & $0.12 $ &   &     2 & $ +0.26$ & $0.09 $ &   &     1 & $ +0.22$ & $0.12 $ \\
\enddata
\end{deluxetable*}
\end{rotatetable*}

\addtocounter{table}{-1}
\begin{rotatetable*}
\begin{deluxetable*}{lrcccrcccrcccrcccrcccrcc}
\centerwidetable
\tabletypesize{\scriptsize}
\tablecolumns{24}
\tablewidth{0pt}
\tablecaption{Abundances ({\it continued})\label{tab:abund2}}
\tablehead{ & \multicolumn{3}{c}{XII-8*         } & & \multicolumn{3}{c}{V-45           } & & \multicolumn{3}{c}{XI-19          } & & \multicolumn{3}{c}{XI-80          } & & \multicolumn{3}{c}{II-70*         } & & \multicolumn{3}{c}{IV-94*         } \\ \cline{2-4}   \cline{6-8}   \cline{10-12} \cline{14-16} \cline{18-20} \cline{22-24}
\colhead{Element} & \colhead{$N$} & \colhead{[X/Fe]} & \colhead{error} & & \colhead{$N$} & \colhead{[X/Fe]} & \colhead{error} & & \colhead{$N$} & \colhead{[X/Fe]} & \colhead{error} & & \colhead{$N$} & \colhead{[X/Fe]} & \colhead{error} & & \colhead{$N$} & \colhead{[X/Fe]} & \colhead{error} & & \colhead{$N$} & \colhead{[X/Fe]} & \colhead{error}}
\startdata
\protect\ion{Fe}{1}         &        53 & $ -2.63$ & $0.06 $ &   &    35 & $ -2.56$ & $0.05 $ &   &    34 & $ -2.43$ & $0.05 $ &   &    45 & $ -2.46$ & $0.06 $ &   &    41 & $ -2.54$ & $0.05 $ &   &    46 & $ -2.56$ & $0.06 $ \\
\protect\ion{Fe}{2}         &        24 & $ -2.48$ & $0.05 $ &   &    26 & $ -2.48$ & $0.04 $ &   &    24 & $ -2.38$ & $0.04 $ &   &    23 & $ -2.37$ & $0.04 $ &   &    21 & $ -2.48$ & $0.04 $ &   &    24 & $ -2.48$ & $0.05 $ \\
\protect $A({\rm Li})$      &           & $<-3.36$ & \nodata &   &       & \nodata  & \nodata &   &       & \nodata  & \nodata &   &       & \nodata  & \nodata &   &       & \nodata  & \nodata &   &       & \nodata  & \nodata \\
\protect CH                 &       syn & $ -0.59$ & $0.12 $ &     & syn & $ -0.46$ & $0.12 $ &     & syn & $ -0.59$ & $0.12 $ &     & syn & $ -0.68$ & $0.13 $ &     & syn & $ -0.40$ & $0.14 $ &     & syn & $ -0.69$ & $0.15 $ \\
\protect\ion{O}{1}          &         2 & $ +0.85$ & $0.09 $ &   &     1 & $ +0.80$ & $0.12 $ &   &     2 & $ +0.49$ & $0.11 $ &   &     1 & $ +0.27$ & $0.14 $ &   &     1 & $<+0.92$ & \nodata &   &     1 & $<+0.68$ & \nodata \\
\protect\ion{Na}{1}         &         1 & $ -0.01$ & $0.11 $ &   &     0 & \nodata  & \nodata &   &     2 & $ +0.44$ & $0.08 $ &   &     4 & $ +0.38$ & $0.07 $ &   &     2 & $ -0.30$ & $0.11 $ &   &     3 & $ +0.25$ & $0.08 $ \\
\protect\ion{Mg}{1}         &         7 & $ +0.56$ & $0.05 $ &   &     8 & $ +0.49$ & $0.04 $ &   &     9 & $ +0.27$ & $0.11 $ &   &     8 & $ +0.38$ & $0.06 $ &   &     7 & $ +0.51$ & $0.05 $ &   &     8 & $ +0.42$ & $0.05 $ \\
\protect\ion{Al}{1}         &         0 & \nodata  & \nodata &   &     0 & \nodata  & \nodata &   &     0 & \nodata  & \nodata &   &     2 & $ +1.34$ & $0.08 $ &   &     2 & $ +0.28$ & $0.27 $ &   &     2 & $ +1.17$ & $0.10 $ \\
\protect\ion{Si}{1}         &         5 & $ +0.70$ & $0.13 $ &   &     5 & $ +0.59$ & $0.09 $ &   &     9 & $ +0.67$ & $0.06 $ &   &     7 & $ +0.70$ & $0.08 $ &   &     2 & $ +0.45$ & $0.09 $ &   &     2 & $ +0.40$ & $0.12 $ \\
\protect\ion{K}{1}          &         1 & $ +0.65$ & $0.11 $ &   &     1 & $ +0.56$ & $0.11 $ &   &     1 & $ +0.66$ & $0.11 $ &   &     1 & $ +0.67$ & $0.11 $ &   &     0 & \nodata  & \nodata &   &     0 & \nodata  & \nodata \\
\protect\ion{Ca}{1}         &        15 & $ +0.45$ & $0.04 $ &   &    18 & $ +0.37$ & $0.04 $ &   &    17 & $ +0.37$ & $0.03 $ &   &    17 & $ +0.41$ & $0.04 $ &   &    16 & $ +0.35$ & $0.04 $ &   &    21 & $ +0.30$ & $0.05 $ \\
\protect\ion{Sc}{2}         &         7 & $ -0.00$ & $0.08 $ &   &     8 & $ +0.05$ & $0.06 $ &   &     7 & $ +0.06$ & $0.06 $ &   &     7 & $ +0.07$ & $0.06 $ &   &     7 & $ +0.12$ & $0.06 $ &   &     7 & $ +0.04$ & $0.07 $ \\
\protect\ion{Ti}{1}         &        37 & $ +0.45$ & $0.05 $ &   &    34 & $ +0.36$ & $0.04 $ &   &    30 & $ +0.39$ & $0.04 $ &   &    21 & $ +0.39$ & $0.04 $ &   &    33 & $ +0.36$ & $0.05 $ &   &    24 & $ +0.30$ & $0.04 $ \\
\protect\ion{Ti}{2}         &        28 & $ +0.34$ & $0.04 $ &   &    35 & $ +0.29$ & $0.04 $ &   &    47 & $ +0.31$ & $0.05 $ &   &    29 & $ +0.35$ & $0.04 $ &   &    34 & $ +0.35$ & $0.04 $ &   &    33 & $ +0.39$ & $0.05 $ \\
\protect\ion{V}{1}          &         2 & $ +0.09$ & $0.23 $ &   &     2 & $ +0.13$ & $0.23 $ &   &     2 & $ +0.26$ & $0.30 $ &   &     2 & $ +0.22$ & $0.25 $ &   &     2 & $ +0.24$ & $0.31 $ &   &     2 & $ +0.14$ & $0.27 $ \\
\protect\ion{Cr}{1}         &        13 & $ -0.11$ & $0.05 $ &   &    18 & $ -0.04$ & $0.10 $ &   &    12 & $ -0.13$ & $0.05 $ &   &    12 & $ -0.06$ & $0.05 $ &   &    12 & $ -0.13$ & $0.05 $ &   &    11 & $ -0.09$ & $0.05 $ \\
\protect\ion{Cr}{2}         &         6 & $ -0.03$ & $0.05 $ &   &     6 & $ +0.00$ & $0.05 $ &   &     6 & $ -0.01$ & $0.05 $ &   &     6 & $ +0.02$ & $0.05 $ &   &     5 & $ +0.01$ & $0.06 $ &   &     5 & $ +0.06$ & $0.06 $ \\
\protect\ion{Mn}{1}         &         6 & $ -0.34$ & $0.05 $ &   &     6 & $ -0.35$ & $0.05 $ &   &     6 & $ -0.34$ & $0.05 $ &   &     6 & $ -0.33$ & $0.05 $ &   &     6 & $ -0.34$ & $0.05 $ &   &     5 & $ -0.30$ & $0.05 $ \\
\protect\ion{Co}{1}         &         6 & $ +0.21$ & $0.14 $ &   &     7 & $ +0.22$ & $0.11 $ &   &     7 & $ +0.29$ & $0.12 $ &   &     7 & $ +0.25$ & $0.11 $ &   &     6 & $ +0.28$ & $0.12 $ &   &     6 & $ +0.25$ & $0.12 $ \\
\protect\ion{Ni}{1}         &        15 & $ +0.12$ & $0.04 $ &   &    14 & $ +0.07$ & $0.04 $ &   &    16 & $ +0.05$ & $0.03 $ &   &    17 & $ +0.08$ & $0.04 $ &   &    19 & $ +0.10$ & $0.04 $ &   &    14 & $ +0.06$ & $0.04 $ \\
\protect\ion{Zn}{1}         &         2 & $ +0.33$ & $0.08 $ &   &     2 & $ +0.25$ & $0.08 $ &   &     2 & $ +0.20$ & $0.08 $ &   &     2 & $ +0.22$ & $0.08 $ &   &     2 & $ +0.21$ & $0.08 $ &   &     2 & $ +0.30$ & $0.08 $ \\
\protect\ion{Sr}{2}         &         1 & $ -0.32$ & $0.17 $ &   &     2 & $ -0.34$ & $0.10 $ &   &     2 & $ -0.27$ & $0.09 $ &   &     2 & $ -0.28$ & $0.10 $ &   &     2 & $ -0.41$ & $0.11 $ &   &     2 & $ -0.35$ & $0.12 $ \\
\protect\ion{Y}{2}          &         6 & $ -0.06$ & $0.06 $ &   &     7 & $ -0.18$ & $0.05 $ &   &     7 & $ -0.14$ & $0.05 $ &   &     7 & $ -0.15$ & $0.05 $ &   &     7 & $ -0.10$ & $0.06 $ &   &     8 & $ -0.10$ & $0.06 $ \\
\protect\ion{Zr}{2}         &         1 & $ +0.28$ & $0.11 $ &   &     1 & $ +0.23$ & $0.11 $ &   &     1 & $ +0.27$ & $0.11 $ &   &     1 & $ +0.26$ & $0.11 $ &   &     1 & $ +0.35$ & $0.11 $ &   &     1 & $ +0.36$ & $0.11 $ \\
\protect\ion{Ba}{2}         &         4 & $ -0.17$ & $0.10 $ &   &     4 & $ -0.34$ & $0.08 $ &   &     5 & $ -0.36$ & $0.08 $ &   &     5 & $ -0.41$ & $0.09 $ &   &     5 & $ -0.25$ & $0.09 $ &   &     5 & $ -0.29$ & $0.08 $ \\
\protect\ion{La}{2}         &         3 & $ +0.35$ & $0.07 $ &   &     4 & $ +0.14$ & $0.07 $ &   &     4 & $ +0.15$ & $0.07 $ &   &     4 & $ +0.15$ & $0.07 $ &   &     4 & $ +0.27$ & $0.07 $ &   &     3 & $ +0.27$ & $0.08 $ \\
\protect\ion{Ce}{2}         &         6 & $ +0.00$ & $0.08 $ &   &     6 & $ -0.13$ & $0.08 $ &   &     5 & $ -0.09$ & $0.07 $ &   &     4 & $ -0.21$ & $0.07 $ &   &     4 & $ +0.01$ & $0.08 $ &   &     2 & $ -0.03$ & $0.09 $ \\
\protect\ion{Nd}{2}         &        28 & $ +0.27$ & $0.05 $ &   &    22 & $ +0.13$ & $0.05 $ &   &    25 & $ +0.08$ & $0.05 $ &   &    19 & $ +0.07$ & $0.05 $ &   &    16 & $ +0.34$ & $0.05 $ &   &    16 & $ +0.35$ & $0.06 $ \\
\protect\ion{Sm}{2}         &         9 & $ +0.51$ & $0.06 $ &   &     5 & $ +0.34$ & $0.07 $ &   &     6 & $ +0.26$ & $0.07 $ &   &     4 & $ +0.26$ & $0.07 $ &   &     3 & $ +0.47$ & $0.09 $ &   &     2 & $ +0.55$ & $0.12 $ \\
\protect\ion{Eu}{2}         &         3 & $ +0.57$ & $0.07 $ &   &     2 & $ +0.32$ & $0.08 $ &   &     2 & $ +0.34$ & $0.08 $ &   &     2 & $ +0.29$ & $0.09 $ &   &     2 & $ +0.49$ & $0.08 $ &   &     2 & $ +0.52$ & $0.09 $ \\
\protect\ion{Dy}{2}         &         1 & $ +0.53$ & $0.11 $ &   &     2 & $ +0.21$ & $0.09 $ &   &     3 & $ +0.24$ & $0.08 $ &   &     1 & $ +0.19$ & $0.11 $ &   &     2 & $ +0.40$ & $0.10 $ &   &     2 & $ +0.48$ & $0.10 $ \\
\enddata
\end{deluxetable*}
\end{rotatetable*}

\addtocounter{table}{-1}
\begin{rotatetable*}
\begin{deluxetable*}{lrcccrcccrcccrcccrcccrcc}
\centerwidetable
\tabletypesize{\scriptsize}
\tablecolumns{24}
\tablewidth{0pt}
\tablecaption{Abundances ({\it continued})\label{tab:abund3}}
\tablehead{ & \multicolumn{3}{c}{I-67           } & & \multicolumn{3}{c}{III-82         } & & \multicolumn{3}{c}{IV-10          } & & \multicolumn{3}{c}{XII-34*        } & & \multicolumn{3}{c}{IV-79          } & & \multicolumn{3}{c}{IX-13          } \\ \cline{2-4}   \cline{6-8}   \cline{10-12} \cline{14-16} \cline{18-20} \cline{22-24}
\colhead{Element} & \colhead{$N$} & \colhead{[X/Fe]} & \colhead{error} & & \colhead{$N$} & \colhead{[X/Fe]} & \colhead{error} & & \colhead{$N$} & \colhead{[X/Fe]} & \colhead{error} & & \colhead{$N$} & \colhead{[X/Fe]} & \colhead{error} & & \colhead{$N$} & \colhead{[X/Fe]} & \colhead{error} & & \colhead{$N$} & \colhead{[X/Fe]} & \colhead{error}}
\startdata
\protect\ion{Fe}{1}         &        37 & $ -2.62$ & $0.06 $ &   &    31 & $ -2.48$ & $0.05 $ &   &    38 & $ -2.43$ & $0.06 $ &   &    49 & $ -2.50$ & $0.05 $ &   &    42 & $ -2.54$ & $0.05 $ &   &    44 & $ -2.43$ & $0.06 $ \\
\protect\ion{Fe}{2}         &        23 & $ -2.54$ & $0.05 $ &   &    20 & $ -2.38$ & $0.04 $ &   &    22 & $ -2.38$ & $0.04 $ &   &    23 & $ -2.40$ & $0.05 $ &   &    22 & $ -2.43$ & $0.04 $ &   &    17 & $ -2.35$ & $0.05 $ \\
\protect $A({\rm Li})$      &           & \nodata  & \nodata &   &       & \nodata  & \nodata &   &       & \nodata  & \nodata &   &       & \nodata  & \nodata &   &       & \nodata  & \nodata &   &       & \nodata  & \nodata \\
\protect CH                 &       syn & $ -0.21$ & $0.14 $ &     & syn & $ -0.48$ & $0.14 $ &     & syn & $ -0.31$ & $0.13 $ &     & syn & $ -0.65$ & $0.16 $ &     & syn & $ -0.35$ & $0.14 $ &     & syn & $ -0.50$ & $0.14 $ \\
\protect\ion{O}{1}          &         0 & \nodata  & \nodata &   &     1 & $<+0.44$ & \nodata &   &     1 & $ +0.87$ & $0.12 $ &   &     1 & $<+0.47$ & \nodata &   &     1 & $ +0.52$ & $0.13 $ &   &     0 & \nodata  & \nodata \\
\protect\ion{Na}{1}         &         0 & \nodata  & \nodata &   &     4 & $ +0.42$ & $0.06 $ &   &     2 & $ -0.22$ & $0.09 $ &   &     3 & $ +0.35$ & $0.07 $ &   &     4 & $ +0.27$ & $0.07 $ &   &     0 & \nodata  & \nodata \\
\protect\ion{Mg}{1}         &         6 & $ +0.53$ & $0.05 $ &   &     8 & $ +0.38$ & $0.05 $ &   &     8 & $ +0.63$ & $0.06 $ &   &     8 & $ +0.34$ & $0.05 $ &   &     8 & $ +0.51$ & $0.04 $ &   &     4 & $ -0.01$ & $0.06 $ \\
\protect\ion{Al}{1}         &         2 & $ +0.34$ & $0.39 $ &   &     2 & $ +1.01$ & $0.13 $ &   &     2 & $ +0.41$ & $0.35 $ &   &     2 & $ +1.41$ & $0.08 $ &   &     2 & $ +0.91$ & $0.21 $ &   &     2 & $ +1.35$ & $0.10 $ \\
\protect\ion{Si}{1}         &         2 & $ +0.48$ & $0.16 $ &   &     7 & $ +0.70$ & $0.07 $ &   &     3 & $ +0.73$ & $0.07 $ &   &     4 & $ +0.77$ & $0.07 $ &   &     4 & $ +0.63$ & $0.06 $ &   &     2 & $ +0.51$ & $0.08 $ \\
\protect\ion{K}{1}          &         0 & \nodata  & \nodata &   &     0 & \nodata  & \nodata &   &     1 & $ +0.63$ & $0.11 $ &   &     1 & $ +0.55$ & $0.10 $ &   &     1 & $ +0.64$ & $0.10 $ &   &     0 & \nodata  & \nodata \\
\protect\ion{Ca}{1}         &         7 & $ +0.33$ & $0.05 $ &   &    17 & $ +0.38$ & $0.04 $ &   &    18 & $ +0.47$ & $0.04 $ &   &    20 & $ +0.40$ & $0.03 $ &   &    18 & $ +0.39$ & $0.03 $ &   &     7 & $ +0.29$ & $0.05 $ \\
\protect\ion{Sc}{2}         &         6 & $ +0.05$ & $0.09 $ &   &     7 & $ +0.04$ & $0.07 $ &   &     8 & $ +0.06$ & $0.07 $ &   &     8 & $ +0.09$ & $0.07 $ &   &     7 & $ +0.06$ & $0.06 $ &   &     6 & $ +0.10$ & $0.07 $ \\
\protect\ion{Ti}{1}         &        30 & $ +0.36$ & $0.05 $ &   &    29 & $ +0.38$ & $0.04 $ &   &    28 & $ +0.45$ & $0.04 $ &   &    33 & $ +0.43$ & $0.04 $ &   &    27 & $ +0.36$ & $0.04 $ &   &    32 & $ +0.40$ & $0.04 $ \\
\protect\ion{Ti}{2}         &        31 & $ +0.38$ & $0.04 $ &   &    38 & $ +0.29$ & $0.04 $ &   &    33 & $ +0.41$ & $0.04 $ &   &    38 & $ +0.32$ & $0.04 $ &   &    35 & $ +0.32$ & $0.04 $ &   &    37 & $ +0.34$ & $0.04 $ \\
\protect\ion{V}{1}          &         2 & $ +0.28$ & $0.26 $ &   &     2 & $ +0.16$ & $0.26 $ &   &     2 & $ +0.26$ & $0.21 $ &   &     2 & $ +0.25$ & $0.22 $ &   &     2 & $ +0.23$ & $0.24 $ &   &     1 & $ -0.09$ & $0.11 $ \\
\protect\ion{Cr}{1}         &        12 & $ -0.15$ & $0.05 $ &   &    12 & $ -0.07$ & $0.05 $ &   &    11 & $ -0.07$ & $0.05 $ &   &    12 & $ -0.10$ & $0.04 $ &   &    12 & $ -0.11$ & $0.04 $ &   &    13 & $ -0.10$ & $0.04 $ \\
\protect\ion{Cr}{2}         &         5 & $ +0.04$ & $0.07 $ &   &     5 & $ +0.02$ & $0.05 $ &   &     6 & $ +0.07$ & $0.06 $ &   &     5 & $ +0.01$ & $0.06 $ &   &     6 & $ +0.05$ & $0.05 $ &   &     5 & $ +0.04$ & $0.06 $ \\
\protect\ion{Mn}{1}         &         6 & $ -0.37$ & $0.05 $ &   &     6 & $ -0.37$ & $0.05 $ &   &     6 & $ -0.36$ & $0.05 $ &   &     6 & $ -0.34$ & $0.05 $ &   &     5 & $ -0.36$ & $0.05 $ &   &     6 & $ -0.31$ & $0.05 $ \\
\protect\ion{Co}{1}         &         6 & $ +0.25$ & $0.11 $ &   &     6 & $ +0.27$ & $0.12 $ &   &     6 & $ +0.22$ & $0.10 $ &   &     6 & $ +0.25$ & $0.10 $ &   &     6 & $ +0.22$ & $0.10 $ &   &     6 & $ +0.21$ & $0.09 $ \\
\protect\ion{Ni}{1}         &        12 & $ +0.12$ & $0.04 $ &   &    14 & $ +0.07$ & $0.04 $ &   &    14 & $ +0.07$ & $0.04 $ &   &    17 & $ +0.05$ & $0.04 $ &   &    16 & $ +0.07$ & $0.04 $ &   &    14 & $ +0.04$ & $0.04 $ \\
\protect\ion{Zn}{1}         &         2 & $ +0.28$ & $0.08 $ &   &     2 & $ +0.26$ & $0.08 $ &   &     2 & $ +0.23$ & $0.08 $ &   &     2 & $ +0.31$ & $0.08 $ &   &     2 & $ +0.22$ & $0.08 $ &   &     2 & $ +0.19$ & $0.08 $ \\
\protect\ion{Sr}{2}         &         2 & $ -0.24$ & $0.11 $ &   &     1 & $ -0.22$ & $0.13 $ &   &     2 & $ -0.23$ & $0.11 $ &   &     2 & $ -0.25$ & $0.11 $ &   &     2 & $ -0.38$ & $0.15 $ &   &     2 & $ -0.21$ & $0.11 $ \\
\protect\ion{Y}{2}          &         7 & $ +0.00$ & $0.06 $ &   &     7 & $ -0.18$ & $0.05 $ &   &     8 & $ -0.15$ & $0.05 $ &   &     8 & $ -0.12$ & $0.05 $ &   &     8 & $ -0.18$ & $0.05 $ &   &     7 & $ -0.19$ & $0.06 $ \\
\protect\ion{Zr}{2}         &         1 & $ +0.47$ & $0.11 $ &   &     1 & $ +0.21$ & $0.11 $ &   &     1 & $ +0.23$ & $0.11 $ &   &     1 & $ +0.33$ & $0.11 $ &   &     1 & $ +0.19$ & $0.11 $ &   &     1 & $ +0.21$ & $0.11 $ \\
\protect\ion{Ba}{2}         &         2 & $ -0.05$ & $0.12 $ &   &     5 & $ -0.38$ & $0.07 $ &   &     5 & $ -0.40$ & $0.08 $ &   &     5 & $ -0.29$ & $0.08 $ &   &     5 & $ -0.26$ & $0.08 $ &   &     2 & $ -0.43$ & $0.11 $ \\
\protect\ion{La}{2}         &         4 & $ +0.49$ & $0.07 $ &   &     3 & $ +0.14$ & $0.07 $ &   &     3 & $ +0.15$ & $0.08 $ &   &     4 & $ +0.07$ & $0.07 $ &   &     4 & $ +0.15$ & $0.07 $ &   &     4 & $ +0.10$ & $0.08 $ \\
\protect\ion{Ce}{2}         &         5 & $ +0.28$ & $0.09 $ &   &     5 & $ -0.07$ & $0.09 $ &   &     2 & $ +0.01$ & $0.11 $ &   &     2 & $ -0.05$ & $0.12 $ &   &     4 & $ -0.00$ & $0.08 $ &   &     1 & $<-0.27$ & \nodata \\
\protect\ion{Nd}{2}         &        18 & $ +0.52$ & $0.06 $ &   &    17 & $ +0.16$ & $0.06 $ &   &    10 & $ +0.21$ & $0.08 $ &   &    12 & $ +0.16$ & $0.06 $ &   &    19 & $ +0.19$ & $0.06 $ &   &     4 & $ +0.23$ & $0.09 $ \\
\protect\ion{Sm}{2}         &         6 & $ +0.66$ & $0.08 $ &   &     6 & $ +0.30$ & $0.06 $ &   &     1 & $<-0.44$ & \nodata &   &     1 & $<+0.16$ & \nodata &   &     4 & $ +0.45$ & $0.08 $ &   &     1 & $ +0.31$ & $0.19 $ \\
\protect\ion{Eu}{2}         &         2 & $ +0.68$ & $0.09 $ &   &     2 & $ +0.32$ & $0.08 $ &   &     2 & $ +0.37$ & $0.09 $ &   &     2 & $ +0.39$ & $0.08 $ &   &     3 & $ +0.46$ & $0.08 $ &   &     1 & $ +0.31$ & $0.12 $ \\
\protect\ion{Dy}{2}         &         3 & $ +0.70$ & $0.08 $ &   &     0 & \nodata  & \nodata &   &     1 & $ +0.14$ & $0.14 $ &   &     2 & $ +0.14$ & $0.15 $ &   &     3 & $ +0.31$ & $0.11 $ &   &     3 & $ +0.39$ & $0.11 $ \\
\enddata
\end{deluxetable*}
\end{rotatetable*}

\addtocounter{table}{-1}
\begin{rotatetable*}
\begin{deluxetable*}{lrcccrcccrcccrcccrcccrcc}
\centerwidetable
\tabletypesize{\scriptsize}
\tablecolumns{24}
\tablewidth{0pt}
\tablecaption{Abundances ({\it continued})\label{tab:abund4}}
\tablehead{ & \multicolumn{3}{c}{VIII-24*       } & & \multicolumn{3}{c}{X-20           } & & \multicolumn{3}{c}{S2710          } & & \multicolumn{3}{c}{VI-90          } & & \multicolumn{3}{c}{S2265          } & & \multicolumn{3}{c}{VIII-45        } \\ \cline{2-4}   \cline{6-8}   \cline{10-12} \cline{14-16} \cline{18-20} \cline{22-24}
\colhead{Element} & \colhead{$N$} & \colhead{[X/Fe]} & \colhead{error} & & \colhead{$N$} & \colhead{[X/Fe]} & \colhead{error} & & \colhead{$N$} & \colhead{[X/Fe]} & \colhead{error} & & \colhead{$N$} & \colhead{[X/Fe]} & \colhead{error} & & \colhead{$N$} & \colhead{[X/Fe]} & \colhead{error} & & \colhead{$N$} & \colhead{[X/Fe]} & \colhead{error}}
\startdata
\protect\ion{Fe}{1}         &        40 & $ -2.49$ & $0.05 $ &   &    44 & $ -2.45$ & $0.05 $ &   &    52 & $ -2.41$ & $0.05 $ &   &    41 & $ -2.50$ & $0.05 $ &   &    47 & $ -2.51$ & $0.06 $ &   &    45 & $ -2.57$ & $0.06 $ \\
\protect\ion{Fe}{2}         &        18 & $ -2.40$ & $0.05 $ &   &    19 & $ -2.40$ & $0.07 $ &   &    18 & $ -2.40$ & $0.06 $ &   &    20 & $ -2.41$ & $0.06 $ &   &    22 & $ -2.42$ & $0.05 $ &   &    17 & $ -2.45$ & $0.06 $ \\
\protect $A({\rm Li})$      &           & \nodata  & \nodata &   &       & \nodata  & \nodata &   &       & \nodata  & \nodata &   &       & \nodata  & \nodata &   &       & \nodata  & \nodata &   &       & \nodata  & \nodata \\
\protect CH                 &       syn & $ -0.39$ & $0.17 $ &     & syn & $ +0.00$ & $0.13 $ &     & syn & $ +0.34$ & $0.12 $ &     & syn & $ -0.05$ & $0.16 $ &     & syn & $ +0.45$ & $0.16 $ &     & syn & $ +0.39$ & $0.16 $ \\
\protect\ion{O}{1}          &         0 & \nodata  & \nodata &   &     1 & $<+0.87$ & \nodata &   &     1 & $<+1.11$ & \nodata &   &     1 & $<+0.83$ & \nodata &   &     1 & $<+1.34$ & \nodata &   &     1 & $<+0.66$ & \nodata \\
\protect\ion{Na}{1}         &         0 & \nodata  & \nodata &   &     3 & $ +0.30$ & $0.08 $ &   &     2 & $ -0.22$ & $0.08 $ &   &     4 & $ +0.40$ & $0.07 $ &   &     2 & $ -0.13$ & $0.09 $ &   &     2 & $ -0.11$ & $0.09 $ \\
\protect\ion{Mg}{1}         &         6 & $ +0.59$ & $0.05 $ &   &     8 & $ +0.30$ & $0.05 $ &   &     8 & $ +0.61$ & $0.05 $ &   &     7 & $ -0.02$ & $0.06 $ &   &     7 & $ +0.58$ & $0.05 $ &   &     8 & $ +0.64$ & $0.05 $ \\
\protect\ion{Al}{1}         &         2 & $ +0.53$ & $0.17 $ &   &     2 & $ +1.01$ & $0.25 $ &   &     2 & $ +0.62$ & $0.39 $ &   &     2 & $ +0.64$ & $0.20 $ &   &     2 & $ +0.71$ & $0.41 $ &   &     2 & $ +0.33$ & $0.30 $ \\
\protect\ion{Si}{1}         &         2 & $ +0.46$ & $0.10 $ &   &     3 & $ +0.64$ & $0.08 $ &   &     4 & $ +0.56$ & $0.08 $ &   &     3 & $ +0.84$ & $0.08 $ &   &     3 & $ +0.64$ & $0.09 $ &   &     3 & $ +0.51$ & $0.08 $ \\
\protect\ion{K}{1}          &         0 & \nodata  & \nodata &   &     1 & $ +0.85$ & $0.14 $ &   &     1 & $ +0.58$ & $0.11 $ &   &     1 & $ +0.62$ & $0.11 $ &   &     1 & $ +0.49$ & $0.11 $ &   &     1 & $ +0.54$ & $0.11 $ \\
\protect\ion{Ca}{1}         &         7 & $ +0.33$ & $0.05 $ &   &    20 & $ +0.35$ & $0.04 $ &   &    20 & $ +0.35$ & $0.04 $ &   &    20 & $ +0.39$ & $0.04 $ &   &    18 & $ +0.37$ & $0.04 $ &   &    20 & $ +0.43$ & $0.04 $ \\
\protect\ion{Sc}{2}         &         6 & $ +0.17$ & $0.06 $ &   &     6 & $ +0.17$ & $0.07 $ &   &     8 & $ +0.23$ & $0.08 $ &   &     7 & $ +0.10$ & $0.06 $ &   &     6 & $ +0.07$ & $0.06 $ &   &     7 & $ +0.03$ & $0.06 $ \\
\protect\ion{Ti}{1}         &        23 & $ +0.37$ & $0.04 $ &   &    26 & $ +0.34$ & $0.04 $ &   &    27 & $ +0.38$ & $0.04 $ &   &    26 & $ +0.43$ & $0.04 $ &   &    24 & $ +0.33$ & $0.05 $ &   &    32 & $ +0.49$ & $0.05 $ \\
\protect\ion{Ti}{2}         &        34 & $ +0.37$ & $0.04 $ &   &    33 & $ +0.38$ & $0.05 $ &   &    31 & $ +0.43$ & $0.04 $ &   &    37 & $ +0.39$ & $0.05 $ &   &    32 & $ +0.34$ & $0.05 $ &   &    32 & $ +0.33$ & $0.04 $ \\
\protect\ion{V}{1}          &         2 & $ +0.24$ & $0.19 $ &   &     1 & $ -0.15$ & $0.16 $ &   &     1 & $ -0.03$ & $0.12 $ &   &     2 & $ +0.20$ & $0.16 $ &   &     2 & $ +0.25$ & $0.13 $ &   &     2 & $ +0.21$ & $0.16 $ \\
\protect\ion{Cr}{1}         &        12 & $ -0.12$ & $0.05 $ &   &     8 & $ -0.17$ & $0.06 $ &   &    11 & $ -0.04$ & $0.05 $ &   &    12 & $ -0.13$ & $0.05 $ &   &    11 & $ -0.05$ & $0.06 $ &   &    13 & $ +0.05$ & $0.13 $ \\
\protect\ion{Cr}{2}         &         6 & $ +0.05$ & $0.06 $ &   &     3 & $ -0.02$ & $0.11 $ &   &     4 & $ +0.07$ & $0.07 $ &   &     3 & $ +0.05$ & $0.08 $ &   &     3 & $ -0.05$ & $0.09 $ &   &     3 & $ +0.07$ & $0.08 $ \\
\protect\ion{Mn}{1}         &         4 & $ -0.39$ & $0.06 $ &   &     3 & $ -0.34$ & $0.09 $ &   &     4 & $ -0.37$ & $0.06 $ &   &     4 & $ -0.26$ & $0.07 $ &   &     4 & $ -0.32$ & $0.07 $ &   &     4 & $ -0.30$ & $0.07 $ \\
\protect\ion{Co}{1}         &         5 & $ +0.24$ & $0.08 $ &   &     6 & $ +0.27$ & $0.09 $ &   &     5 & $ +0.22$ & $0.08 $ &   &     5 & $ +0.26$ & $0.07 $ &   &     6 & $ +0.32$ & $0.11 $ &   &     6 & $ +0.26$ & $0.07 $ \\
\protect\ion{Ni}{1}         &        12 & $ +0.02$ & $0.05 $ &   &    14 & $ +0.09$ & $0.05 $ &   &    17 & $ +0.07$ & $0.05 $ &   &    15 & $ +0.08$ & $0.05 $ &   &    10 & $ +0.03$ & $0.06 $ &   &    15 & $ +0.08$ & $0.05 $ \\
\protect\ion{Zn}{1}         &         2 & $ +0.25$ & $0.08 $ &   &     1 & $ +0.17$ & $0.14 $ &   &     2 & $ +0.19$ & $0.09 $ &   &     2 & $ +0.21$ & $0.11 $ &   &     2 & $ +0.25$ & $0.16 $ &   &     2 & $ +0.31$ & $0.09 $ \\
\protect\ion{Sr}{2}         &         2 & $ -0.07$ & $0.13 $ &   &     2 & $ -0.13$ & $0.13 $ &   &     2 & $ -0.33$ & $0.12 $ &   &     2 & $ -0.23$ & $0.13 $ &   &     2 & $ -0.17$ & $0.12 $ &   &     2 & $ -0.13$ & $0.13 $ \\
\protect\ion{Y}{2}          &         7 & $ -0.13$ & $0.07 $ &   &     3 & $ -0.04$ & $0.10 $ &   &     4 & $ -0.14$ & $0.10 $ &   &     5 & $ -0.20$ & $0.08 $ &   &     4 & $ -0.23$ & $0.09 $ &   &     6 & $ -0.12$ & $0.08 $ \\
\protect\ion{Zr}{2}         &         1 & $ +0.38$ & $0.11 $ &   &     1 & $ +0.28$ & $0.15 $ &   &     1 & $ +0.23$ & $0.13 $ &   &     1 & $ +0.31$ & $0.13 $ &   &     1 & $ +0.35$ & $0.17 $ &   &     1 & $ +0.38$ & $0.13 $ \\
\protect\ion{Ba}{2}         &         2 & $ -0.30$ & $0.12 $ &   &     5 & $ -0.39$ & $0.09 $ &   &     5 & $ -0.41$ & $0.07 $ &   &     4 & $ -0.27$ & $0.08 $ &   &     5 & $ -0.37$ & $0.08 $ &   &     4 & $ -0.17$ & $0.08 $ \\
\protect\ion{La}{2}         &         4 & $ +0.21$ & $0.07 $ &   &     1 & $ +0.42$ & $0.16 $ &   &     1 & $ -0.13$ & $0.20 $ &   &     3 & $ +0.16$ & $0.11 $ &   &     1 & $ +0.27$ & $0.16 $ &   &     3 & $ +0.30$ & $0.09 $ \\
\protect\ion{Ce}{2}         &         1 & $ +0.02$ & $0.16 $ &   &     1 & $<+0.43$ & \nodata &   &     1 & $<+0.18$ & \nodata &   &     1 & $<-0.11$ & \nodata &   &     1 & $<-0.31$ & \nodata &   &     1 & $<-0.36$ & \nodata \\
\protect\ion{Nd}{2}         &         2 & $ +0.12$ & $0.11 $ &   &     1 & $<+0.11$ & \nodata &   &     1 & $<-0.12$ & \nodata &   &     1 & $<+0.13$ & \nodata &   &     1 & $<+0.27$ & \nodata &   &     1 & $<+0.12$ & \nodata \\
\protect\ion{Sm}{2}         &         1 & $<+0.29$ & \nodata &   &     1 & $<+0.30$ & \nodata &   &     1 & $<+0.33$ & \nodata &   &     1 & $<+0.19$ & \nodata &   &     1 & $<+0.49$ & \nodata &   &     1 & $<+0.78$ & \nodata \\
\protect\ion{Eu}{2}         &         2 & $ +0.34$ & $0.09 $ &   &     2 & $ +0.42$ & $0.12 $ &   &     1 & $ +0.19$ & $0.16 $ &   &     2 & $ +0.39$ & $0.09 $ &   &     2 & $ +0.32$ & $0.10 $ &   &     2 & $ +0.53$ & $0.10 $ \\
\protect\ion{Dy}{2}         &         1 & $<+0.44$ & \nodata &   &     1 & $ +0.44$ & $0.20 $ &   &     1 & $<+0.96$ & \nodata &   &     1 & $<+0.89$ & \nodata &   &     1 & $<+0.52$ & \nodata &   &     1 & $<+1.25$ & \nodata \\
\enddata
\end{deluxetable*}
\end{rotatetable*}

\addtocounter{table}{-1}
\begin{rotatetable*}
\begin{deluxetable*}{lrcccrcccrcccrcccrcccrcc}
\centerwidetable
\tabletypesize{\scriptsize}
\tablecolumns{24}
\tablewidth{0pt}
\tablecaption{Abundances ({\it continued})\label{tab:abund5}}
\tablehead{ & \multicolumn{3}{c}{VII-28         } & & \multicolumn{3}{c}{G17181\_0638   } & & \multicolumn{3}{c}{C17333\_0832   } & & \multicolumn{3}{c}{S3108          } & & \multicolumn{3}{c}{S652           } & & \multicolumn{3}{c}{S19            } \\ \cline{2-4}   \cline{6-8}   \cline{10-12} \cline{14-16} \cline{18-20} \cline{22-24}
\colhead{Element} & \colhead{$N$} & \colhead{[X/Fe]} & \colhead{error} & & \colhead{$N$} & \colhead{[X/Fe]} & \colhead{error} & & \colhead{$N$} & \colhead{[X/Fe]} & \colhead{error} & & \colhead{$N$} & \colhead{[X/Fe]} & \colhead{error} & & \colhead{$N$} & \colhead{[X/Fe]} & \colhead{error} & & \colhead{$N$} & \colhead{[X/Fe]} & \colhead{error}}
\startdata
\protect\ion{Fe}{1}         &        41 & $ -2.48$ & $0.06 $ &   &    26 & $ -2.57$ & $0.08 $ &   &    46 & $ -2.47$ & $0.07 $ &   &    37 & $ -2.53$ & $0.07 $ &   &    33 & $ -2.50$ & $0.08 $ &   &    32 & $ -2.49$ & $0.06 $ \\
\protect\ion{Fe}{2}         &        19 & $ -2.37$ & $0.06 $ &   &     6 & $ -2.53$ & $0.12 $ &   &    17 & $ -2.40$ & $0.06 $ &   &    14 & $ -2.50$ & $0.07 $ &   &    17 & $ -2.37$ & $0.07 $ &   &    13 & $ -2.46$ & $0.08 $ \\
\protect $A({\rm Li})$      &           & \nodata  & \nodata &   &       & \nodata  & \nodata &   &       & \nodata  & \nodata &   &       & \nodata  & \nodata &   &       & \nodata  & \nodata &     & syn & $ +1.95$ & $0.13 $ \\
\protect CH                 &       syn & $ +0.01$ & $0.19 $ &   &       & \nodata  & \nodata &     & syn & $ +0.17$ & $0.17 $ &     & syn & $ +0.37$ & $0.18 $ &     & syn & $ +0.24$ & $0.16 $ &     & syn & $ +0.49$ & $0.13 $ \\
\protect\ion{O}{1}          &         1 & $<+0.65$ & \nodata &   &     1 & $<+1.02$ & \nodata &   &     1 & $<+1.19$ & \nodata &   &     1 & $<+1.85$ & \nodata &   &     1 & $<+1.37$ & \nodata &   &     1 & $<+2.42$ & \nodata \\
\protect\ion{Na}{1}         &         4 & $ +0.34$ & $0.07 $ &   &     2 & $ -0.28$ & $0.12 $ &   &     3 & $ +0.30$ & $0.07 $ &   &     2 & $ -0.29$ & $0.09 $ &   &     2 & $ -0.12$ & $0.10 $ &   &     2 & $ +0.06$ & $0.09 $ \\
\protect\ion{Mg}{1}         &         6 & $ +0.21$ & $0.05 $ &   &     4 & $ +0.56$ & $0.07 $ &   &     7 & $ +0.40$ & $0.06 $ &   &     7 & $ +0.54$ & $0.06 $ &   &     6 & $ +0.50$ & $0.06 $ &   &     7 & $ +0.61$ & $0.06 $ \\
\protect\ion{Al}{1}         &         2 & $ +0.66$ & $0.15 $ &   &     0 & \nodata  & \nodata &   &     2 & $ +0.47$ & $0.13 $ &   &     2 & $ +0.48$ & $0.35 $ &   &     2 & $ +0.40$ & $0.22 $ &   &     2 & $ +0.26$ & $0.17 $ \\
\protect\ion{Si}{1}         &         4 & $ +0.65$ & $0.08 $ &   &     0 & \nodata  & \nodata &   &     2 & $ +0.51$ & $0.16 $ &   &     2 & $ +0.56$ & $0.20 $ &   &     3 & $ +0.76$ & $0.14 $ &   &     2 & $ +0.42$ & $0.12 $ \\
\protect\ion{K}{1}          &         1 & $ +0.62$ & $0.11 $ &   &     0 & \nodata  & \nodata &   &     1 & $ +0.49$ & $0.12 $ &   &     1 & $ +0.40$ & $0.13 $ &   &     1 & $ +0.54$ & $0.13 $ &   &     1 & $<+0.20$ & \nodata \\
\protect\ion{Ca}{1}         &        17 & $ +0.42$ & $0.04 $ &   &    11 & $ +0.34$ & $0.06 $ &   &    19 & $ +0.37$ & $0.04 $ &   &    17 & $ +0.38$ & $0.05 $ &   &    16 & $ +0.38$ & $0.05 $ &   &    14 & $ +0.42$ & $0.05 $ \\
\protect\ion{Sc}{2}         &         7 & $ +0.09$ & $0.07 $ &   &     1 & $ +0.17$ & $0.19 $ &   &     7 & $ +0.02$ & $0.07 $ &   &     4 & $ +0.12$ & $0.09 $ &   &     5 & $ +0.06$ & $0.08 $ &   &     5 & $ +0.09$ & $0.09 $ \\
\protect\ion{Ti}{1}         &        22 & $ +0.42$ & $0.05 $ &   &    10 & $ +0.43$ & $0.07 $ &   &    26 & $ +0.37$ & $0.05 $ &   &    16 & $ +0.39$ & $0.06 $ &   &    16 & $ +0.32$ & $0.06 $ &   &     9 & $ +0.41$ & $0.06 $ \\
\protect\ion{Ti}{2}         &        33 & $ +0.37$ & $0.05 $ &   &     6 & $ +0.37$ & $0.12 $ &   &    25 & $ +0.32$ & $0.05 $ &   &    23 & $ +0.39$ & $0.07 $ &   &    18 & $ +0.28$ & $0.07 $ &   &    17 & $ +0.32$ & $0.07 $ \\
\protect\ion{V}{1}          &         1 & $ +0.05$ & $0.12 $ &   &     0 & \nodata  & \nodata &   &     2 & $ +0.12$ & $0.10 $ &   &     0 & \nodata  & \nodata &   &     0 & \nodata  & \nodata &   &     1 & $<-0.28$ & \nodata \\
\protect\ion{Cr}{1}         &        10 & $ -0.04$ & $0.06 $ &   &     6 & $ -0.10$ & $0.08 $ &   &     6 & $ -0.09$ & $0.07 $ &   &     4 & $ -0.17$ & $0.08 $ &   &     5 & $ -0.15$ & $0.09 $ &   &     2 & $ -0.13$ & $0.09 $ \\
\protect\ion{Cr}{2}         &         4 & $ -0.01$ & $0.07 $ &   &     1 & $ +0.31$ & $0.21 $ &   &     3 & $ -0.09$ & $0.09 $ &   &     0 & \nodata  & \nodata &   &     1 & $ +0.02$ & $0.16 $ &   &     1 & $<-0.13$ & \nodata \\
\protect\ion{Mn}{1}         &         4 & $ -0.40$ & $0.07 $ &   &     1 & $ -0.17$ & $0.21 $ &   &     4 & $ -0.43$ & $0.08 $ &   &     1 & $ -0.52$ & $0.16 $ &   &     3 & $ -0.32$ & $0.09 $ &   &     1 & $<-0.24$ & \nodata \\
\protect\ion{Co}{1}         &         6 & $ +0.37$ & $0.09 $ &   &     0 & \nodata  & \nodata &   &     5 & $ +0.32$ & $0.16 $ &   &     4 & $ +0.20$ & $0.09 $ &   &     3 & $ +0.17$ & $0.10 $ &   &     4 & $ +0.27$ & $0.08 $ \\
\protect\ion{Ni}{1}         &        15 & $ +0.04$ & $0.05 $ &   &     7 & $ +0.21$ & $0.08 $ &   &    11 & $ +0.03$ & $0.06 $ &   &     7 & $ +0.05$ & $0.07 $ &   &     3 & $ +0.02$ & $0.10 $ &   &     4 & $ +0.15$ & $0.08 $ \\
\protect\ion{Zn}{1}         &         2 & $ +0.29$ & $0.09 $ &   &     1 & $<+0.35$ & \nodata &   &     1 & $ +0.19$ & $0.20 $ &   &     1 & $ +0.16$ & $0.18 $ &   &     1 & $ +0.28$ & $0.16 $ &   &     1 & $ +0.36$ & $0.19 $ \\
\protect\ion{Sr}{2}         &         2 & $ -0.17$ & $0.12 $ &   &     0 & \nodata  & \nodata &   &     2 & $ -0.27$ & $0.13 $ &   &     2 & $ -0.32$ & $0.16 $ &   &     2 & $ -0.45$ & $0.18 $ &   &     2 & $ -0.42$ & $0.15 $ \\
\protect\ion{Y}{2}          &         7 & $ -0.21$ & $0.07 $ &   &     2 & $ +0.05$ & $0.17 $ &   &     3 & $ -0.09$ & $0.10 $ &   &     1 & $ -0.04$ & $0.16 $ &   &     1 & $<-0.25$ & \nodata &   &     1 & $<-0.22$ & \nodata \\
\protect\ion{Zr}{2}         &         1 & $ +0.27$ & $0.13 $ &   &     0 & \nodata  & \nodata &   &     1 & $ +0.22$ & $0.15 $ &   &     1 & $<+0.71$ & \nodata &   &     1 & $<+0.69$ & \nodata &   &     1 & $<+0.83$ & \nodata \\
\protect\ion{Ba}{2}         &         4 & $ -0.32$ & $0.08 $ &   &     4 & $ -0.11$ & $0.12 $ &   &     4 & $ -0.43$ & $0.08 $ &   &     5 & $ -0.12$ & $0.09 $ &   &     4 & $ -0.49$ & $0.09 $ &   &     2 & $ -0.50$ & $0.10 $ \\
\protect\ion{La}{2}         &         1 & $<+0.17$ & \nodata &   &     0 & \nodata  & \nodata &   &     1 & $<+0.36$ & \nodata &   &     1 & $<+0.58$ & \nodata &   &     1 & $<+0.44$ & \nodata &   &     1 & $<+0.86$ & \nodata \\
\protect\ion{Ce}{2}         &         1 & $<-0.07$ & \nodata &   &     0 & \nodata  & \nodata &   &     1 & $<+0.04$ & \nodata &   &     1 & $<+0.25$ & \nodata &   &     1 & $<+0.55$ & \nodata &   &     1 & $<+0.74$ & \nodata \\
\protect\ion{Nd}{2}         &         1 & $<-0.01$ & \nodata &   &     1 & $<+0.60$ & \nodata &   &     1 & $<-0.42$ & \nodata &   &     1 & $<-0.02$ & \nodata &   &     1 & $<+0.25$ & \nodata &   &     1 & $<+0.81$ & \nodata \\
\protect\ion{Sm}{2}         &         1 & $<+0.40$ & \nodata &   &     1 & $<+1.20$ & \nodata &   &     1 & $<+0.67$ & \nodata &   &     1 & $<+0.82$ & \nodata &   &     1 & $<+0.59$ & \nodata &   &     1 & $<+1.04$ & \nodata \\
\protect\ion{Eu}{2}         &         1 & $ +0.30$ & $0.15 $ &   &     1 & $<+1.06$ & \nodata &   &     1 & $<+0.38$ & \nodata &   &     1 & $ +0.64$ & $0.14 $ &   &     1 & $<+0.50$ & \nodata &   &     1 & $<+0.76$ & \nodata \\
\protect\ion{Dy}{2}         &         1 & $<+0.88$ & \nodata &   &     0 & \nodata  & \nodata &   &     1 & $<+1.15$ & \nodata &   &     1 & $<+1.32$ & \nodata &   &     1 & $<+1.32$ & \nodata &   &     1 & $<+1.86$ & \nodata \\
\enddata
\end{deluxetable*}
\end{rotatetable*}

\addtocounter{table}{-1}
\begin{rotatetable*}
\begin{deluxetable*}{lrcccrcccrcccrcccrcc}
\centerwidetable
\tabletypesize{\scriptsize}
\tablecolumns{20}
\tablewidth{0pt}
\tablecaption{Abundances ({\it continued})\label{tab:abund6}}
\tablehead{ & \multicolumn{3}{c}{D21            } & & \multicolumn{3}{c}{S3880          } & & \multicolumn{3}{c}{S4038          } & & \multicolumn{3}{c}{S61            } & & \multicolumn{3}{c}{S162           } \\ \cline{2-4}   \cline{6-8}   \cline{10-12} \cline{14-16} \cline{18-20}
\colhead{Element} & \colhead{$N$} & \colhead{[X/Fe]} & \colhead{error} & & \colhead{$N$} & \colhead{[X/Fe]} & \colhead{error} & & \colhead{$N$} & \colhead{[X/Fe]} & \colhead{error} & & \colhead{$N$} & \colhead{[X/Fe]} & \colhead{error} & & \colhead{$N$} & \colhead{[X/Fe]} & \colhead{error}}
\startdata
\protect\ion{Fe}{1}         &        26 & $ -2.36$ & $0.07 $ &   &    41 & $ -2.46$ & $0.07 $ &   &    31 & $ -2.42$ & $0.07 $ &   &    15 & $ -2.63$ & $0.10 $ &   &    16 & $ -2.37$ & $0.08 $ \\
\protect\ion{Fe}{2}         &         9 & $ -2.37$ & $0.08 $ &   &    15 & $ -2.39$ & $0.08 $ &   &    14 & $ -2.44$ & $0.08 $ &   &     6 & $ -2.54$ & $0.12 $ &   &    11 & $ -2.44$ & $0.09 $ \\
\protect $A({\rm Li})$      &       syn & $ +2.01$ & $0.13 $ &     & syn & $ +2.16$ & $0.13 $ &     & syn & $ +2.08$ & $0.13 $ &     & syn & $ +2.23$ & $0.13 $ &     & syn & $ +2.13$ & $0.16 $ \\
\protect CH                 &       syn & $ +0.06$ & $0.16 $ &     & syn & $ +0.26$ & $0.14 $ &     & syn & $ +0.08$ & $0.16 $ &   &       & \nodata  & \nodata &     & syn & $ +0.23$ & $0.16 $ \\
\protect\ion{O}{1}          &         1 & $<+1.62$ & \nodata &   &     1 & $<+2.33$ & \nodata &   &     1 & $<+2.00$ & \nodata &   &     1 & $<+2.07$ & \nodata &   &     1 & $<+2.69$ & \nodata \\
\protect\ion{Na}{1}         &         2 & $ +0.26$ & $0.08 $ &   &     2 & $ +0.19$ & $0.09 $ &   &     2 & $ +0.18$ & $0.12 $ &   &     2 & $ -0.34$ & $0.12 $ &   &     2 & $ +0.18$ & $0.14 $ \\
\protect\ion{Mg}{1}         &         3 & $ +0.06$ & $0.09 $ &   &     5 & $ +0.54$ & $0.07 $ &   &     2 & $ -0.04$ & $0.12 $ &   &     4 & $ +0.64$ & $0.09 $ &   &     2 & $ -0.09$ & $0.10 $ \\
\protect\ion{Al}{1}         &         2 & $ +0.30$ & $0.17 $ &   &     2 & $ +0.21$ & $0.24 $ &   &     2 & $ +0.31$ & $0.19 $ &   &     2 & $ -0.11$ & $0.13 $ &   &     2 & $ +0.55$ & $0.16 $ \\
\protect\ion{Si}{1}         &         3 & $ +0.88$ & $0.11 $ &   &     1 & $ +0.22$ & $0.18 $ &   &     1 & $ +0.57$ & $0.18 $ &   &     1 & $ +0.57$ & $0.33 $ &   &     1 & $ +0.65$ & $0.24 $ \\
\protect\ion{K}{1}          &         1 & $<+0.58$ & \nodata &   &     1 & $<+0.32$ & \nodata &   &     1 & $<+0.51$ & \nodata &   &     1 & $<+0.58$ & \nodata &   &     1 & $<+0.59$ & \nodata \\
\protect\ion{Ca}{1}         &        11 & $ +0.32$ & $0.05 $ &   &    12 & $ +0.37$ & $0.06 $ &   &    14 & $ +0.38$ & $0.05 $ &   &     9 & $ +0.48$ & $0.08 $ &   &    10 & $ +0.46$ & $0.06 $ \\
\protect\ion{Sc}{2}         &         5 & $ +0.07$ & $0.09 $ &   &     5 & $ -0.09$ & $0.09 $ &   &     3 & $ +0.12$ & $0.10 $ &   &     2 & $ +0.04$ & $0.14 $ &   &     5 & $ +0.24$ & $0.10 $ \\
\protect\ion{Ti}{1}         &         5 & $ +0.66$ & $0.23 $ &   &     6 & $ +0.44$ & $0.08 $ &   &     5 & $ +0.42$ & $0.08 $ &   &     2 & $ +1.94$ & $0.15 $ &   &     0 & \nodata  & \nodata \\
\protect\ion{Ti}{2}         &        14 & $ +0.30$ & $0.07 $ &   &    15 & $ +0.22$ & $0.07 $ &   &    12 & $ +0.32$ & $0.08 $ &   &     8 & $ +0.21$ & $0.10 $ &   &    14 & $ +0.38$ & $0.11 $ \\
\protect\ion{V}{1}          &         1 & $<+0.30$ & \nodata &   &     1 & $<+0.24$ & \nodata &   &     1 & $<+0.21$ & \nodata &   &     1 & $<+0.46$ & \nodata &   &     1 & $<+0.24$ & \nodata \\
\protect\ion{Cr}{1}         &         1 & $ -0.17$ & $0.12 $ &   &     1 & $ -0.12$ & $0.13 $ &   &     1 & $ -0.15$ & $0.13 $ &   &     2 & $ +0.32$ & $0.35 $ &   &     1 & $ -0.02$ & $0.13 $ \\
\protect\ion{Cr}{2}         &         0 & \nodata  & \nodata &   &     1 & $<-0.11$ & \nodata &   &     1 & $ -0.22$ & $0.18 $ &   &     1 & $<-0.25$ & \nodata &   &     1 & $<-0.13$ & \nodata \\
\protect\ion{Mn}{1}         &         1 & $ -0.19$ & $0.18 $ &   &     1 & $<-0.98$ & \nodata &   &     1 & $<-0.26$ & \nodata &   &     1 & $<-0.31$ & \nodata &   &     1 & $<+0.10$ & \nodata \\
\protect\ion{Co}{1}         &         2 & $ +0.24$ & $0.12 $ &   &     3 & $ +0.14$ & $0.10 $ &   &     4 & $ +0.35$ & $0.10 $ &   &     1 & $ +0.24$ & $0.23 $ &   &     2 & $ +0.59$ & $0.13 $ \\
\protect\ion{Ni}{1}         &         2 & $ +0.23$ & $0.17 $ &   &     2 & $ -0.00$ & $0.11 $ &   &     2 & $ +0.16$ & $0.12 $ &   &     1 & $<-0.13$ & \nodata &   &     2 & $ +0.61$ & $0.16 $ \\
\protect\ion{Zn}{1}         &         1 & $<+0.41$ & \nodata &   &     1 & $<+0.36$ & \nodata &   &     1 & $<+0.38$ & \nodata &   &     1 & $<+0.47$ & \nodata &   &     1 & $<+0.36$ & \nodata \\
\protect\ion{Sr}{2}         &         2 & $ -0.11$ & $0.16 $ &   &     2 & $ -0.37$ & $0.18 $ &   &     2 & $ -0.36$ & $0.23 $ &   &     2 & $ -0.52$ & $0.22 $ &   &     2 & $ -0.49$ & $0.23 $ \\
\protect\ion{Y}{2}          &         1 & $<-0.21$ & \nodata &   &     1 & $<-0.33$ & \nodata &   &     1 & $<+0.03$ & \nodata &   &     1 & $<+0.44$ & \nodata &   &     1 & $<+0.47$ & \nodata \\
\protect\ion{Zr}{2}         &         1 & $<+0.94$ & \nodata &   &     1 & $<+0.75$ & \nodata &   &     1 & $<+1.06$ & \nodata &   &     1 & $<+1.29$ & \nodata &   &     1 & $<+1.08$ & \nodata \\
\protect\ion{Ba}{2}         &         4 & $ -0.39$ & $0.10 $ &   &     2 & $ -0.61$ & $0.10 $ &   &     2 & $ -0.59$ & $0.11 $ &   &     1 & $<-0.74$ & \nodata &   &     1 & $ -0.61$ & $0.14 $ \\
\protect\ion{La}{2}         &         1 & $<+0.76$ & \nodata &   &     1 & $<+1.04$ & \nodata &   &     1 & $<+1.07$ & \nodata &   &     1 & $<+1.12$ & \nodata &   &     1 & $<+1.37$ & \nodata \\
\protect\ion{Ce}{2}         &         1 & $<+0.80$ & \nodata &   &     1 & $<+0.85$ & \nodata &   &     1 & $<+1.06$ & \nodata &   &     1 & $<+1.55$ & \nodata &   &     1 & $<+1.62$ & \nodata \\
\protect\ion{Nd}{2}         &         1 & $<+1.02$ & \nodata &   &     1 & $<+0.87$ & \nodata &   &     1 & $<+0.82$ & \nodata &   &     1 & $<+1.19$ & \nodata &   &     1 & $<+1.62$ & \nodata \\
\protect\ion{Sm}{2}         &         1 & $<+1.49$ & \nodata &   &     1 & $<+1.59$ & \nodata &   &     1 & $<+1.16$ & \nodata &   &     1 & $<+1.66$ & \nodata &   &     1 & $<+1.67$ & \nodata \\
\protect\ion{Eu}{2}         &         1 & $<+0.98$ & \nodata &   &     1 & $<+0.94$ & \nodata &   &     1 & $<+0.94$ & \nodata &   &     1 & $<+1.32$ & \nodata &   &     1 & $<+1.26$ & \nodata \\
\protect\ion{Dy}{2}         &         1 & $<+2.25$ & \nodata &   &     1 & $<+1.90$ & \nodata &   &     1 & $<+2.12$ & \nodata &   &     1 & $<+2.46$ & \nodata &   &     1 & $<+2.79$ & \nodata \\
\enddata
\end{deluxetable*}
\end{rotatetable*}

\noindent Giant stars have deep convective envelopes that re-mix the abundances to levels close to their initial surface values.  Observations of metal-poor globular clusters show only small differences in the abundances of heavy metals, like Fe, between the main sequence turn-off and the giant branch \citep{coh05a,kor07,lin08}.  As a result, we do not expect to see any trends in abundance for elements heavier than oxygen for the giants in our sample.

Nonetheless, we do observe abundance trends with luminosity.  We interpret these trends as artifacts of our measurements.  For example, they could be uncorrected non-LTE effects.  The main goal of our study is to quantify star-to-star abundance variations.  Therefore, we take the conservative approach of removing the trend with luminosity.  We do not fit the luminosity trend of elements that are known to vary in globular clusters: Li, C, O, Na, Mg, Al, and K\@.  For other elements, we fit the abundance of each element as a function of luminosity.  For most elements, the function is a quadratic.  For elements measured over a limited range of luminosity (Ce, Nd, and Sm), the fit is linear.  The luminosity correction is a subtraction of this line or quadratic, normalized such that the correction is zero at $M_{G,0} = 0$.  We used \leopy\ \citep{fel19} to perform a ``censored'' fit, which takes into account upper limits.  Figure~\ref{fig:lumcorr} shows the fitted luminosity trends.  We applied the luminosity correction only to elements with red trend lines in the figure.  Elements not shown in the figure were not corrected.

\subsection{Error Analysis}
\label{sec:errors}

Error analysis in high-resolution spectroscopy can sometimes be simplistic or arbitrary.  For species with multiple absorption lines, one common approach is to report the standard deviation divided by the square root of the number of lines.  For species with just one or a few lines, sometimes arbitrary uncertainties---such as 0.1 or 0.2~dex---are reported.  Some of this guesswork is unavoidable because the uncertainties on abundances inherit difficult-to-quantify uncertainties on predecessor variables, such as oscillator strengths, deficiencies in the model atmosphere, or non-LTE corrections.

\citet{mcw95,mcw13} improved on the standard error analysis by accounting for uncertainties in atmospheric parameters, including covariance from correlated parameters, like $T_{\rm eff}$ and $\log g$.  Accounting for covariance lessens the severity of the spurious correlations between abundance ratios that could be introduced by errors in atmospheric parameters \citep{roe15}.  \citet{ji20} extended this framework to the abundance measurements themselves (the mean abundance from all the lines of a single species) rather than merely their uncertainties.  \citeauthor{ji20}\ documented their approach in their Appendix~B, which we followed in this work.

The framework requires that we know the correlations between the atmospheric parameters.  The correlation between $T_{\rm eff}$ and $\log g$ is a direct consequence of the way in which surface gravity is computed.  Specifically, $g \propto T_{\rm eff}^4$ (Equation~\ref{eq:g}).  We computed the Pearson correlation coefficients between the three unique pairs of $T_{\rm eff}$, $\log g$, and $v_t$ from their measured values (Table~\ref{tab:atmpars}).  The value of [M/H] was identical for all stars in our sample, so we copied the correlation coefficients that involve [M/H] from \citet{ji20}.

Following Appendix~B of \citet{ji20}, the correlation matrix ($\rho$) is multiplied into a vector ($\delta$) containing the uncertainties in the atmospheric parameters for the star.  Table~\ref{tab:atmpars} gives the uncertainties in $T_{\rm eff}$, $\log g$, and $v_t$.  We used 0.05~dex as the uncertainty in [M/H]\@.  This number is the weighted standard deviation of the \ion{Fe}{2} abundances.  The matrix $\rho$ and the vector $\delta$ can be used together with the $e_i$ (the abundance uncertainties propagated from the EW uncertainties) to compute a matrix called $\widetilde{\Sigma}$:

\begin{equation}
\widetilde{\Sigma} = {\rm diag}(e_i^2 + s_{\rm X}^2) + \delta \rho \delta^T \label{eq:Sigma}
\end{equation}

\noindent The variable $s_{\rm X}$ is a systematic uncertainty applied to every line of a given species.  We calculated the $s_{\rm X}$ values according to the procedure described by \citeauthor{ji20}  As a modification to that procedure, we forced the minimum value of $s_{\rm X}$ to be 0.1.  This minimum value approximates the hard-to-estimate systematic errors we discussed at the beginning of Section~\ref{sec:errors}.  The difference in our procedure is that we apply the minimum systematic error to each line rather than the final abundance measurements, which have no forced minimum.

This framework permits negative weights.  In effect, a negative weight on an absorption line pushes the average abundance {\it away} from the abundance measured for that line.  Negative weights are a result of the assumption that line strengths behave linearly with changes in abundance and in stellar parameters.  This assumption is not necessarily true for strong lines.  We resolve this conundrum by iteratively omitting lines with negative weights, then recomputing $\widetilde{\Sigma}$ and the new weights.

Table~\ref{tab:ew} includes the weights for each absorption line.  The table omits lines that were rejected for having negative weights.

\citeauthor{ji20}\ also described how to calculate the error in abundance ratios.  Most lines in our study vary in the same direction with changes in atmospheric parameters.  As a result, the error on most abundance ratios is smaller than simply adding the individual elements' errors in quadrature.  We also calculated the covariance between two different abundance ratios.  The figures in Section~\ref{sec:trends} show the covariance as ellipses.

Table~\ref{tab:abund1} gives the abundances and errors for each star.  \ion{Fe}{1} and \ion{Fe}{2} are given as [Fe/H]\@.  The other elements are shown as ratios to iron ([X/Fe]).  The neutral species are shown relative to \ion{Fe}{1}, and the ionized species are shown relative to \ion{Fe}{2}\@.  The columns labeled $N$ give the number of lines used in the abundance determinations.  Elements measured with spectral synthesis are reported with ``syn'' in this column.


\section{Abundance Patterns}
\label{sec:trends}

Globular clusters have fascinating abundance patterns, some of which we discussed in Section~\ref{sec:intro}.  In this section, we examine some of the well-known abundance patterns, like the Na--O anti-correlation, as well as some lesser-known patterns, such as the potassium distribution, in M92.  Most interestingly, we discuss the variation of neutron-capture abundances.

\subsection{Trends with Luminosity}
\label{sec:lumtrend}

Stellar evolution along the RGB for low-mass stars can alter the surface abundances of some elements.  The main mechanisms for these alterations are the first dredge-up at the base of the RGB and thermohaline mixing at the luminosity function bump \citep{cha07}.

The abundances of most elements heavier than O are not expected to show evolution on the RGB\@.  Nonetheless, we observed trends with luminosity for many elements heavier than O, which we interpreted as systematic errors in our LTE analysis (Section~\ref{sec:lumcorr}).  We applied corrections to some of those elements but not to Li, C, or O, which we discuss here.

\begin{figure}
\centering
\includegraphics[width=\linewidth]{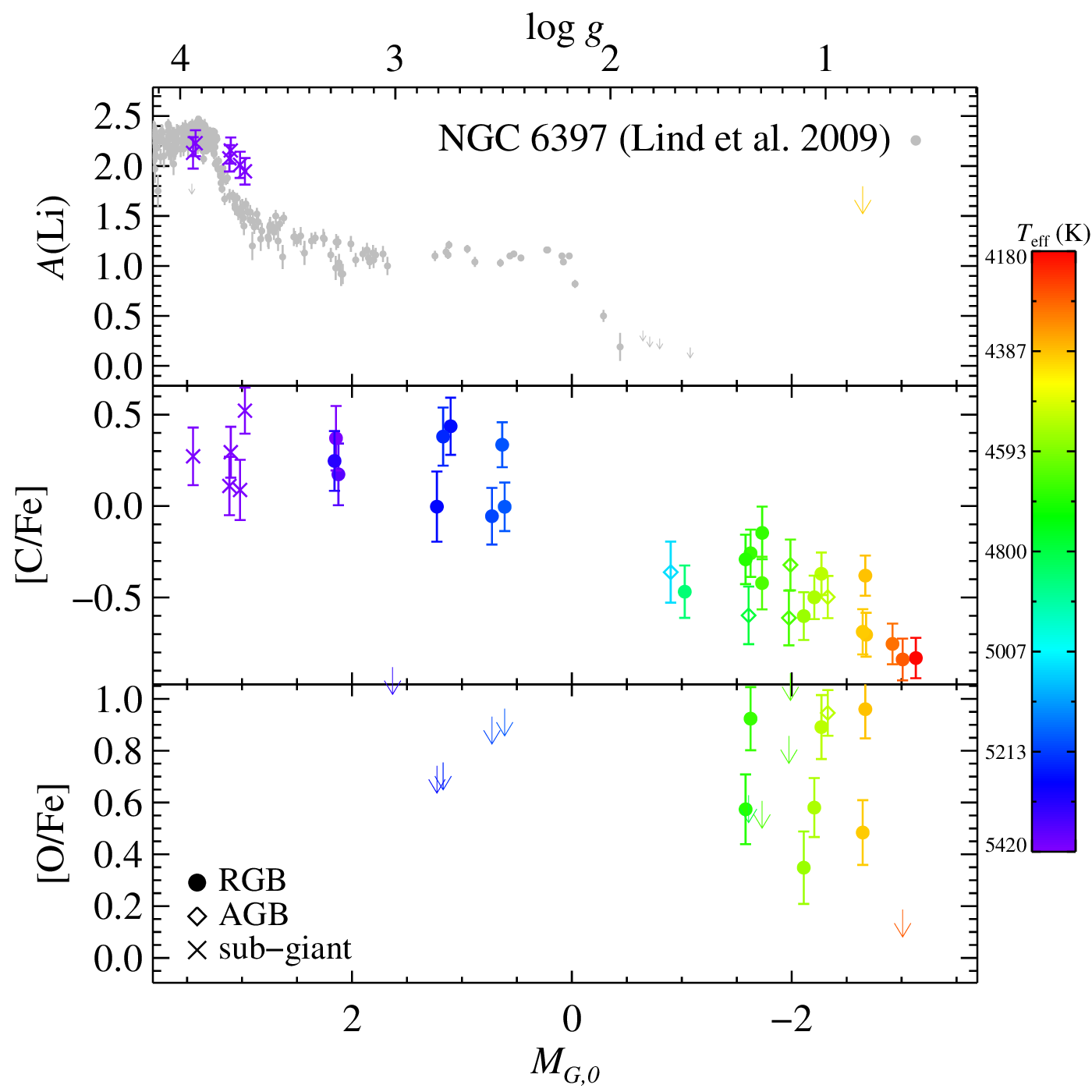}
\caption{The evolution of Li, C, and O abundances in M92 with stellar luminosity.  The Li abundances in M92 are compared to those in NGC~6397 \citep{lin09li}.\label{fig:lum}}
\end{figure}

Lithium is the element that most dramatically exhibits changes on the RGB\@.  The top panel of Figure~\ref{fig:lum} shows Li abundance measurements in M92 compared with those in NGC 6397 \citep{lin09li}, a slightly more metal-rich globular cluster.  Unlike the exquisite observations of \citeauthor{lin09li}, our observations were not tailored to measure Li abundances.  Therefore, the only stars in our sample that show Li detections are near the main sequence turn-off, where the first dredge-up has not yet depleted Li to the fullest extent.  Previous measurements of Li abundance in M92 include those by \citet{del95} and \citet{boe98}.  They observed Li abundances up to four times higher than we observed.  As they pointed out, there appears to be a dispersion of $A({\rm Li})$ on the sub-giant branch, even at fixed $T_{\rm eff}$.

Whereas the first dredge-up dilutes Li by mixing the surface into layers of the star that were once hot enough to burn Li, thermohaline mixing actively destroys Li by mixing it into layers that are presently hot enough to burn Li.  The same mechanism also mixes the surface material to layers that are hot enough to participate in the CNO cycle \citep{smi96,smi02,smi05,smi06,ang11}.  As a result, the surface abundances of C and O are depleted, and the surface abundance of N is enhanced.  Figure~\ref{fig:lum} shows that the C abundances decline around an approximate absolute magnitude of zero.  Our measurements of O on the lower RGB are not sufficiently sensitive to observe a decline in O abundances.  There is a dispersion in C and O abundances even at fixed magnitude.  GCs are well-known to show such dispersions.

\subsection{Light elements}

M92 has been known to exhibit a dispersion in light elements since \citet{coh79} measured abundances of red giants with the echelle spectrograph at the Kitt Peak/Mayall telescope.  She found a star-to-star scatter of 0.8~dex in Na.  \citet{nor85} found that Na abundances were correlated with N abundances, consistent with today's concordant view of light-element abundance variations in GCs.  \citet{sne91} first quantified the Na--O anti-correlation in M92, which \citet{kra93} and \citet{sne94} placed in context with the Na--O anti-correlations in other globular clusters.  At the time, there was debate over whether the anti-correlation was primordial or a consequence of mixing during evolution on the RGB \citep{kra94}.  At last, \citet{gra01} ruled out mixing as a source of the abundance variations observing the Na--O anti-correlation on the sub-giant branch of two slightly more metal-rich GCs.

\begin{figure*}
\centering
\includegraphics[width=\linewidth]{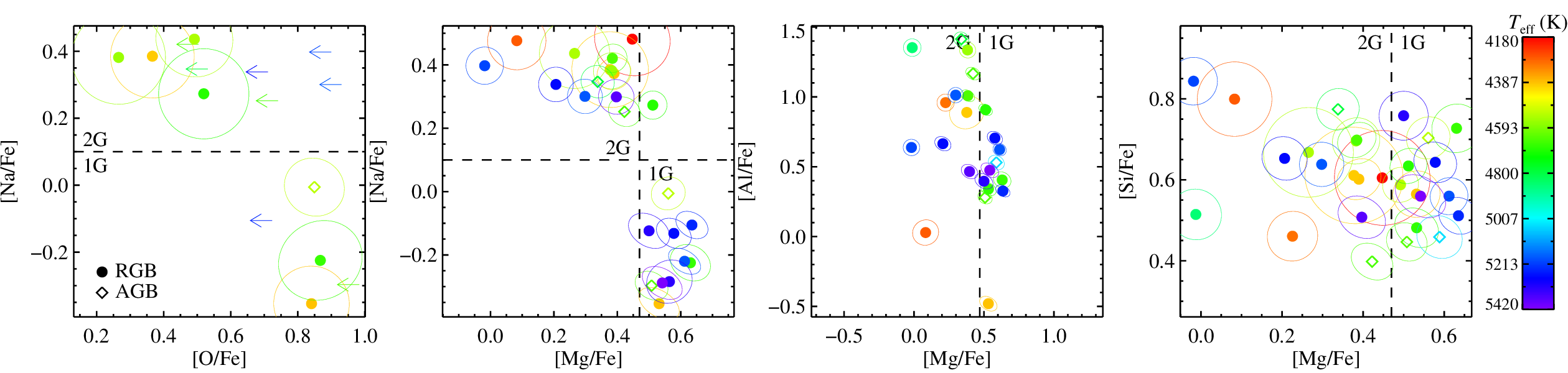}
\caption{Light element abundance anti-correlations for giant stars in M92.  The ellipses represent the $1\sigma$ uncertainties, including covariance.  The dashed lines in the indicate the division between first- and second-generation stars (1G and 2G)\@.\label{fig:light}}
\end{figure*}

Figure~\ref{fig:light} shows measurements of the Na--O (first panel) and Mg--Al (third panel) anti-correlations in M92.  This figure and subsequent figures show only giant stars (RGB and AGB) to limit the effect of systematic errors.  The error ellipses in the figure include covariance between the element ratios (Section~\ref{sec:errors}).  The measurements are color-coded by $T_{\rm eff}$ in order to investigate whether abundances change with stellar parameters.  Trends between abundance and $T_{\rm eff}$ could reflect abundance changes with evolution for some elements, like O\@.  However, trends with $T_{\rm eff}$ for most elements would indicate a systematic error in the abundance analysis.  We do not observe any clear pattern with $T_{\rm eff}$.  In fact, the luminosity correction to abundances (Section~\ref{sec:lumcorr}) virtually ensures that there would be no such trend.

The light element abundance correlations are notoriously difficult to observe in optical spectra.  The atomic lines of O in the optical are very weak ([\ion{O}{1}]$\;\lambda\lambda 6300,6364$, dipole-forbidden and split by $J$ degeneracy of the lower level) or highly subject to NLTE corrections (\ion{O}{1}$\;\lambda\lambda\lambda 7772,7774,7775$, triply split by $J$ degeneracy of the upper level).  Likewise, the optical Al lines are either weak (\ion{Al}{1}$\;\lambda\lambda 6696,6699$, split by $J$ degeneracy of the upper level) or very strong and therefore subject to NLTE corrections and hypersensitivity to microturbulent velocity (\ion{Al}{1}$\;\lambda\lambda 3944,3962$, split by $J$ degeneracy of the lower level).  On the other hand, Na and Mg are comparatively easy to measure.  The second panel of Figure~\ref{fig:light} shows the Na--Mg anti-correlation.  The relationship between Na and Mg is not as direct as between Mg and Al.  High-temperature proton capture effectively converts Ne into Na and Mg into Al.  The conversion of Mg into Al makes it sensible to show the Mg--Al anti-correlation, but the ease of measuring Na and Mg prompts us to show the Na--Mg anti-correlation instead.

The separation into two populations is particularly clear in the Na--Mg anti-correlation.  Population 1G has halo-like abundance patterns: sub-solar [Na/Fe] and elevated [Mg/Fe].  On the other hand, population 2G shows the signatures of high-temperature proton burning: enhanced [Na/Fe] and depleted [Mg/Fe].  \citet{mil17} quantified the degree to which various GCs separated into multiple populations using the ``chromosome map'' based on \textit{HST} wide-band filters.  M92 is a fairly typical cluster in the chromosome map.  However, it may stand out in Na and Al abundances (see APOGEE2 discussion below).

We classified each star as belonging to 1G or 2G based on its Na or Mg abundance.  For classification by Na abundance, we drew a dividing line at ${\rm [Na/Fe]} = +0.1$, with 1G below the line and 2G above the line.  There are no giants with Na abundances in the range $-0.01 < {\rm [Na/Fe]} < +0.25$.  Therefore, classification by Na abundance is unambiguous.  We drew a separate dividing line at ${\rm [Mg/Fe]} = +0.47$.  However, the classification by [Mg/Fe] is not as clear as the classification by [Na/Fe] because one star (IV-79)---classified as 2G from its Na abundance---is on the 1G side of the [Mg/Fe] dividing line.  Figures~\ref{fig:light}--\ref{fig:namgk} and \ref{fig:firstpeak}--\ref{fig:secondpeakb} show these dividing lines.

\citet{yon05} discovered a Mg--Si anti-correlation in NGC~6752, which could be explained by proton capture onto $^{27}{\rm Al}$.  That reaction has two branches: $^{27}{\rm Al}(p,\alpha)^{24}{\rm Mg}$ and $^{27}{\rm Al}(p,\gamma)^{28}{\rm Si}$.  The ratio of the second branch to the first branch increases with the temperature of the proton burning \citep{pra17}.  The two branches naturally result in a Mg--Si anti-correlation and an Al--Si correlation.  The right panel of Figure~\ref{fig:light} shows the Mg--Si anti-correlation.  While it is clear that Mg and Si both vary from star to star in M92, the anti-correlation is not nearly as obvious as for Na--Mg.

The infrared spectrum is more amenable to measuring O from molecular features and Al from better-behaved atomic lines.  \citet{mas19} and \citet{mes20} used APOGEE2 $H$-band spectra \citep{abo18} to measure O, Mg, and Al abundances in M92.  (The Na lines in the $H$-band are too weak to observe in M92.)  The APOGEE2 version of the Mg--Al (and Al--O) anti-correlations are much more obvious than in our Figure~\ref{fig:light}.  Along with M53, M92 is one of the two GCs with the clearest separation into two populations in the APOGEE2 abundance patterns, especially those involving Al.  \citet{mas19} observed a significant population of stars in M92 and M15 with ${\rm [Mg/Fe]} < 0$.  We do not observe such a population with our optical spectra.  We do not know whether the discrepancy results from a different selection of stars or a difference in analysis techniques.

\citet{and01} found that foreground interstellar \ion{Na}{1} absorption varies over small spatial scales in the direction of M92.  There are two main absorption clouds, separated in velocity by 19~km~s$^{-1}$.  We confirm from a qualitative inspection of the HIRES spectra that there is a large scatter in EW of both interstellar clouds across the face of M92.  The EWs of one cloud do not appear correlated with the EWs of the other cloud.  Fortunately, the interstellar absorption is separated by at least 70~km~s$^{-1}$ from the \ion{Na}{1} lines in M92 stars.  Consequently, the large interstellar variations do not affect the stellar abundances.

\subsection{Potassium}
\label{sec:K}

\begin{figure}
\centering
\includegraphics[width=\linewidth]{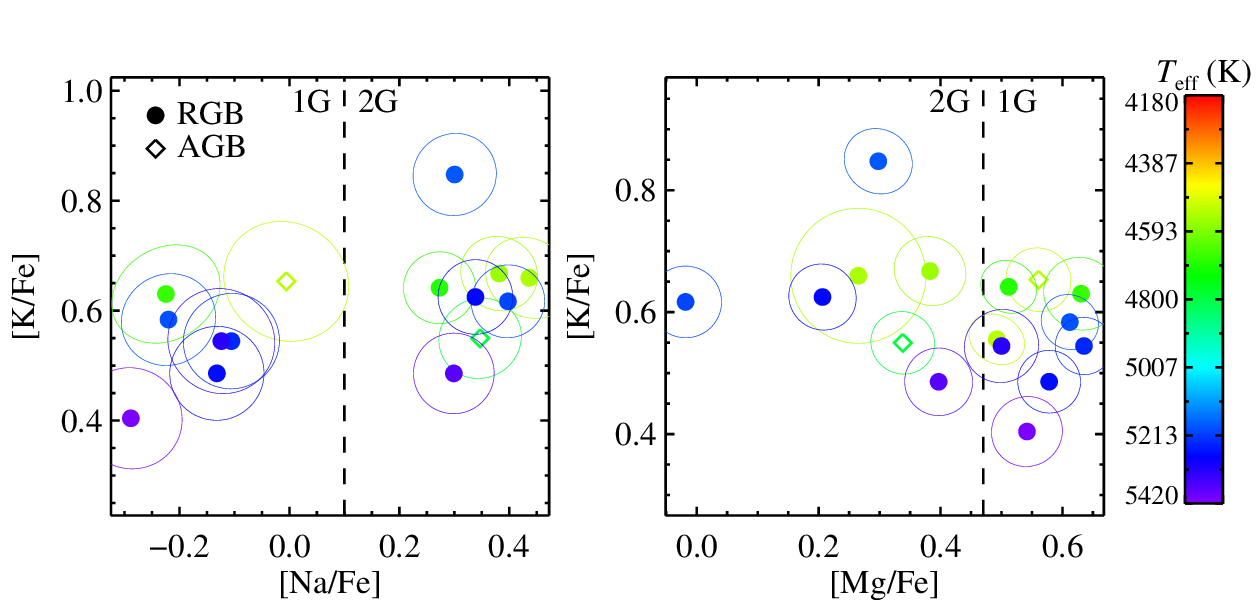}
\caption{Potassium vs.\ sodium (left) and magnesium (right) abundances in M92.  The dashed lines indicate the division between first- and second-generation stars (1G and 2G)\@.\label{fig:namgk}}
\end{figure}

\citet{coh11a}, \citet{coh12}, and \citet{muc12} found that NGC~2419, an outer halo GC, exhibits an unusual variation in K abundances.  The abundances are unusual for showing a dispersion, whereas most other GCs were not known to show a dispersion \citep{car13}.  Furthermore, the K abundances in NGC~2419 are strongly anti-correlated with Mg.  The anti-correlation suggests a similar nucleosynthetic pathway to the Na--O and Mg--Al anti-correlations: high-temperature hydrogen burning \citep{ventura12,ili16}.  In the last decade, at least seven more GCs were found to show potassium abundance variations, with some clusters showing a Mg--K anti-correlation \citep{muc15,muc17,car21,car22,alv22}.

Figure~\ref{fig:namgk} shows the K abundances in M92 vs.\ Na and Mg abundances.  There is no apparent correlation or anti-correlation between K and Na or Mg.  All stars have [K/Fe] in the range +0.2 to +0.7.  The scatter in [K/Fe] is slightly more than expected from the measurement uncertainty if they all had the same [K/Fe] value.  The potassium abundance is only slightly lower for 1G than for 2G\@.  Although the significance of this result is not high, it is qualitatively consistent with the Mg--K anti-correlation observed in NGC~2419 and other clusters.

We initially measured a high K abundance in star X-20, but we had doubts about the telluric correction.  Our colleagues from the California Planet Survey re-observed the star (Section~\ref{sec:CPS}) so that we could check this measurement.  The EW of \ion{K}{1}$\;\lambda 7699$ was 32\% smaller in the new spectrum than in the archival spectrum, probably due to difficulties in the telluric correction.  X-20 still has the highest K abundance among our sample, but it is no longer a highly significant outlier.

\subsection{Iron}

The abundances of iron and the iron-peak elements in M92 behave like most other GCs.  Specifically, there is little star-to-star dispersion.  However, \citet{lan98} reported that one star, XI-19, has stronger iron lines than XII-8 and V-45, which are very close in the CMD\@.  They found that [Fe/H] in XI-19 was $0.18 \pm 0.01$ higher than the other two stars.  Broadband \citep{leg22,lar22} and narrow-band \citep{lee23} photometry also supports the presence of multiple metallicities in M92.

All three of these stars are in our sample.  In fact, we confirm that [\ion{Fe}{1}/H] in XI-19 is $0.19 \pm 0.05$ higher than XII-8 and $0.13 \pm 0.04$ higher than V-45.  The abundance of [\ion{Fe}{2}/H] is higher in XI-19 than the other two stars by $0.10 \pm 0.06$.  The abundances of other iron-peak elements scale with Fe.  Our sample also contains XI-80, which is also extremely close in the CMD to the other three stars.  The abundances of Ti and Fe in XI-80 are nearly identical to XI-19, which is to say that they are higher than XII-8 and V-45.

Subtle iron abundance variations have been detected in ``normal'' GCs \citep[e.g.,][]{mar09}, but the findings are sometimes controversial \citep[e.g.,][]{muc15b}.  The question is thorny because of subtleties, like NLTE effects and atmosphere modeling differences between the RGB and AGB, that can cause the appearance of iron abundance variations.  Although we corroborate \citeposs{lan98} detection of an iron abundance variation in M92, our sample is not ideal to explore this question in detail because it spans the entire RGB\@.  The ideal sample would be confined to a tight locus in the CMD\@.  Thus, we save this question for a future sample (see Section~\ref{sec:summary}).

\subsection{Neutron-capture elements}

\begin{figure*}
\centering
\includegraphics[width=\linewidth]{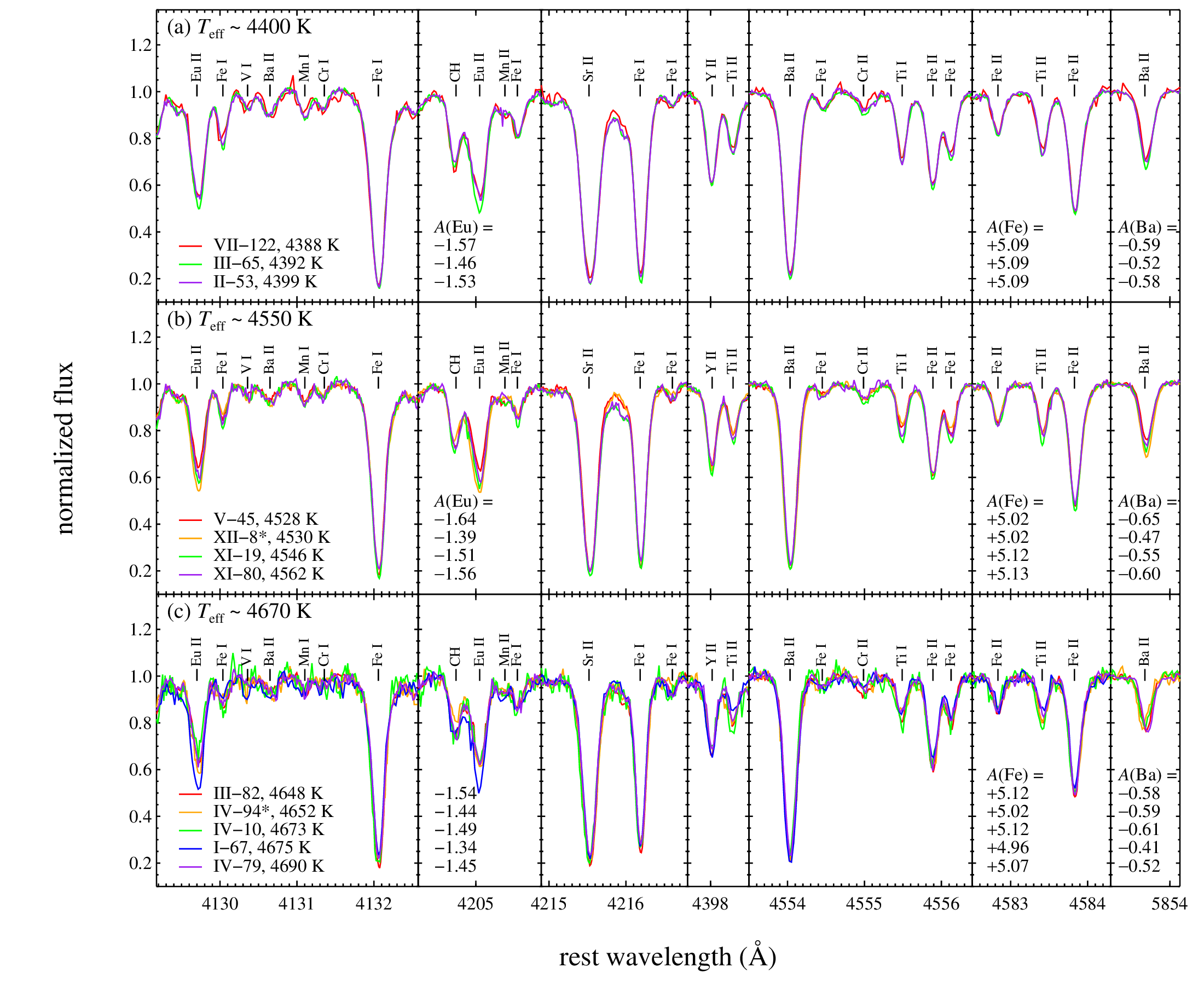}
\caption{Keck/HIRES archival spectra of giant stars in M92.  The spectra in each row are grouped by $T_{\rm eff}$.  The stars have $T_{\rm eff}$ within 25~K of 4400~K ({\it top}), 4550~K ({\it middle}), and 4670~K ({\it bottom}).  AGB stars are indicated with asterisks in their names.  Abundances of Fe, Ba, and Eu are given.  The absorption lines \ion{Eu}{2}$\;\lambda 4130$, \ion{Eu}{2}$\;\lambda 4205$, and \ion{Ba}{2}$\;\lambda 5854$ are particularly variable.\label{fig:spectra}}
\end{figure*}

The neutron-capture abundances in M92 have been controversial since \citet{roe11a} and \citet{roe11b} claimed a significant star-to-star variation, followed by \citeposs{coh11b} refutation of that claim.  In this subsection, we re-examine the variation of the neutron-capture abundances.  Our sample includes all of the stars analyzed by \citet{coh11b} because she observed them with Keck/HIRES\@.  Hence, those spectra are in our HIRES archival sample.

The simplest test for abundance variations is to compare spectra with a limited range of atmospheric parameters.  Figure~\ref{fig:spectra} shows spectra in three bins of $T_{\rm eff}$.  The stars represented in each panel were chosen to have $T_{\rm eff}$ within 25~K of a central value.  The actual ranges of $T_{\rm eff}$ from top to bottom are 11~K, 34~K, and 42~K\@.  The lines that vary the most from star to star are \ion{Eu}{2}$\;\lambda 4130$, \ion{Eu}{2}$\;\lambda 4205$, and \ion{Ba}{2}$\;\lambda 5854$.  Lines of lighter elements, like Ti and Fe, are not perfectly identical, but they are less variable than the neutron-capture lines.  Furthermore, the variations in the transition metals do not correlate with the variations in the neutron-capture elements.  We conclude that there is a genuine dispersion in neutron-capture abundances in M92.

\begin{figure*}
\centering
\includegraphics[width=0.7\linewidth]{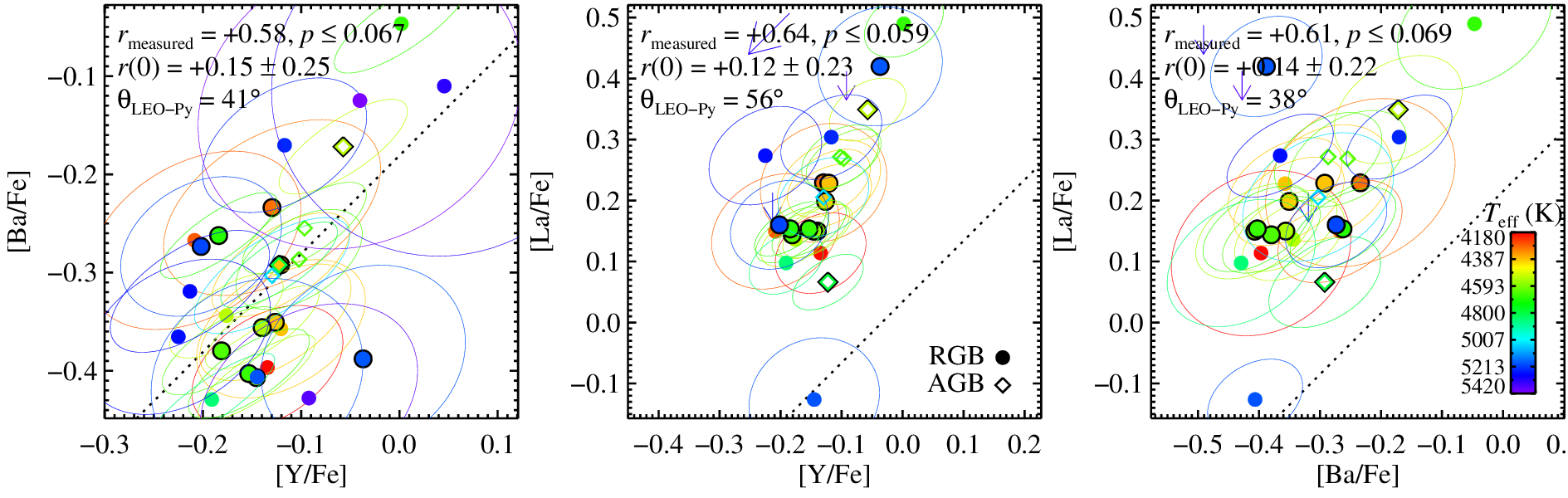}
\includegraphics[width=0.7\linewidth]{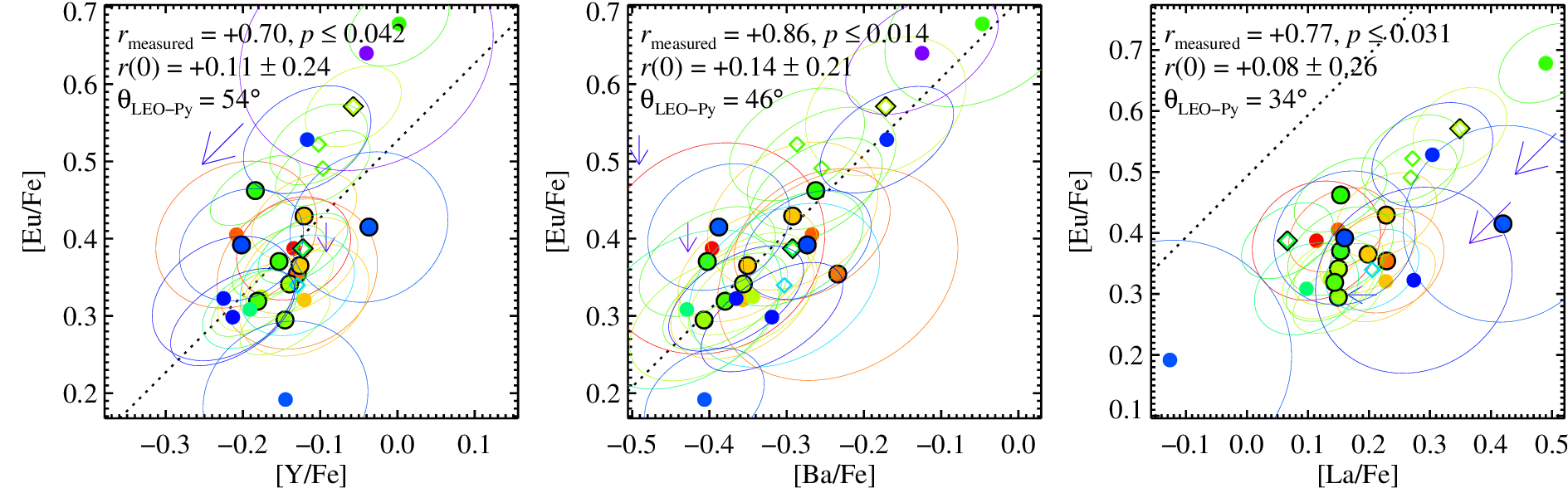}
\includegraphics[width=0.7\linewidth]{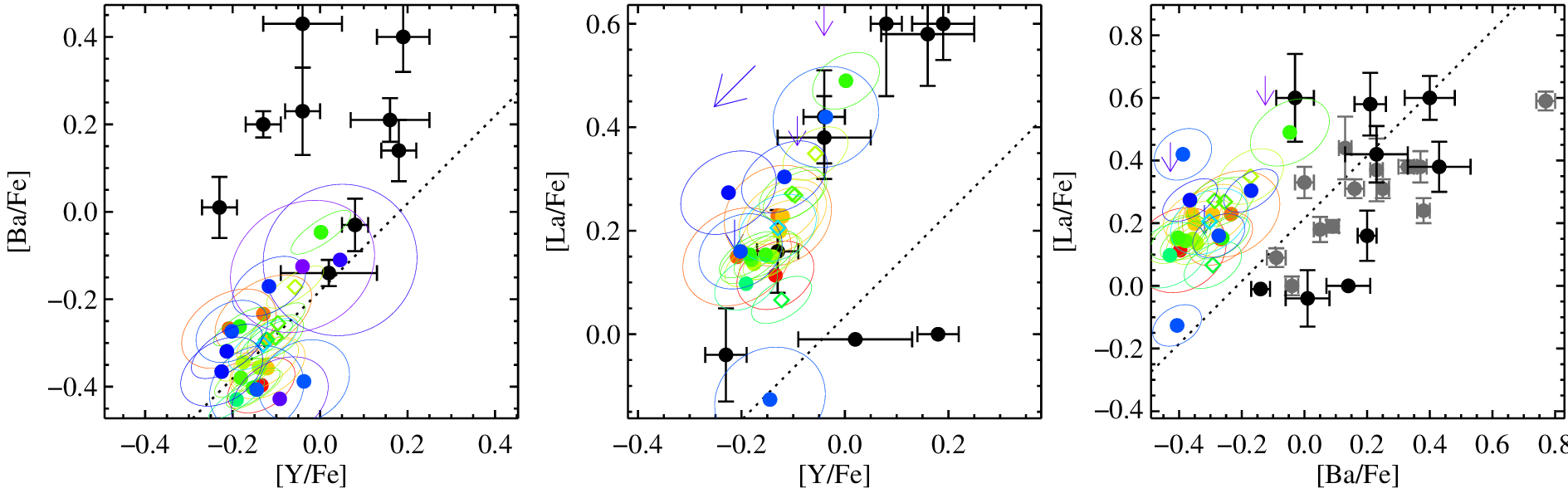}
\includegraphics[width=0.7\linewidth]{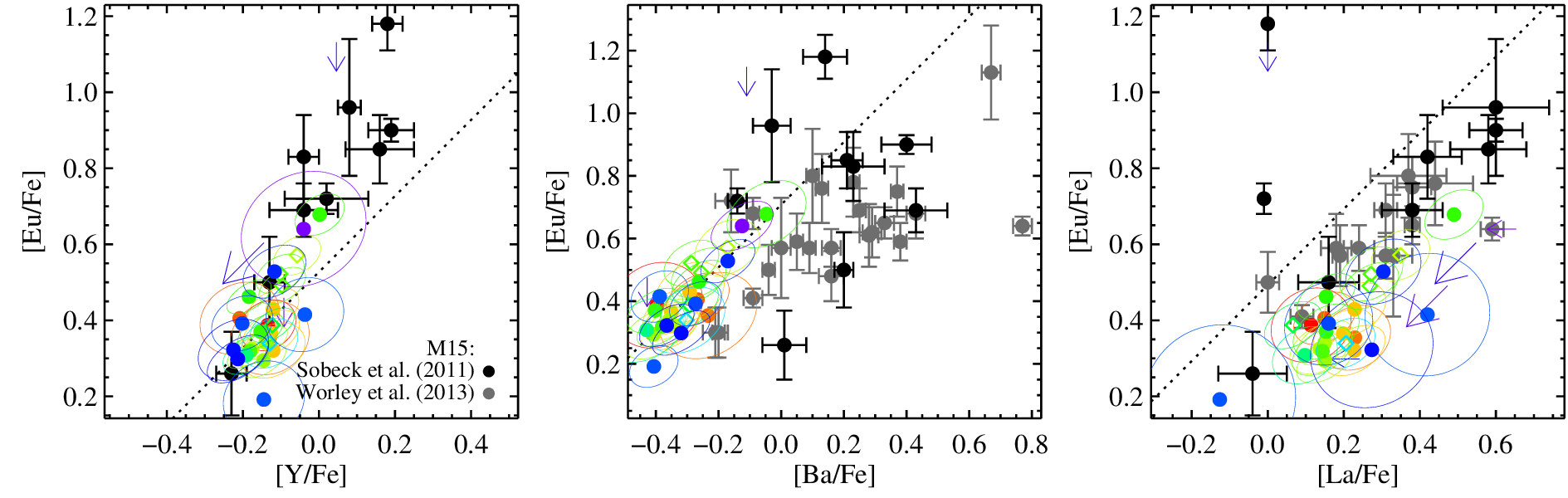}
\caption{Correlations between the neutron-capture elements Y, Ba, La, and Eu.  The top two rows show M92 by itself.  The linear Pearson correlation coefficient, $r_{\rm measured}$, is given in the upper left of each panel.  Each panel also includes $r(0)$, an estimate of the correlation coefficient if the scatter in the points was due entirely to their individual covariances, represented by the ellipses.  Another test of correlation is $\theta_{\rm LEO-Py}$, the slope of the best-fit line taking covariance into account.  The bottom two rows show M92 with measurements in M15 from the literature \citep{sob11,wor13}.  The dashed lines show the solar $r$-process pattern \citep{sim04}.  The symbols (evolutionary state) and colors ($T_{\rm eff}$) have the same meaning as in Figure~\ref{fig:cmd}.  M92 stars in the sample of \citet{coh11b} are outlined in black in the top two rows.\label{fig:ncapture_corr}}
\end{figure*}

Two quantities can be correlated only if there is a dispersion in both quantities.  Figure~\ref{fig:ncapture_corr} shows the correlations between the different permutations of Y, Ba, La, and Eu.  The abundances of each pair are correlated.  We judge the significance of the correlation with the linear Pearson correlation coefficient ($r_{\rm measured}$, given in the figure), where zero is uncorrelated and one is perfectly correlated.  We also show the $p$ value, which gives the probability that a correlation at least as significant as $r_{\rm measured}$ would appear by chance.

The Pearson correlation coefficient does not take into account the covariance for each data point.  Therefore, we also calculated the Pearson correlation coefficient, $r(0)$, for the null test where the true values for all the points are identical.  The true value of $r(0)$ in this case is exactly zero.  We sampled $10^4$ ``observed'' values by taking the covariance ellipses as probability distributions.  We computed the mean and standard deviation of the resulting correlation coefficients, which are given by $r(0)$ in the figure.  The difference between $r(0)$ and zero is the correlation expected from covariance alone.  In all cases, $r_{\rm measured}$ is at least 3.8 times $r(0)$, indicating that the correlations are significant, even when accounting for covariance.

We also judged the degree of correlation by fitting a straight line to the data while accounting for both the variance and covariance in the uncertainties.  We parameterized the line as $y = x \tan \theta + b$. 
 Such a parameterization allows for a flat prior in the slope $\theta$ without giving undue weight to large absolute values of the slope \citep[a common problem when the line is parameterized as $y = mx + b$;][]{hog10}.  We used \leopy\ \citep{fel19} to fit for $\theta_{\rm LEO-Py}$, taking into account covariance.  Figure~\ref{fig:ncapture_corr} shows the resulting values.  We do not include uncertainties on $\theta_{\rm LEO-Py}$ because the formal uncertainties are less than 1\%.  A perfect correlation would result in $\theta_{\rm LEO-Py} = 45^{\circ}$, and no correlation would result in $0^{\circ}$ or $90^{\circ}$.  All of the slopes are in the range $(45 \pm 11)^{\circ}$.
 
 Our statistical tests show that the correlations between neutron-capture abundances are significant, even when accounting for covariance in the abundance uncertainties.  We conclude that there is a significant dispersion in the neutron-capture abundances in M92.

We further test whether the dispersion is a result of systematic errors by ruling out any residual trends with $T_{\rm eff}$ or evolutionary state.  The colors in Figure~\ref{fig:ncapture_corr} correspond to $T_{\rm eff}$, and the symbols distinguish RGB and AGB stars, as in Figure~\ref{fig:cmd}.  There is no apparent trend with either $T_{\rm eff}$ or evolutionary state.

We also confirm that the abundances are consistent with the $r$-process.  There is copious evidence that the ``main'' $r$-process abundance pattern is universal, such that the ratio of any two neutron-capture elements will be constant in a star whose neutron-capture nucleosynthesis was dominated by the $r$-process \citep[e.g.,][]{sne96}.  It is typical to compare neutron-capture abundances to the solar system $r$-process abundance pattern.  Figure~\ref{fig:ncapture_corr} shows the solar-scaled pattern computed by \citet[][also reported by \citealt{sne08}]{sim04}.  The abundances in M92 agree very well with the solar system $r$-process pattern except for the abundance of La, which is larger in M92.  \citet{sne08} reported that the typical metal-poor star has a higher [La/Eu] ratio than the solar-system $r$-process value, but the difference is only about 0.05~dex, whereas we observe an offset of $\sim 0.3$~dex.  Our luminosity correction (Section~\ref{sec:lumcorr}) could explain the offset.  The luminosity correction for La is among the larger corrections.

We compare the neutron-capture abundances in M92 with those in M15 in the bottom two rows of Figure~\ref{fig:ncapture_corr}.  M15 shows a large dispersion in the $r$-process, as has been widely reported.  The comparatively smaller dispersion in M92 is one reason why it has been difficult to conclude that is has a definitive dispersion.  For some elements, like La and Eu, the abundances in M15 are consistent with the $r$-process.  The abundances of other elements are not so clearly associated with a pure $r$-process.  M15's Y abundances are quite scattered, as pointed out by \citet{ots06}, who concluded that neutron-capture processes other than the main $r$-process were at work in M15.  The Ba abundances are also more scattered than Eu and La, perhaps because Ba is often measured from strong lines, which are more sensitive than weak lines to errors in atmospheric parameters, like microturbulent velocity.  Like M92, the La abundances in M15 are higher than the solar system $r$-process pattern reported by \citet{sim04}.

\begin{deluxetable*}{lccccccccc}
\tablecolumns{10}
\tablewidth{0pt}
\tablecaption{Abundance Dispersions in 1G and 2G\label{tab:dispersions}}
\tablehead{
 & \multicolumn{4}{c}{Na} & & \multicolumn{4}{c}{Mg} \\ \cline{2-5} \cline{7-10}
\colhead{Element} & \colhead{$\sigma({\rm 1G})$} & \colhead{$\sigma({\rm 2G})$} & \colhead{$\chi_r^2({\rm 1G})$} & \colhead{$\chi_r^2({\rm 2G})$} & & \colhead{$\sigma({\rm 1G})$} & \colhead{$\sigma({\rm 2G})$} & \colhead{$\chi_r^2({\rm 1G})$} & \colhead{$\chi_r^2({\rm 2G})$}
}
\startdata
Sr &                $<0.06$ & $0.05_{-0.03}^{+0.04}$ & 0.95 & 1.32 & & $0.11_{-0.03}^{+0.04}$ &        $0.04 \pm 0.03$ & 3.70 & 1.22 \\
 Y &                $<0.04$ &                $<0.02$ & 0.41 & 0.36 & &        $0.04 \pm 0.03$ &                $<0.02$ & 1.29 & 0.52 \\
Zr &                $<0.05$ &                $<0.02$ & 0.32 & 0.39 & & $0.07_{-0.03}^{+0.04}$ &                $<0.03$ & 2.14 & 0.61 \\
Ba & $0.09_{-0.05}^{+0.06}$ &                $<0.02$ & 1.39 & 0.49 & & $0.12_{-0.03}^{+0.04}$ &                $<0.03$ & 3.99 & 0.79 \\
La & $0.18_{-0.06}^{+0.10}$ & $0.04_{-0.03}^{+0.04}$ & 2.68 & 1.06 & & $0.16_{-0.04}^{+0.05}$ & $0.08_{-0.03}^{+0.04}$ & 6.95 & 1.74 \\
Ce &                $<0.08$ &                $<0.03$ & 0.09 & 0.64 & & $0.13_{-0.04}^{+0.06}$ &                $<0.04$ & 3.84 & 0.67 \\
Nd &                $<0.10$ &                $<0.04$ & 0.44 & 0.75 & & $0.16_{-0.04}^{+0.07}$ &        $0.07 \pm 0.04$ & 5.19 & 1.37 \\
Sm &                $<0.26$ & $0.07_{-0.04}^{+0.06}$ & 0.84 & 1.38 & & $0.18_{-0.06}^{+0.11}$ & $0.09_{-0.04}^{+0.05}$ & 5.34 & 1.81 \\
Eu & $0.14_{-0.06}^{+0.08}$ &                $<0.03$ & 2.10 & 0.61 & & $0.15_{-0.04}^{+0.05}$ &                $<0.04$ & 6.11 & 0.97 \\
Dy & $0.22_{-0.12}^{+0.29}$ &        $0.09 \pm 0.05$ & 1.64 & 1.68 & & $0.22_{-0.06}^{+0.11}$ & $0.11_{-0.03}^{+0.05}$ & 9.19 & 2.65 \\
\enddata
\end{deluxetable*}

\begin{figure}
\centering
\includegraphics[width=\linewidth]{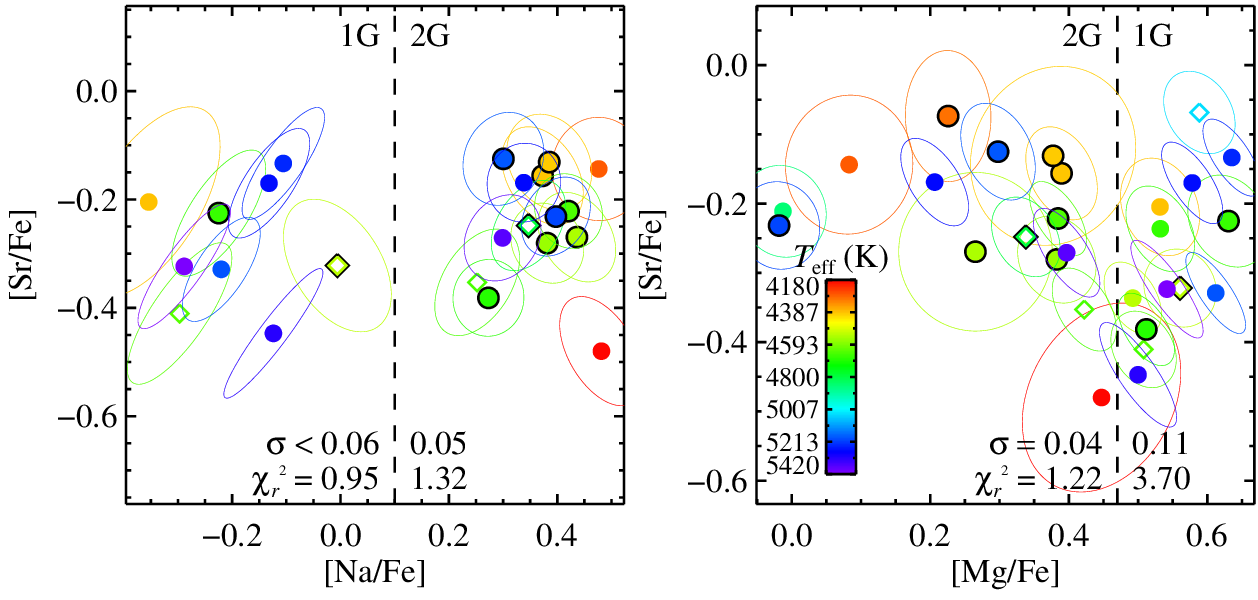}
\includegraphics[width=\linewidth]{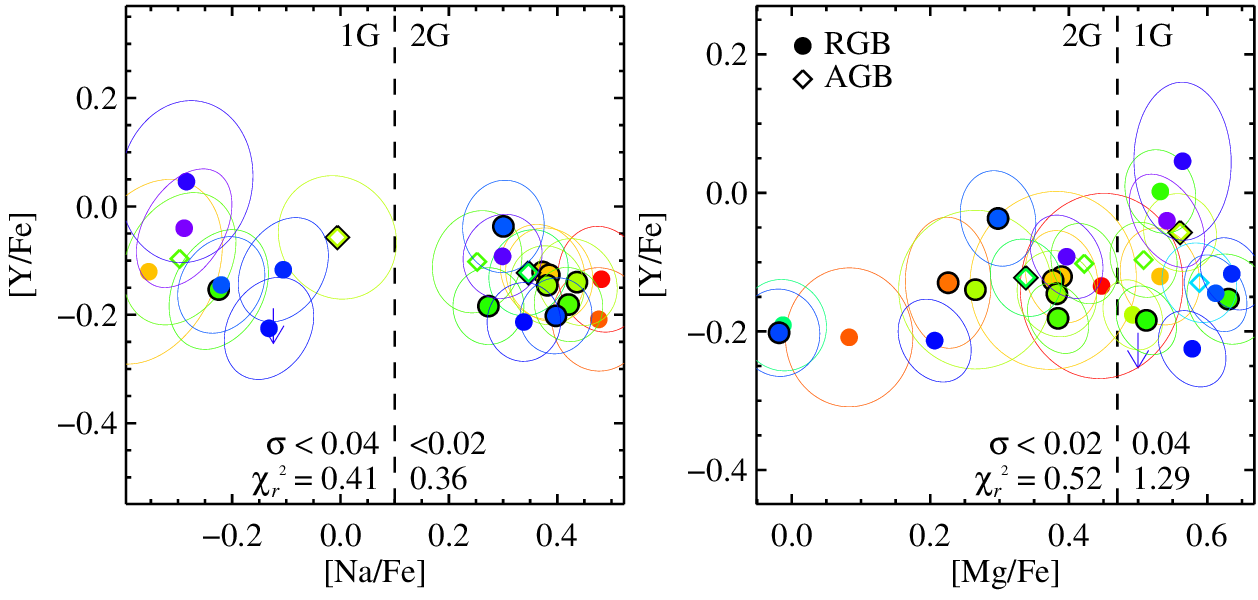}
\includegraphics[width=\linewidth]{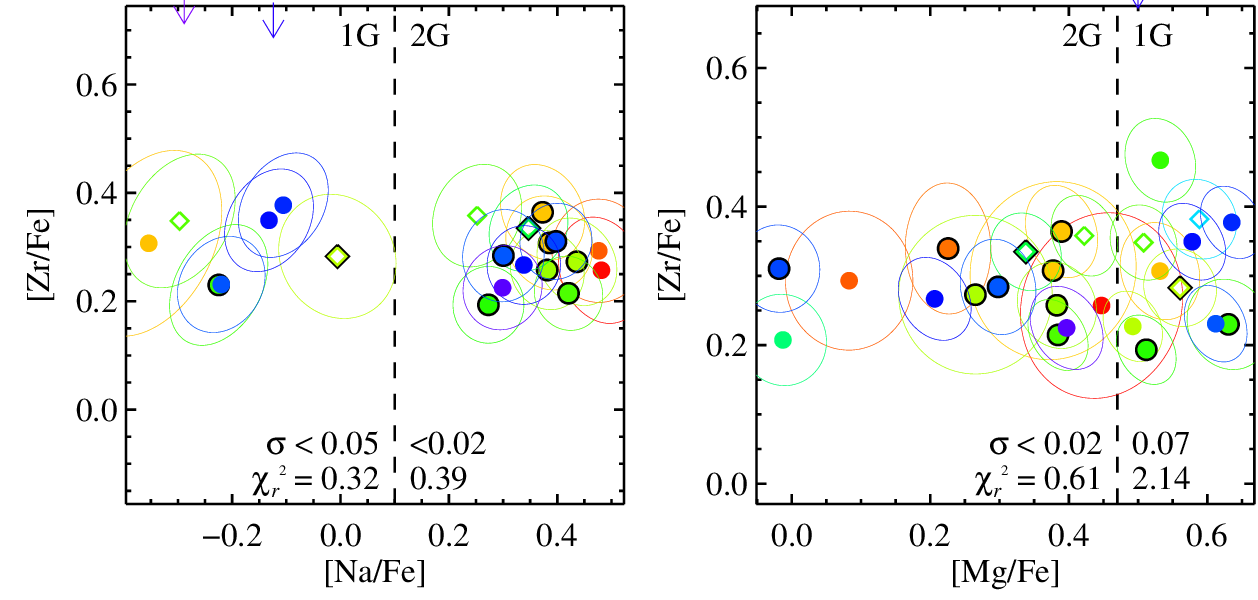}
\caption{First-peak neutron-capture abundances (Sr, Y, and Zr) vs.\ light element abundances (Na and Mg).  The dashed lines indicate the division between first- and second-generation stars (1G and 2G)\@.  Stars included in \citeposs{coh11b} sample are outlined in black, but the abundance measurements are ours.\label{fig:firstpeak}}
\end{figure}

\begin{figure}
\centering
\includegraphics[width=\linewidth]{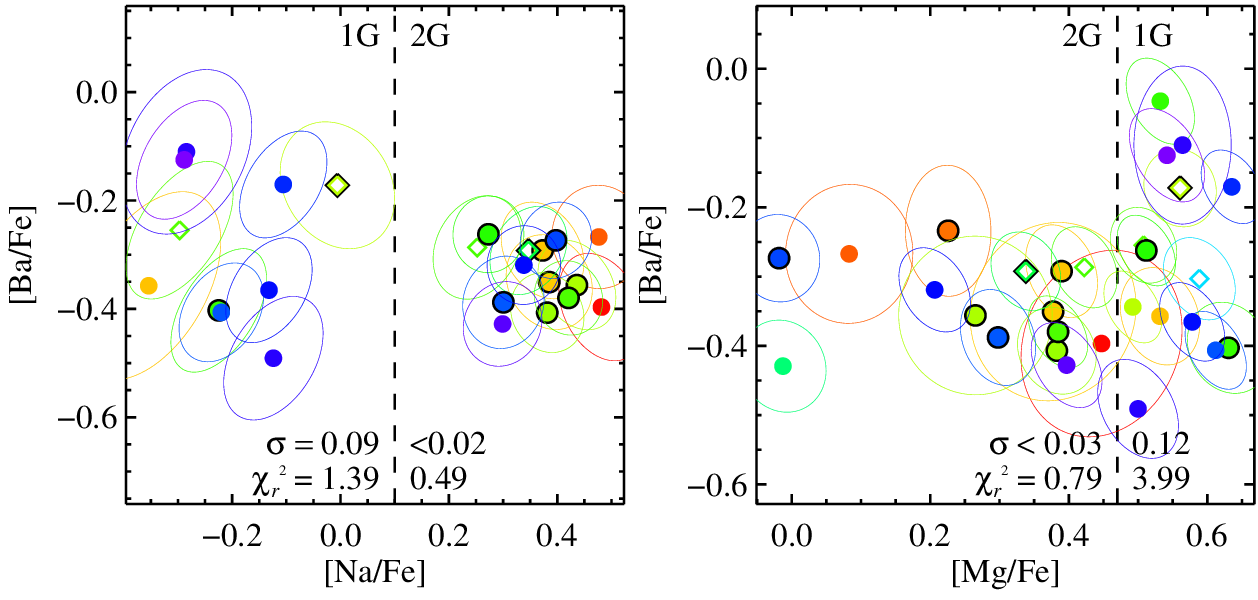}
\includegraphics[width=\linewidth]{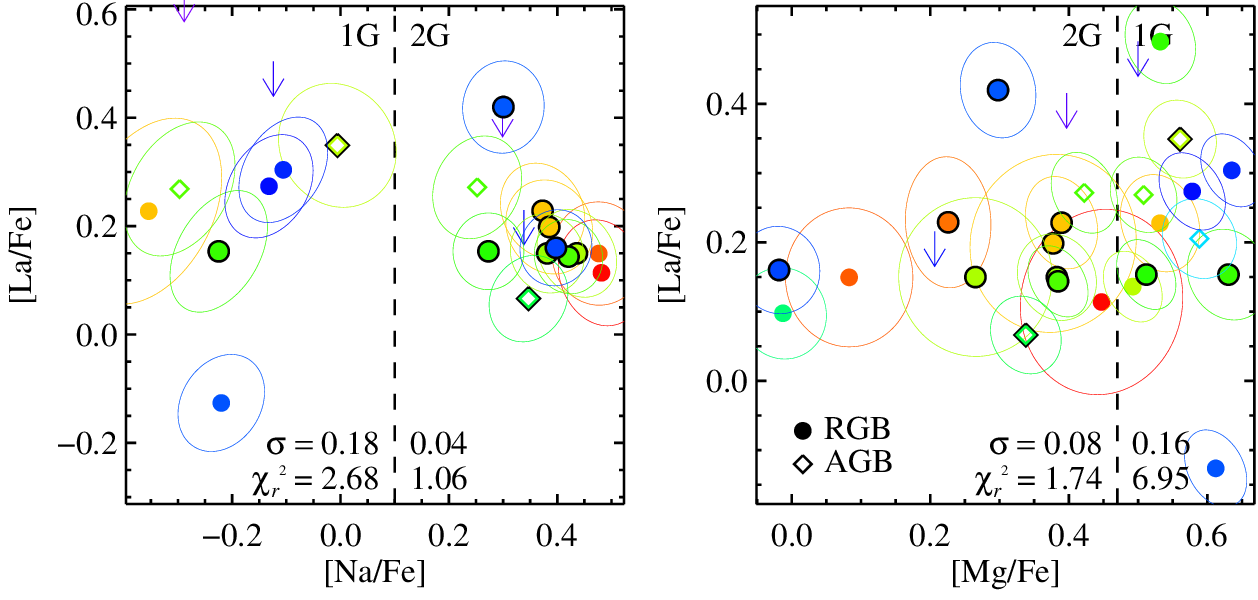}
\includegraphics[width=\linewidth]{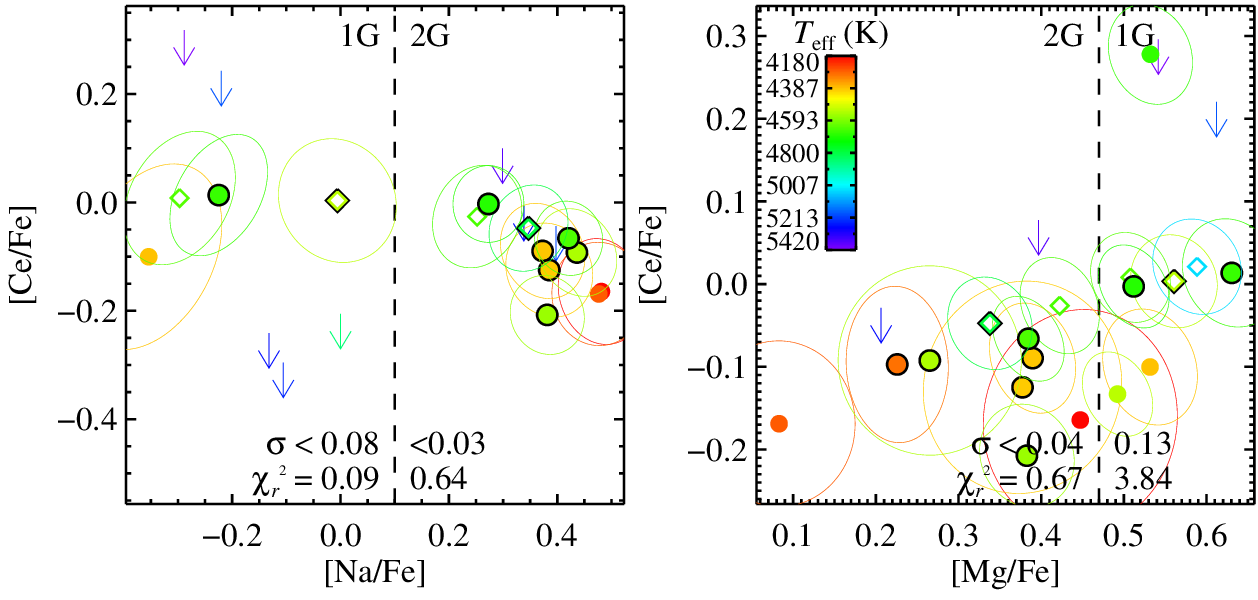}
\caption{Same as Figure~\ref{fig:firstpeak} but for barium and some of the lanthanides (Ba, La, Ce).\label{fig:secondpeaka}}
\end{figure}

\begin{figure}
\centering
\includegraphics[width=\linewidth]{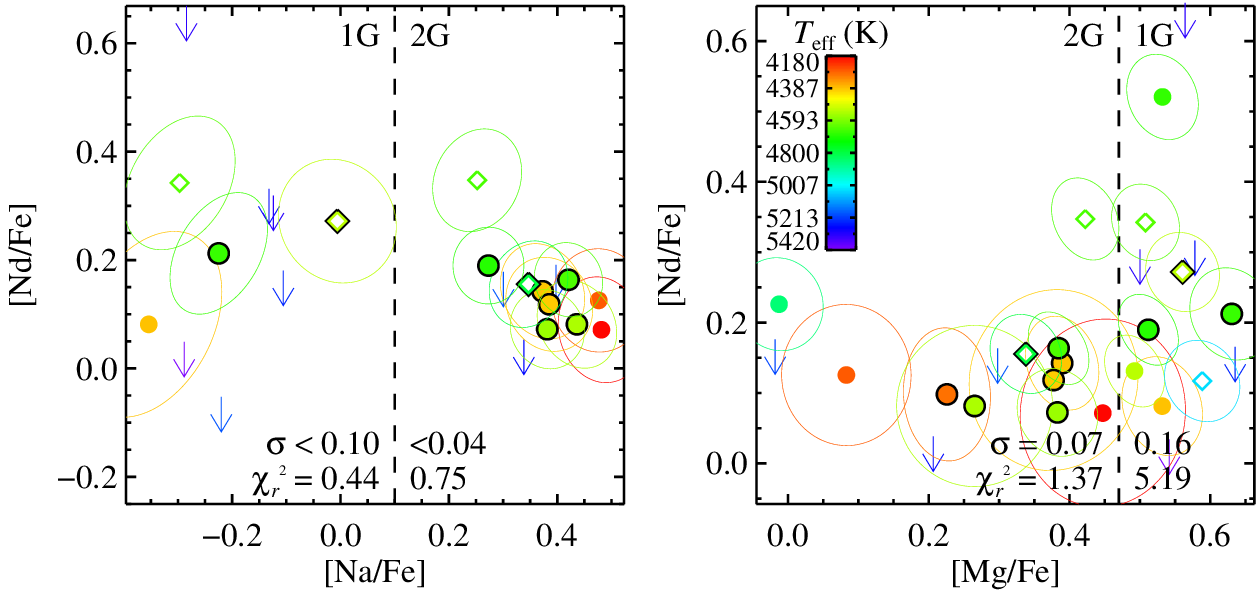}
\includegraphics[width=\linewidth]{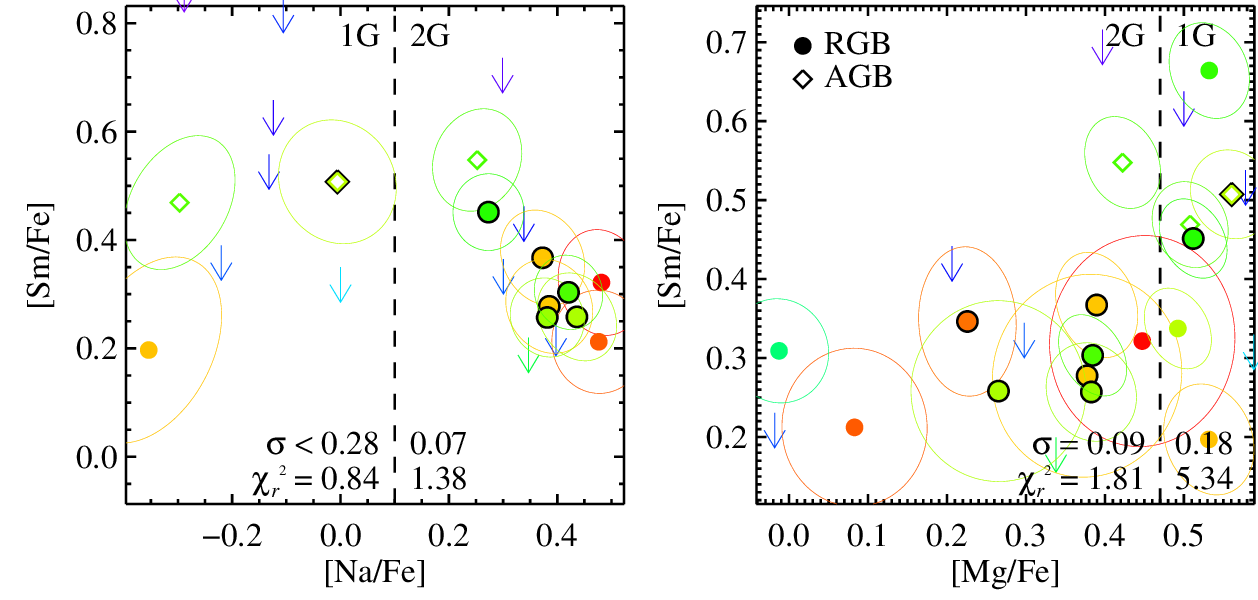}
\includegraphics[width=\linewidth]{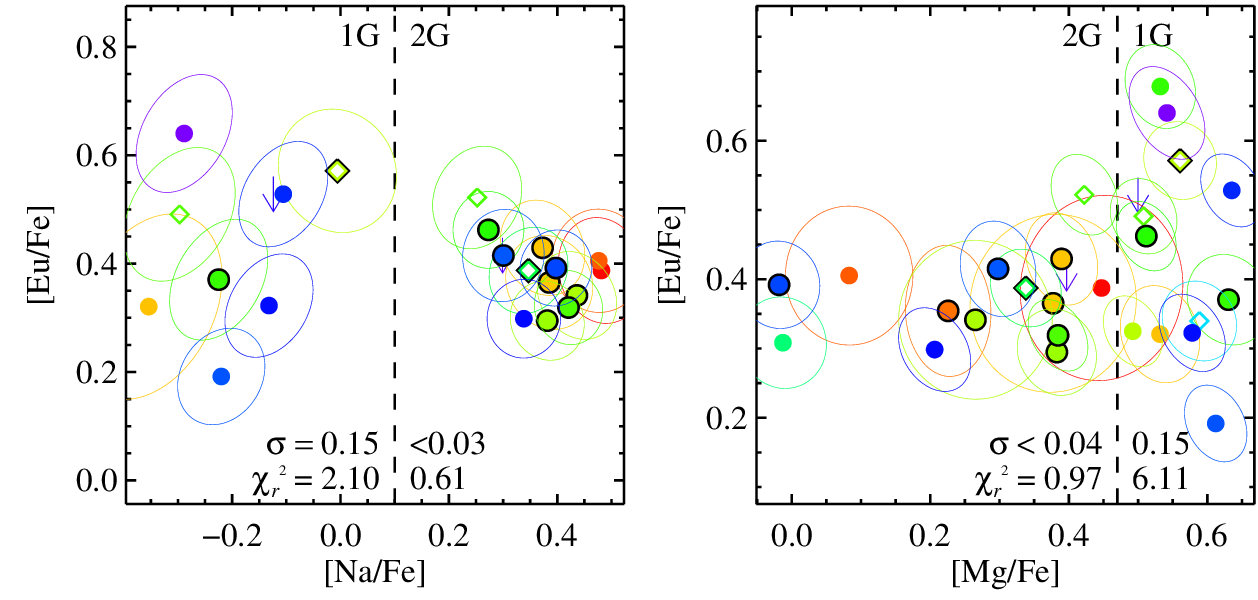}
\includegraphics[width=\linewidth]{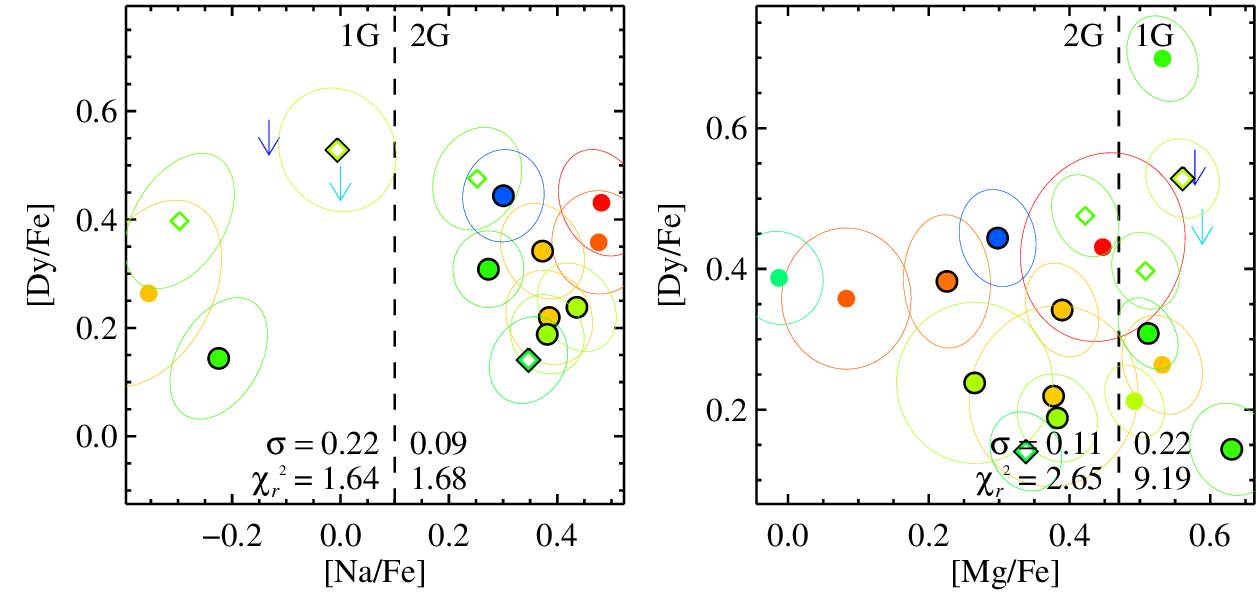}
\caption{Same as Figure~\ref{fig:secondpeaka} but for other lanthanides (Nd, Sm, Eu, and Dy).\label{fig:secondpeakb}}
\end{figure}

It is agreed in the literature that there is no relation between light elements and neutron-capture elements in M15 \citep[e.g.,][]{sne97,sne00a,wor13}.  Figures~\ref{fig:firstpeak}, \ref{fig:secondpeaka}, and \ref{fig:secondpeakb} explore whether such a relation exists in M92.  The left panels show neutron-capture abundances (Sr, Y, Zr, Ba, La, Ce, Nd, Eu, and Dy) vs.\ Na, and the right panel shows the same elements vs.\ Mg.  Like in M15, the average neutron-capture abundance does not trend with the abundances of the light elements.

However, the dispersion in neutron-capture elements is significantly larger in 1G (low Na, high Mg) stars than in 2G (high Na, low Mg) stars.  Table~\ref{tab:dispersions} and each panel of Figures~\ref{fig:firstpeak}--\ref{fig:secondpeakb} give the standard deviation ($\sigma$) and the reduced chi-squared ($\chi_r^2$) for each generation.  Both measurements take into account measurement uncertainties.  The standard deviation is found by maximizing a Gaussian likelihood function:

\begin{equation}
L_i = \frac{1}{\sigma \sqrt{2 \pi}} \exp \left(\frac{\left({\rm [X/Fe]}_i - \langle {\rm [X/Fe]} \rangle\right)^2}{\delta_{{\rm [X/Fe]},i}^2 + \sigma^2} \right)
\end{equation}

\noindent where $\delta_{{\rm [X/Fe]},i}$ is the error on [X/Fe] in star $i$, taking into account covariance between the abundance of X and the abundance of Fe (Section~\ref{sec:errors}).  The total likelihood ($\prod_i L_i$) is maximized by varying the mean $\langle {\rm [X/Fe]} \rangle$ and $\sigma$ through a Monte Carlo Markov chain (MCMC)\@.  The errors reported on $\sigma$ in Table~\ref{tab:dispersions} are the 16$^{\rm th}$ and 84$^{\rm th}$ percentiles of the MCMC\@.  The reduced chi-squared is

\begin{equation}
\chi_r^2 = \sum_i \left(\frac{{\rm [X/Fe]}_i - \langle{\rm [X/Fe]}\rangle}{\delta_{{\rm [X/Fe]},i}}\right)^2
\end{equation}

\noindent where $\langle{\rm [X/Fe]}\rangle$ is the mean of [X/Fe] weighted by $\delta_{\rm [X/Fe]}^{-2}$.  In the cases where $\chi_r^2 < 1$, we report upper limits for $\sigma$.  The separation by [Na/Fe] is more accurate because none of the stars have ambiguous identifications.  On the other hand, more stars have Mg measurements than Na measurements because some spectra did not include the Na~D doublet.  In all cases, $\sigma$ and $\chi_r^2$ for all ten neutron-capture elements are larger for 1G than for 2G when the generations are divided by Mg abundance.  Some elements (Sr, Zr, Ce, Nd, Sm) do not show a larger dispersion in 1G when the generations are divided by Na abundance, but in some of these cases, the dispersion is measured from just a few stars.

There is some evidence that the difference between 1G and 2G is more pronounced for the first-peak $r$-process elements (Sr, Y, Ba) than for Ba and the lanthanides (Ba, La, Ce, Nd, Sm, Eu, and Dy).  When 1G and 2G are divided by Mg abundance, $\chi_r^2$ in 1G exceeds 3.8 for Ba and the lanthanides, but it is less than 3.8 for the first-peak $r$-process elements.  We discuss the implications of this distinction in Section~\ref{sec:peaks}.

\subsubsection{Comparison to Roederer \& Sneden (2011) and Cohen (2011)}

Although we observed a neutron-capture dispersion, our measurement of the dispersion is not as large as reported by \citet{roe11b}.  Using WIYN/Hydra spectra, they found standard deviations of 0.12, 0.17, and 0.23 for [Y/Fe], [La/Fe], and [Eu/Fe], respectively.  Those values are slightly larger than what we observed for 1G and significantly larger than for 2G\@.  It is possible that the smaller dispersion that we report is a result of the higher spectral resolution and generally higher S/N of the HIRES spectra compared to the Hydra spectra.

The Hydra spectra did not include useful absorption lines of Na or Mg.  Instead, \citeauthor{roe11b}\ cross-matched the Hydra spectra with the Na abundances of \citet{sne00b}, derived from high-resolution spectroscopy.  They did not observe the pattern that we observe, wherein the dispersions in [La/Fe] and [Eu/Fe] are larger for stars with higher [Na/Fe].  The reason for this discrepancy is unclear, but the smaller uncertainties in our measurements of $r$-process abundances would make our sample more sensitive to differences in abundance dispersions.

On the other hand, we do observe a dispersion, in contrast with the conclusion of \citet{coh11b}. \citeauthor{coh11b} concluded that the apparent dispersion observed by \citeauthor{roe11b} was a result of lower data quality.  However, \citet{coh11b} did not analyze all of the available HIRES spectra.  Figures~\ref{fig:ncapture_corr}--\ref{fig:secondpeakb} indicate the spectra that were part of her sample.  By design, all 12 of her spectra are part of our sample because they are all in the KOA\@.  Of the 12 stars, two were members of 1G, and 10 were members of 2G\@.  A random sample of stars in M92 would indeed include more 2G stars than 1G stars because 2G is more populous \citep[e.g.,][]{mas19}.  We were able to detect a dispersion because we analyzed a sample that included more 1G stars.  If we were to restrict our sample to the same stars analyzed by \citeauthor{coh11b}, we also would have concluded that there is no significant dispersion in neutron-capture abundances.

\section{Discussion}
\label{sec:discussion}

The most interesting result of this study is the decrease in $r$-process abundance dispersion from the first to the second generation of stars in M92.  This is the first discovery of a relation between the light elements and neutron-capture elements in a GC\@.  As a result, this is also the first linkage between $r$-process evolution and the evolutionary history of a GC\@.

We propose that an $r$-process event occurred shortly before or concurrently with the formation of 1G in M92.  The $r$-process event polluted the cluster gas inhomogeneously at first.  The stars in 1G formed before the gas had time to mix evenly.  The mixing timescale for the gas would be similar to the crossing time ($t_{\rm cross} = R/v$).  If we assume that the typical length $R$ is the scale length \citep[$2\arcmin$ or 5~pc;][]{dru07}, and the relevant velocity $v$ is the velocity dispersion \citep[6.3~km~s$^{-1}$;][]{dru07}, then $t_{\rm cross}$ is 0.8~Myr.  Regardless, 0.8~Myr is shorter than the $\sim 30$~Myr timescale for AGB stars to produce the products of high-temperature hydrogen burning, like enhanced Na and depleted Mg \citep[e.g.,][]{der08,bas18}.  By the time the gas was polluted with 2G material, the $r$-process material was already well mixed.  This explains why there is little dispersion in the $r$-process in 2G\@.

The preceding estimate is based on the current structural parameters of M92, which might have been different in the past.  In fact, the cluster might not have been in dynamical equilibrium during its formation, so a dynamical timescale might not be relevant.  Simulations of cluster formation show that the gas is far from a spherical distribution.  In fact, it is highly asymmetric and filamentary \citep[e.g.,][]{li19,gru21}.  In summary, the timescale based on today's crossing time is more of a convenience rather than a rigorous estimate.  Nonetheless, we will now consider the implications of a $\gtrsim 1$~Myr delay between the start of formation of 1G and the start of formation of 2G\@.

The low-mass stars we observe today in M92 required tens of millions of years (the Kelvin--Helmholtz timescale) to collapse from protostars into main sequence stars.  Therefore, our scenario posits that the protostars in 1G had already begun to collapse, locking in the $r$-process material, before the gas became well-mixed.  The protostars in 2G formed after the gas mixed and homogenized.  However, the low-mass protostars in neither population reached the main sequence until tens of millions of years after this homogenization.

Our proposed scenario requires a very prompt source of the $r$-process if the source was a star that formed concurrently with 1G in M92.  Although neutron star mergers (NSMs) and their associated kilonovae are the only confirmed sources of the $r$-process \citep[e.g.,][]{cho17}, their delay times \citep[30~Myr or longer\footnote{It is possible that some binary neutron stars could merge as soon as 1~Myr after the core collapse supernovae that created them \citep{ben19,saf19}.  However, these NSM candidates are extremely rare.  Furthermore, the relevant timescale is the time from stellar birth to NSM, including the hydrostatic burning lifetime of at least several Myr.  We conclude that a NSM could not be born in a GC and also enrich that GC in less than 0.8~Myr.},][]{kal01} exceed the requirement that the $r$-process be produced on the mixing timescale of $t_{\rm cross} = 0.8$~Myr.  \citet{bek17} and \citet{zev19} independently explored scenarios where a fast-merging NSM enriches a proto-GC\@.  Both models result in 2G being preferentially enriched in the $r$-process relative to 1G, which is inconsistent with our observations.

Instead, a short delay time between 1G and 2G could imply that the $r$-process source is a massive star that formed concurrently with 1G\@.  Some proposed mechanisms include magnetorotational supernovae \citep{nis15} and collapsars \citep{sie19}.

On the other hand, the $r$-process source could have been a star born before M92 was formed.  Our scenario requires that the star \textit{exploded} during or shortly before the formation of 1G\@.  However, the precursor star could have been \textit{born} shortly after the Big Bang.  In principle, the progenitor to a NSM could have been born about 100~Myr after the Big Bang and then exploded approximately 1~Gyr later, when M92 formed.

\citet{tar21} proposed an alternative scenario for M15, wherein an $r$-process event occurred near but external to the GC\@.  The cluster's natal gas cloud was inhomogeneously polluted with $r$-process material, as it would be if the event happened inside the cluster.  However, the external $r$-process material could continue to fall onto the cluster for the duration of star formation in 1G and 2G\@.  The extended pollution time is necessary to explain why the $r$-process dispersion persists in both stellar generations in M15.  However, that requirement is not necessary in M92, where only 1G shows the dispersion.

One problem with making the $r$-process with massive stars is that massive stars also synthesize Fe \citep{mac19}.  Any inhomogeneity in the $r$-process should be matched by an inhomogeneity in Fe.  However, the inhomogeneity in Fe will be washed out by the many other core collapse supernovae in M92 that also produced Fe.\footnote{M92 has a current stellar mass of $2.7 \times 10^5~M_{\sun}$ \citep{bau20}.  Corrected for tidal stripping, its initial mass was about $1.2 \times 10^6~M_{\sun}$ \citep{bau19}.  For a \citet{kro01} initial mass function, 1\% of stars explode as core collapse supernovae. 
 Therefore, M92 experienced about $10^4$ core collapse supernovae.}  The $r$-process event would have been rare.  In fact, there might have been just a single event.\footnote{One event per $10^6$ stars is slightly lower than the frequency of long gamma ray bursts and one-tenth the frequency of $r$-process events required to explain the scatter of $r$-process abundances in metal-poor stars \citep{bra21}.}  We observe a standard deviation in [\ion{Fe}{2}/H] of 0.05.  The standard deviation of [Eu/Fe] in 1G is 0.15.  If the $r$-process event inhomogeneously polluted the cluster with $0.1~M_{\sun}$ of Fe such that the dispersion in [Fe/H] was 0.15, only two more events that evenly polluted the cluster with $0.1~M_{\sun}$ of Fe each would be required to reduce the dispersion to 0.05.  Naturally, individual supernovae would not evenly pollute the cluster on timescales less than $t_{\rm cross}$.  However, there would be $\sim 10^4$ core collapse supernovae that produced Fe but not $r$-process.  The Fe inhomogeneities would be averaged out, whereas the inhomogeneity of the rare $r$-process event would persist for at least $t_{\rm cross}$.

M15 and M92 are the GCs that display the most obvious dispersions of $r$-process abundances.  They are also the most metal-poor ``classical'' GCs.  The association of low metallicity and $r$-process dispersion is probably not coincidental.  Consider M5, which has 13 times the iron abundance of M92 \citep{har96}.  If M5 experienced the same type and frequency of $r$-process events as M92, then those events would contribute the same mass of $r$-process ejecta to M5's natal gas.  The $r$-process abundance of the gas that forms 1G in M5 would be $[(r_{\rm new} + r_{\rm natal}))/{\rm Fe}_{\rm natal}]$.  However, the natal abundance of Fe in M5 would be 13 times the value of M92.  As a result, the $r$-process dispersion in M5's 1G would be 13 times smaller.  Such a dispersion would be undetectable.  Therefore, the abundances of metal-poor GCs would more obviously show the effect of rare events.

\subsection{First-Peak $r$-process Compared to Lanthanides}
\label{sec:peaks}

Sr, Y, and Zr do not follow the same abundance pattern as the heavier, ``main'' $r$-process in metal-poor stars.  As a result, \citet{tra04} posited the existence of a lighter element primary process (LEPP).  One candidate for the LEPP is the ``limited'' (or ``weak'') $r$-process that occurs in neutrino-driven winds from proto-neutron stars formed during core collapse supernovae \citep{fro06,pru06,wan06,wan13,arc13}.  On the other hand, the main $r$-process requires higher neutron densities, which might be found in neutron star mergers \citep{lat74}, jet-driven, magnetorotational supernovae \citep{nis15}, or supernovae from massive, rapidly rotating stars \citep[collapsars,][]{sie19}.  The prevalence and existence of each of these sites is hotly debated.  Nonetheless, it is clear that the first-peak $r$-process has multiple formation channels.

We found that the first-peak $r$-process elements exhibit a smaller distinction between 1G and 2G than Ba and the lanthanides (see Table~\ref{tab:dispersions} and Figures~\ref{fig:firstpeak}--\ref{fig:secondpeakb}).  One possibility is that the limited $r$-process site is common, whereas the main $r$-process site is rare.  For example, consider that commonplace, low-mass core collapse supernovae could synthesize Sr, Y, and Zr.  The large number ($\sim 10^4$) of these events during the early formation of M92 would cause the gas to converge on a single abundance of these lighter $r$-process elements.  On the other hand, if Ba and the lanthanides were created by few events---perhaps a single event---then the gas would not converge on a single abundance of these heavier elements until it was evenly mixed by hydrodynamic processes.

There is copious evidence that the main $r$-process source is rare and prolific \citep[e.g.,][]{ji16,brauer19}.  In other words, the events must happen infrequently but must produce a great deal of $r$-process elements.  The large dispersion of the main $r$-process and the smaller dispersion of the limited $r$-process in M92 add further support to the rare and prolific nature of the main $r$-process.  Neutron star mergers, magnetorotational supernovae, and collapsars are all rare and prolific producers of the main $r$-process.

The Milky Way also contains some evidence that the limited $r$-process happens more frequently and earlier in the Galaxy's history than the main $r$-process.  For example, \citet{hol20} found that limited-$r$ stars (those with ${\rm [Sr/Ba]} > 0.5$) are more prevalent at lower metallicities.  One explanation is the main $r$-process sources---those that produce Ba---are less frequent.  Stars that formed at low metallicities sampled the ejecta of fewer events.  If the limited $r$-process is more common than the main $r$-process, then some metal-poor stars could be enriched in Sr but poor in Ba.

\section{Summary}
\label{sec:summary}

We measured detailed abundances of \nstars\ stars in M92 from archival Keck/HIRES spectroscopy.  We stacked all available spectra to achieve the maximum S/N possible.  Our analysis takes into account covariance between stellar parameters, like temperature and surface gravity.  We made the following observations:

\begin{itemize}
\item M92 shows typical light-element abundance variations, like the Na--Mg anti-correlation.  The cluster has a clear separation into first and second generations, which is especially apparent in the Na--Mg diagram.

\item M92 does not display an obvious Mg--K anti-correlation, such as the one observed in NGC~2419.

\item The neutron-capture abundance pattern in M92 is consistent with the $r$-process, with the exception that La appears overabundant relative to the solar-system $r$-process pattern.  The same phenomenon has previously been observed in M15.

\item We affirm that M92 has a dispersion in $r$-process abundances.  The existence of a dispersion was previously reported by \citet{roe11b}.

\item The dispersion in the $r$-process is limited to the first generation of stars (low Na, high Mg).  So far, M92 is unique in showing any relation between light element and neutron-capture abundances.

\item The dispersion is smaller for Sr, Y, and Zr than for Ba and the lanthanides.  We posited that the higher frequency of limited $r$-process events compared to main $r$-process events explains the difference in dispersions.
\end{itemize}

We proposed a scenario wherein a source of the main $r$-process polluted M92 as the first stars began to form.  The first generation formed faster than a crossing time, which caused the $r$-process abundances to be inhomogeneous.  The gas homogenized by the time the second generation formed, resulting in a negligible dispersion in $r$-process among the second generation.  From rough estimates of the current crossing time in M92, we conclude that the formation starting times of the first and second generations in M92 were separated by at least 0.8~Myr.  Although this timescale is shorter than other relevant timescales, like the lifetime of a massive AGB star, it could constrain GC formation theories where the two populations form simultaneously (e.g., early disk accretion, \citealt{bas13}) or nearly simultaneously (e.g., very massive stars, \citealt{gie18}).

Our archival spectroscopic sample was not designed to measure abundance variations.  The stars were selected from the full magnitude range of the RGB and even the main sequence turn-off.  The ideal sample of stars to detect abundance dispersions would span a narrow range of temperature and surface gravity (i.e., color and magnitude).  In such a sample, absorption line strengths would correspond almost directly to abundance.  We have already begun to acquire such samples in M15 and M92.  These samples will provide more precise quantifications of dispersions among the $r$-process abundances and the relation between the $r$-process and lighter elements.

\begin{acknowledgments}

We are very grateful to Andrew Howard, Howard Isaacson, and the California Planet Search consortium for observing star X-20 with HIRES\@.  We thank Eric Bell, Ivanna Escala, Oleg Gnedin, Keith Hawkins, J.\ Chris Howk, Rebecca Surman, Ralph Wijers, and especially Ian Roederer for insightful conversations.  This research has made use of the Keck Observatory Archive (KOA), which is operated by the W. M. Keck Observatory and the NASA Exoplanet Science Institute (NExScI), under contract with the National Aeronautics and Space Administration.  This publication makes use of data products from the Two Micron All Sky Survey, which is a joint project of the University of Massachusetts and the Infrared Processing and Analysis Center/California Institute of Technology, funded by the National Aeronautics and Space Administration and the National Science Foundation (NSF)\@.  This research has made use of the SIMBAD database, operated at CDS, Strasbourg, France.  E.N.K.\ acknowledges support from NSF CAREER grant AST-2233781.  A.P.J.\ acknowledges support from the U.S. National Science Foundation (NSF) grant AST-2206264.  This work was performed in part at the Aspen Center for Physics, which is supported by NSF grant PHY-1607611.

\end{acknowledgments}

\facility{Keck:I (HIRES)}
\software{\moog\ \citep{sne73,sne12}, \atlas\ \citep{kur93}, \texttt{XIDL}, \mpfit\ \citep{mar12}, \leopy\ \citep{fel19}, \linemake\ \citep{pla21a,pla21b}}

\bibliography{m92}
\bibliographystyle{apj}

\end{document}